\documentclass[12pt]{article}
\usepackage{amsmath, amssymb}

\usepackage{epsfig}

\newtheorem{theorem}{Theorem}[section]
\newtheorem{proposition}[theorem]{Proposition}
\newtheorem{corollary}[theorem]{Corollary}
\newtheorem{lemma}[theorem]{Lemma}
\newtheorem{definition}[theorem]{Definition}

\newcommand{\rd}{{\rm d}}
\newcommand{\be}{\begin{equation}}
\newcommand{\ee}{\end{equation}}
\newcommand{\bey}{\begin{eqnarray}}
\newcommand{\eey}{\end{eqnarray}}

\newcommand{\bef}{\begin{figure}}
\newcommand{\eef}{\end{figure}}
\newcommand{\bec}{\begin{center}}
\newcommand{\eec}{\end{center}}

\newcommand{\tri}{| \! |\!|}

\newcommand{\eqn}{\begin{eqnarray}}
\newcommand{\eeqn}{\end{eqnarray}}

\newcommand{\bP}{{\bf P}}

\newcommand{\bA}{{\bf A}}

\newcommand{\bw}{{\bf w}}
\newcommand{\bp}{{\bf p}}
\newcommand{\bq}{{\bf q}}

\newcommand{\bv}{{\bf v}}
\newcommand{\bu}{{\bf u}}

\newcommand{\Tor}{{\bf T}}

\newcommand{\tp}{{\tilde p}}

\newcommand{\tu}{{\tilde u}}

\newcommand{\tsi}{{\tilde\sigma}}

\newcommand{\tgamma}{{\tilde\gamma}}
\newcommand{\tGamma}{{\tilde\Gamma}}

\newcommand{\tbp}{{\tilde{\bf p}}}

\newcommand{\tbu}{{\tilde{\bf  u}}}

\renewcommand{\a}{\alpha}

\newcommand{\e}{\varepsilon}
\newcommand{\s}{\sigma}

\newcommand{\om}{{\omega}}

\newcommand{\bE}{{\bf E}}
\newcommand{\cU}{{\cal U}}
\newcommand{\cM}{{\cal M}}
\newcommand{\cX}{{\cal X}}
\renewcommand{\iint}{\int \!\! \int}
\newcommand{\bR}{{\mathbb R}}
\newcommand{\bC}{{\bf C}}
\newcommand{\bN}{{\bf N}}
\newcommand{\bZ}{{\mathbb Z}}

\newcommand{\wt}{\widetilde}
\newcommand{\wh}{\widehat}
\newcommand{\ov}{\overline}
\newcommand{\balpha}{\mbox{\boldmath $\alpha$}}

\newcommand{\cI}{{\cal I}}

\newcommand{\cG}{{\cal G}}

\newcommand{\cA}{{\cal A}}
\newcommand{\cB}{{\cal B}}
\newcommand{\cE}{{\cal E}}
\newcommand{\cP}{{\cal P}}

\newcommand{\cV}{{\cal V}}

\newcommand{\cL}{{\cal L}}
\newcommand{\cN}{{\cal N}}
\newcommand{\cO}{{\cal O}}
\newcommand{\cT}{{\cal T}}

\newcommand{\Om}{\Omega}

\newcommand{\de}{ \mbox{\rm \scriptsize deg}}
\newcommand{\dee}{ \mbox{\rm \tiny deg}}
\input epsf

\def\fS{{\frak S}}
\def\fb{{\frak b}}

\date{Apr 5, 2007}

\begin{document}

\title{Quantum diffusion of the random Schr\"odinger evolution 
\\
in the scaling limit }
\author{L\'aszl\'o Erd\H os${}^1$\thanks{Partially
supported by NSF grant DMS-0200235 and EU-IHP Network 
``Analysis and Quantum'' HPRN-CT-2002-0027.}
\\ Manfred  Salmhofer${}^2$
\\ Horng-Tzer Yau${}^3$\thanks{Partially supported by NSF grant
DMS-0307295 and MacArthur Fellowship.} \\
\\
\small
${}^1\;$Institute of Mathematics, University of Munich, 
\\
\small
Theresienstr. 39, D-80333 Munich, Germany
\\
\small
${}^2\;$Max--Planck Institute for Mathematics, Inselstr.\  22, 04103 Leipzig,
and
\\ 
\small
Theoretical Physics, University of Leipzig, Postfach 100920, 04009 Leipzig, 
Germany
\\
\small
${}^3\;$Department of Mathematics, Harvard University, Cambridge
MA-02138, USA \\ }

\maketitle

\abstract{ We consider  random Schr\"odinger equations
on $\bR^d$ for $d\ge 3$
with a homogeneous Anderson-Poisson type
random potential.
Denote by $\lambda$ the coupling constant  and $\psi_t$ the solution
with initial data $\psi_0$. The space and time variables
scale as $x\sim \lambda^{-2 -\kappa/2}, t \sim \lambda^{-2 -\kappa}$
with $0< \kappa < \kappa_0(d)$.
We prove that, in the limit $\lambda \to 0$,
the expectation of the Wigner distribution of $\psi_t$
converges weakly to the
solution of a heat equation
in the space variable $x$ for arbitrary $L^2$  initial data.

The proof is based on analyzing  the phase cancellations
of multiple scatterings on  the random potential by expanding
the propagator into a sum of
Feynman graphs.
In this paper we consider the non-recollision graphs and
prove  that  the amplitude  of the {\it non-ladder} diagrams
is smaller than their ``naive size" 
by an extra $\lambda^c$ factor {\em
per non-(anti)ladder vertex} for some $c > 0$.
This is the first rigorous result showing that
the improvement over the naive estimates on the  Feynman graphs
grows as a power of the small parameter with the exponent
depending linearly on the number of vertices. This estimate
allows us to prove the convergence of the perturbation series. }


\section{Introduction}

The fundamental equations governing the basic laws of physics,
the Newton and the Schr\"odinger
equations,  are
time reversible and have no dissipation.
It is remarkable that dissipation is nevertheless ubiquitous
in nature,
so that almost all macroscopic phenomenological equations
are dissipative.
The oldest such example is perhaps the
equation of heat conductance found by
Fourier.

On a microscopic level, Brown observed almost two centuries ago
that the motion of a pollen suspended
in water was erratic \cite{Br}. This led to the kinetic explanation by
Einstein in 1905 \cite{Ei} that such a motion was created by the
constant ``kicks" on the relatively heavy pollen  by the
light water molecules. It should be noted that
at that time even the atomic-molecular structure of
matter was not universally accepted.
Einstein's theory was strongly supported by
Boltzmann's kinetic theory, which, however,
was phenomenological and seriously debated at the time.
Finally in 1908 Perrin \cite{Per} experimentally
verified Einstein's theory and used it, among others,  to
give a  precise estimate on the Avogadro number.
These experiments gave the strongest evidence
for atoms and molecules at that time.

In Einstein's kinetic theory
both the  heavy particle (the pollen) and the light particles
(the water molecules) obey Newton's law.
Therefore, Einstein's kinetic theory in fact postulated
the emergence of the Brownian motion from a classical non-dissipative
reversible dynamics.
The key difficulty of a mathematically
rigorous derivation of Brownian motion from reversible dynamics
is similar to the justification of Boltzmann's
molecular chaos assumption (Sto{\ss}zahlansatz); the
dissipative character emerges only in a scaling limit,
as the number of degrees of freedom
goes to infinity.

The first mathematical definition of the Brownian motion was
given in 1923  by Wiener, who  constructed the Brownian motion
as a scaling limit of random walks. This construction
was built upon a stochastic microscopic dynamics
which by itself is dissipative.
The derivation of the Brownian motion
from a time reversible Hamiltonian system, however,
was not seriously considered until more than half a century later.
Kesten-Papanicolaou  \cite{KP} in 1978 proved that the velocity distribution
of a particle moving in an environment consisting of
random scatterers (i.e., Lorenz gas with random scatterers)
converges to a Brownian motion
 in a weak coupling limit for $d\ge 3$.
In this model the bath of light particles is
replaced with  random static impurities.
The same result was obtained later in $d=2$ dimensions by D\"urr, Goldstein and
Lebowitz \cite{DGL2}.  In a  recent work \cite{KR}, Komorowski and Ryzhik
have controlled the same evolution on a longer time scale and proved
the convergence to Brownian motion of the position process as well.
Bunimovich and Sinai \cite{BS} proved the convergence of the
periodic Lorenz gas with a hard core interaction to a Brownian motion
in 1980. In this model the only source of randomness is the distribution
of the initial condition.
Finally, D\"urr, Goldstein and Lebowitz \cite{DGL1} proved that
the velocity process of a
heavy particle in a light ideal gas, which is
a model with a dynamical environment, converges to the
Ornstein-Uhlenbeck process.

Wiener's construction of Brownian motion is based on a random walk.
 The random walk could easily be replaced by the Markovian process
generated by a linear Boltzmann
equation.  The linear Boltzmann equation was rigorously derived from
the classical Lorenz gas
by Gallavotti \cite{G}, Spohn \cite{Sp2} and Boldrighini, 
Bunimovich and Sinai \cite{BBS}.
(The nonlinear  Boltzmann equation was derived by Lanford 
\cite{L} for short time.)
Although Brownian motion was discovered and theorized in the context of
classical dynamics,  we shall prove that it also describes the motion of
a quantum particle in a random environment.

\bigskip

The random Schr\"odinger equation, or the quantum Lorentz
model,  is given  by the evolution equation:
\be\label{sch}
i \partial_t \psi_t(x)=  H \psi_t(x), \qquad
 H=H_\om  = -  \frac 1 2 \Delta_x +  \lambda V_\om(x)\; ,
\ee
where $\lambda$ is the coupling constant and $V_\om$
is the random potential. The first scale with a non-trivial limiting
dynamics is the weak coupling limit, $\lambda\to0$, where the space, time
and the coupling constant
are subject to the kinetic scaling:
\be\label{wc}
t \to t \e^{-1}, \quad x \to x \e^{-1}, \quad \lambda= \sqrt \e \; .
\ee
Under this limit, the appropriately
rescaled  Wigner distribution (see \eqref{wig})
of the solution to
the Schr\"odinger evolution \eqref{sch} converges weakly to a linear
 Boltzmann equation. This was first established by Spohn \cite{Sp1}
for a random potential with Gaussian distribution and
small macroscopic time.
This method was extended to  study higher order correlations
in  \cite{HLW}.
A different method (applicable to the lattice setting
and general random potential, see remarks later on) was developed
in \cite{EY}
where the short time restriction was removed. This method was also extended
to the phonon case \cite{E} and  Lukkarinen and Spohn \cite{LS}
have employed a similar technique for studying
the energy transport in a harmonic crystal with  weakly
perturbed random masses.

Since the long time limit of a Boltzmann equation is
a heat equation, we shall take a time scale  longer (see
\eqref{scale})
 than in the  kinetic scaling limit \eqref{wc}.
Our aim is to prove that the limiting dynamics of the Schr\"odinger evolution
in a random potential under this scaling is governed by a heat equation.
This requires to control the Schr\"odinger dynamics
up to a time scale $\lambda^{-2-\kappa}$, $\kappa>0$.
Quantum correlations that are small
on the kinetic scale and are neglected in the first limit
may contribute on the longer  time scale. The derivation
of the heat equation
 is  thus much more difficult than first deriving the Boltzmann equation
from Schr\"odinger dynamics on the kinetic scale and then showing
that Boltzmann equation converges to a diffusive equation under
a different limiting procedure. Notice that the limit in our approach
is a long time scaling limit which involves  {\it no
semiclassical limit.}

The approach of this paper also applies  to lattice models
and yields a derivation of  Brownian motion from the Anderson model 
\cite{ESY1, ESY2}  and in fact, we present our main technical
steps in a unified framework.
The dynamics of the Anderson model was postulated by Anderson
to be localized  for large coupling constant
$\lambda$ and  extended for small coupling constant
(away from the band edges
and in dimension $d \ge 3$).
The localization conjecture
was first established rigorously by
Goldsheid, Molchanov and Pastur \cite{GMP} in
one dimension, by Fr\"ohlich-Spencer \cite{FS},
and later by Aizenman-Molchanov \cite{AM} in several
dimensions, and many other works have since contributed to this field.
The progress for the extended state conjecture, however, has  been limited. 
It was proved by Klein \cite{Kl}  that
all eigenfuctions are extended on the Bethe lattice (see also 
\cite{ASW, FHS}). In Euclidean space,
Schlag, Shubin  and Wolff \cite{SSW}
proved that the eigenfunctions cannot be localized in a region smaller
than $\lambda^{-2+ \delta}$ for some $\delta > 0$ in $d=2$. Chen \cite{Ch},
partly based on the method \cite{EY}, extended this result to all
dimensions $d\ge 2$ and $\delta=0$ (with logarithmic corrections). Extended states 
for Schr\"odinger equation with a sufficiently decaying random potential were 
proven by Rodnianski and Schlag \cite{RS} and Bourgain \cite{B}  (see also  \cite{D}).

In summary, all known results  for the Anderson
model (or its modifications) in Euclidean space
are  in regions
where the dynamics
have either no effective collision
 or there are typically
 only finitely many of them.
Under the  diffusive scaling of this paper, see (\ref{scale}), 
the number of effective scatterings is a negative fractional power of the
scaling parameter.  In particular, it goes to infinity in the scaling limit,
 as it should be the case if we aim to obtain a Brownian motion.
As in \cite{Ch}, one may derive from our  dynamical result that
the eigenfunctions cannot be localized
in a region smaller than $\lambda^{-2-\kappa/2}$
and dimension $d\ge 3$.

\medskip

{\it Acknowledgement.} The authors are grateful
for the financial support and hospitality of
 the Erwin Schr\"odiger Institut, Vienna, Max Planck Institut, Leipzig,
Stanford University and Harvard University, where part of this
work has been done.  We thank Jani Lukkarinen for 
pointing out a simplification of Lemma \ref{lemma:con}
for the continuum model and we also thank 
the referee for his comments and suggestions.

\section{Statement of the main result}\label{sec:mainresult}
\setcounter{equation}{0}

\subsection{Notations}

Let
\be\label{H}
        H := -\frac{\hbar^2 }{2m}\Delta + \lambda V
\ee
denote a random Schr\"odinger operator acting on $L^2(\bR^d)$, $d\ge 3$,
with a random potential $V=V_\om(x)$ and 
 a small positive coupling constant $\lambda$. We shall choose 
the units so that 
$  h=1$,  $m=1$ so that $\hbar^2/(2m) = [2 (2\pi)^2]^{-1}$. 
The potential is given by
\be
   V_\om (x): = \int_{\bR^d} B(x-y)\rd \mu_\om(y)\; ,
\label{Vom}
\ee
where $B$ is a single site potential profile and
$\mu_\om$ is a Poisson point process on $\bR^d$ with
homogeneous unit density and with independent, identically
distributed random masses. More precisely, for
almost all realizations $\om$ consists of a countable, locally 
finite collection of points, $\{ y_\gamma(\om)\in\bR^d \; : \; 
\gamma=1,2, \ldots \}$,
and random weights $\{ v_\gamma(\om)\in \bR
 \; : \; \gamma=1,2,\ldots \}$ such that
the random measure is given by
\be
      \mu_\om = \sum_{\gamma=1}^\infty v_\gamma(\om) \delta_{y_\gamma(\om)}\; ,
\label{def:muom}
\ee
where $\delta_y$ denotes the Dirac mass at $y\in\bR^d$.
The Poisson process $\{ y_\gamma(\om)\}$ is independent of
the weights $\{ v_\gamma(\om)\}$.  The weights are real
i.i.d. random variables with distribution $\bP_v$
and with moments   $m_k:=\bE_v \, v_\gamma^{k}$ satisfying
\be
m_2=1,\;\;  m_{2d}<\infty,\;\; \;\;
m_1=m_3=m_5=0\, .
\label{momm}
\ee
The expectation with respect to the random process $\{ y_\gamma, v_\gamma\}$
is denoted by $\bE$.

For the single-site potential, we assume that $B$
 is a spherically symmetric Schwarz function
with $0$ in the support of its Fourier transform, i.e.
\be  
   0\in  \mbox{supp}\,( \wh B)\; .
\label{nonzero}
\ee
More precisely, we introduce the norm
$$
   \| f \|_{m,n} := \sum_{|\alpha|\leq n}
\| \langle x \rangle^m  \partial^\alpha f(x)\|_\infty
$$
with $\langle x \rangle : = (1+x^2)^{1/2}$ (here $\alpha$ is a 
multiindex) and we assume
\be
  \| B \|_{k,k} <  C_k \qquad \forall k\in \bN\; .
\label{Bcondi}
\ee
Actually, it is sufficient to assume
 \eqref{Bcondi} for all $k\leq k_0(\kappa)$.

We note that the operator $H_\om$ is not bounded from below
due to the possible large concentration of Poisson points
in some region. Nevertheless, $H_\om$ is self-adjoint
under very general conditions, see \cite{KM}.

We introduce a few notational conventions.
The letters   $x,y,z$ will always denote configuration space variables, while
$p, q, r, u, v, w$ will be reserved for $d$-dimensional momentum
variables. 
The norm without indices, $\| \cdot \|$, will
always denote the standard $L^2(\bR^d)$ norm. The bracket
$(\cdot \, , \cdot)$ denotes the standard scalar product on $L^2(\bR^d)$
and $\langle \cdot \, , \cdot \rangle$ will denote the pairing
between the Schwarz space and its dual 
 on the phase space $\bR^d\times \bR^d$.

Integrals without explicit domains will
always denote integration over $\bR^d$ with respect to the
Lebesgue measure. For any $f\in L^2(\bR^d)$,
the  Fourier transform is given by
\be
    \wh f(p) := \int
  e^{-2\pi ip\cdot x} f(x) \rd x \; , \qquad p\in \bR^d \; .
\label{FF}
\ee
and the inverse Fourier
transform is given by
$$
     g(x)
    =  \int   \wh g(p) e^{ 2\pi i p\cdot x}
    \rd p \; , \qquad x\in \bR^d \; .
$$
For functions defined on the phase space, $f(x,v)$, the
Fourier transform will always be taken only in the space variable, i.e.
$$
   \wh f(\xi, v) :=  \int e^{-2\pi i\xi\cdot x} f(x,v) \rd x \; ,
  \qquad \xi\in \bR^d \; .
$$

The Fourier transform of the kinetic energy operator  is given by
$$
\widehat{\big[-\frac{\hbar^2}{2 m}\Delta f \big]}(p)= 
 \widehat{\big[-\frac{1}{2}\cdot \frac{1}{(2\pi)^2}\Delta f \big]}  \; (p) =   e(p) \wh f(p) \; ,
$$

where 
\be
e(p):= \frac{1}{2} \, p^2
\label{disp}
\ee
 is the dispersion law. The velocity is given by $\frac{1}{2\pi}\nabla 
e(p)= \frac{1}{2\pi} \, p$. 
We note that the additional $ \hbar^2/m= (2\pi)^{-2}$ factor was erroneously
omitted in the definition of the Hamiltonian in (1.2) of \cite{ESY3}.

Define the {\it Wigner transform} of a function $\psi\in
L^2(\bR^d)$ 
\be
        W_\psi(x,v): = \int
        e^{2\pi iv\cdot \eta} \overline{\psi(x+\frac{\eta}{2})}
  \psi(x-\frac{\eta}{2}) \rd \eta\; .
\label{WP}
\ee
The  Fourier transform of $W_\psi(x, v)$ in the $x$ variable is therefore
\be
     \wh W_\psi(\xi, v)
    =\overline{\widehat
     \psi\Big(v-\frac{\xi}{2}\Big)}
        \widehat\psi\Big(v+\frac{\xi}{2}\Big) \; .
\label{FW}
\ee
Define the rescaled Wigner distribution as
\be
        W^\e_\psi (X, V) : = \e^{-d}W_\psi\Big( \frac{X}{\e}, V\Big)\; .
\label{wig}
\ee
Its  Fourier transform in $X$ is given by
$$
     \widehat W^\e_{\psi}(\xi, V) = 
     \overline{\widehat \psi \Big(V-\frac{\e \xi}{2}\Big)}
        \widehat \psi \Big( V+ \frac{\e \xi}{2}\Big) \; .
$$

For any function $h:\bR^d\to\bC$ and energy value $e\ge0$ we introduce the
notation
\be
   [h](e):= \int h(v) \delta(e-e(v))\rd v:=
  \int_{\Sigma_e} h(q) \;\frac{\rd\nu(q)}{|\nabla e(q)|} \; ,
\label{coarea}
\ee
where $\rd\nu(q)$ is the restriction of the Lebesgue measure
onto the energy surface $\Sigma_e:=\{ q \; : \; e(q)=e\}$
that is the ball of radius $\sqrt{2e}$. 
More
explicitly,
$$
  [h](e):=  (2e)^{\frac{d}{2}-1} \int_{S^{d-1}} 
  h( \sqrt{2e} \phi)\rd \phi \; .
$$
Clearly
\be
   \int h(v)\rd v = \int_0^\infty [h](e) \rd e \; .
\label{cc}
\ee
The normalization of the measure $[\cdot ]_e$ is given by
\be
    [1](e):= c_{d-1} (2e)^{\frac{d}{2}-1}\; ,
\label{1edef}
\ee
where $c_{d-1}$ is the volume of the unit sphere $S^{d-1}$.

\subsection{Main Theorem}\label{sec:maintheorem}

The weak coupling limit is defined by the following scaling:
\be \label{wcl}
t=\cT/\e, \quad
x=\cX/\e,  \quad \e= \lambda^2\; .
\ee
The Wigner distribution $W^\e_{\psi_{ \cT/\e}} (\cX, V)$
converges weakly to a function $F_\cT(\cX, V)$ that satisfies
 the Boltzmann equation
\eqn\label{B}
        &&\partial_\cT F_\cT(\cX,V) + 
 V\cdot \nabla_\cX F_\cT(\cX,V)  \qquad\qquad
        \nonumber\\
        &&\hspace{1cm}=
       2\pi \int \rd U |\wh B(U-V)|^2\delta(e(U)-e(V)) \Big[ F_\cT(\cX,U)
         -  F_\cT(\cX,V) \Big]\; .
\eeqn
Note that the Boltzmann equation can be viewed as
 the generator of a Markovian semigroup on phase space.
The proof of \eqref{B} for continuum Gaussian model was given in
\cite{EY}; the $\bZ^d$ lattice case with general i.i.d. random potential was
considered in \cite{Ch}. The derivation
of the Boltzmann equation for  potential
 \eqref{Vom} follows from these two proofs in a straightforward way.

In this paper we consider the long time scaling
\be\label{scale}
t=\lambda^{-\kappa}\big( \lambda^{-2}T\big), \quad 
x= \lambda^{-\kappa/2}\big(\lambda^{-2}X\big) = X/\e, 
\quad \e = \lambda^{\kappa/2+2} \; 
\ee
with some $\kappa > 0$. 
This scaling corresponds to the long time limit of the
Boltzmann equation with diffusive scaling.

 For some energy $e>0$, let
\be
   L_e f(v): = \int \rd u \; \sigma(u, v) [ f(u)-f(v)], \qquad e(v)=e\; ,
\label{Lgen}
\ee
be the generator of the momentum jump process $v(t)$ on $\Sigma_e$
with  collision kernel  
\be
\sigma(u,v):=2\pi|\wh B(u-v)|^2\delta( e(u)-e(v)) \; .
\label{def:sigma}
\ee
Notice that we use $u, v$ etc.  to denote the momenta; the corresponding velocities
are $u/(2 \pi), v/(2 \pi)$ etc.  

A well-known argument shows that
 $B\not\equiv 0$ and the regularity of $B$ guarantees the following
properties. Some details will be given in \cite{ESY3}.

\begin{lemma} \label{lemma:erg}
 For each $e>0$
the Markov process  $\{ v(t) \}_{t\ge 0}$ with generator $L_e$
is uniformly exponentially mixing. The unique invariant
measure 
is the uniform distribution,  $[\, \cdot \, ](e)/[1](e)$,
on the energy surface $\Sigma_e$.
\end{lemma}
Let
$$
   D_{ij}(e): = \frac{1}{(2\pi)^2}
\int_0^\infty \cE_e \big[ v^{(i)}(t) v^{(j)}(0)\big] 
\rd t\; ,\qquad v= (v^{(1)}, \ldots , v^{(d)}), \quad
 i,j=1,2,\ldots d,
$$
be the  velocity autocorrelation matrix, where $\cE_e$
denotes the expectation with respect to this Markov process
in equilibrium.
By the spherical symmetry of $\wh B$ and $e(U)$, 
the autocorrelation matrix is constant times the identity:
\be
     D_{ij}(e) = D_e \; \delta_{ij}, \qquad  D_e: = \frac{1}{(2\pi)^2d} \;
    \int_0^\infty \cE_e \big[ v(t)\cdot v(0) \big] \rd t \; .
\label{diffconst}
\ee
The main result of the paper  is the following theorem.

\begin{theorem}\label{main}
Let $d\ge3$ and $\psi_0 \in L^2(\bR^d)$ be a normalized initial wave function.
Let $\psi(t):= \exp(-itH)\psi_0$ solve the Schr\"odinger equation
(\ref{sch}). Let $\cO(x, v)$ be a Schwarz function on $\bR^d\times \bR^d$.
For almost all energy  $e>0$, $[|\wh \psi_0(v)|^2 ](e)$ is finite and 
 let $f$ be the solution to the heat equation
\be
\partial_T f(T, X, e) = \; D_e \;
\Delta_X f(T, X, e) \label{eq:heat}
\ee
with the initial condition
$$
     f(0, X, e): = \delta(X) \Big[ |\wh \psi_0(v)|^2 \Big](e)\; 
$$
for these energies.
Then there exist $0<\kappa_0(d)\leq 2$ such that
for  $0<\kappa<\kappa_0(d)$ and for $\e$ and $\lambda$ 
related by \eqref{scale},
the rescaled  Wigner distribution  satisfies
\be
\lim_{\lambda \to 0} \int\rd X \! \int \! \rd v \;  \cO(X, v)  \bE
 W^\e_{\psi(\lambda^{-\kappa-2} T)} (X, v)
= \int\rd X \int \!\! \rd v \; \cO(X, v) f(T, X, e(v)) \; ,
\label{fint}
\ee
and the limit is uniform on $ T\in [0, T_0]$ with any fixed $T_0$.
In $d=3$ one can choose $\kappa_0(3) = 1/370$.
\end{theorem}

\noindent

{\it Remark 1.} The coefficient $ \big[ |\wh \psi_0(v)|^2 \big](e)$
in the initial condition $f(0, X, e)$ is finite
for almost all $e$ by using \eqref{cc} for $h=|\wh \psi_0|^2$.

\smallskip

{\it Remark 2.} 
The total cross section of the collision process \eqref{Lgen}, 
\be
   \sigma_0(e): =\int \rd u \; \sigma(u, v)\; \qquad e=e(v)\; ,
\label{si0}
\ee
 is  a function of $e=e(v)$ only.
Assuming  $\wh B(0)\neq 0$, we see that
$\sigma_0(e) \sim [1](e)$ for small $e$, and
$\sigma_0(e)\sim e^{-1/2}$ for large $e$. 
 It follows from Lemma \ref{lemma:erg} and from
standard probability arguments
that the diffusion constant \eqref{diffconst} scales
as $D_e\sim e/\sigma_0(e)$ for small $e\ll 1$ and
$D_e\sim e^2/\sigma_0(e)$ for large $e\gg 1$. 
If $\wh B$ vanishes at 0 (but \eqref{nonzero} still holds),
then the small energy behaviour of $\sigma_0(e)$ and $D_e$
depends on the rate of vanishing of $\wh B$ at 0 in
a straighforward way.

\smallskip

{\it Remark 3.}
 The condition \eqref{nonzero} is not essential, but
the theorem needs to be modified 
if $\wh B$ vanishes on $D(0, \delta)$, a ball of radius $\delta>0$
about the origin. Let $\delta>0$ be the maximal radius
so that $D(0,\delta)\cap \mbox{supp} \, (\wh B)=\emptyset$.
In this case the total cross section $\sigma_0(e)$ is zero for
all energy values $e\leq \delta^2/8$, because the diameter
of the energy surface $\Sigma_e$ is smaller than the
minimal range of $\wh B$. Therefore the evolution  is ballistic
for the part of the initial wave function that is supported on energy
shells $e\leq \delta^2/8$. For the other part of the wave function
the diffusion equation still holds.

\bigskip

Fig.~\ref{fig:3scale} below
shows the three different scales schematically.
On the  Schr\"odinger scale both time and space are of order 1
in atomic units.
On the kinetic scale time and space are rescaled by $\lambda^{-2}$.
The dynamics is given by the Boltzmann equation characterized by
finitely many collisions. On the diffusive scale we rescaled
the time and space by an additional factor $\lambda^{-\kappa}$
and $\lambda^{-\kappa/2}$ respectively. The typical number
of collisions is of order $\lambda^2 t \sim\lambda^{-\kappa}$.

If we assume that the Boltzmann equation holds under all scalings,
Theorem \ref{main} can be easily understood.
{F}rom the Boltzmann equation \eqref{B},
the momentum distribution develops according to
the Markovian generator $L_e$. 
 Therefore, the
Boltzmann equation \eqref{B} describes a process that a particle
travels with a fixed momentum $v$ up to an exponentially
distributed random time with average value $\sigma_0(e(v))^{-1}$,
 then it changes momentum
from $v$ to a new momentum $u$ on the same energy surface $\Sigma_e$
chosen by the probability distribution $P(u)= \sigma(u,v)/\sigma_0(e(v))$.
 The different energy sectors do not
interact. Clearly, this process then converges to a  Brownian motion
in configuration space with a diffusion coefficient given by 
\eqref{diffconst} and with momentum restricted to
a fixed energy shell $\Sigma_e$.

\begin{figure}
\begin{center}
\epsfig{file=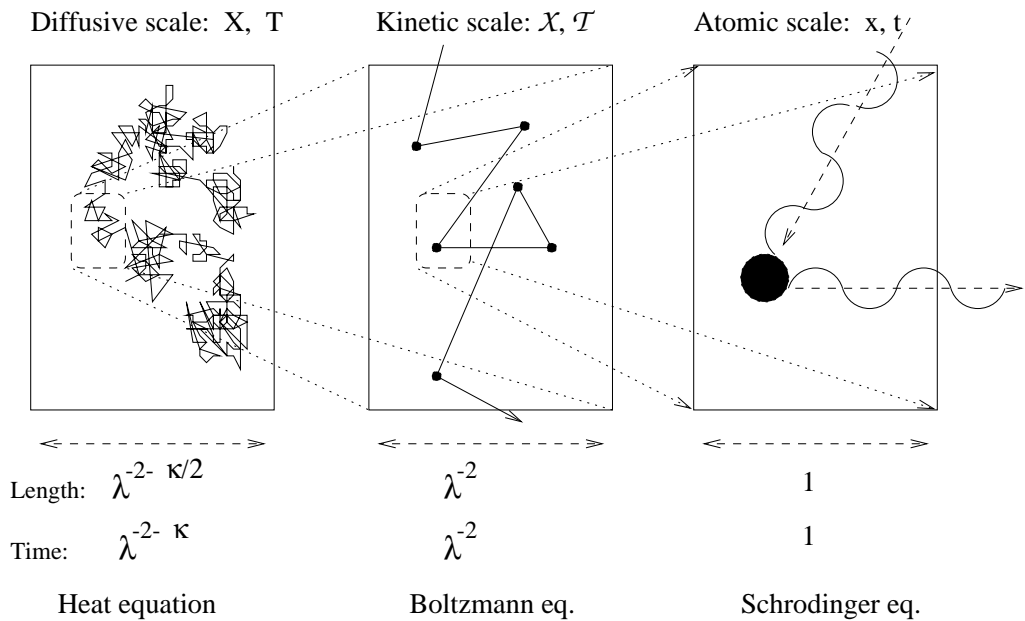,scale=.85}
\end{center}
\caption{Evolution equations on three scales}\label{fig:3scale}
\end{figure}

\bigskip
Under the assumption that the Boltzmann equation is valid for all time,
this argument applies in all dimensions.
The random Schr\"odinger evolution, however,  is expected to be localized
for $d \le 2$ even for small  coupling constant.
Therefore, even though the  Boltzmann approximation was
proved to be valid for $d\ge 2$ \cite{EY}, \cite{Ch}
(it is not valid for $d=1$) in the weak coupling limit,
it will not be valid for all time in $d = 2$.
It is expected that memory effects and quantum correlations
eventually dominate the evolution and ruin the Markovian
character of the Boltzmann picture. Heuristic ideas
show this transition happens at an exponentially large time
(see, e.g. \cite{VW}).

The effects of the quantum correlations and memory 
are not expected to
change the Boltzmann picture drastically in $d\ge 3$, 
but one expects corrections to the diffusion equation and a
transition between different energy shells for $\kappa\ge 2$
(see \cite{LR}).

\medskip

Although Theorem \ref{main} is formulated
for the Euclidean space,  quantum diffusion
on large scales is not restricted to continuum models.
In fact, we prove in \cite{ESY2} an analogous result for the Anderson model
where the Hamiltonian  \eqref{H} is defined on
$\ell^2(\bZ^d)$, $d\ge 3$. The kinetic energy is the
discrete Laplacian and the random potential is given by
\be
   V_\om(x) := \sum_{\gamma\in \bZ^d} v_\gamma \, \delta(x-\gamma),
 \qquad x\in \bZ^d, 
\label{Vomdis}
\ee
where $\{ v_\gamma=v_\gamma(\om)\;: \;\gamma\in \bZ^d\}$
 are i.i.d. random variables. 

The discrete model is technically more involved for two 
reasons. First, the dispersion relation of the
lattice Laplacian lacks convexity which simplifies several 
estimates in the continuum model. We will address this issue in
 \cite{ESY2}.    

Second, our choice
of the  potential in the continuum model 
contains more randomness than in the discrete one.
Comparing \eqref{Vom}--\eqref{def:muom} with \eqref{Vomdis}
we note that in the continuum model both the locations
of the obstacles  and 
the coupling constants  are random variables,
 while in the Anderson model the locations
are deterministic. Formally, the potential \eqref{Vomdis}
corresponds to  \eqref{Vom}--\eqref{def:muom} with the
deterministic choice $y_\gamma=\gamma\in \bZ^d$ and with
 $B(x)$ being the lattice delta function $\delta(x)$.
The random choice of the centers in the continuum model
simplifies the formula in the calculation of the
high moments of the random potential.
Nevertheless, we will formulate our estimates in
a unified setup that can be applied  directly
to the Anderson model in \cite{ESY2} as well.

\subsection{Strategy of the proof}

The above heuristic argument using the Boltzmann equation,
besides being misleading for $d=2$, 
also masks the difficulties in proving Theorem \ref{main},
namely that one has to follow the full {\em quantum mechanical} 
time evolution through infinitely many collisions.
The main tool of our proof is to use the
Duhamel expansion to decompose the wave function into elementary
wave functions characterized by their collision histories.
We then apply two strategies to simplify the expansion:
(i)  renormalization of the propagator, i.e., resumming the
 two legged subdiagrams;
(ii) stopping rules to control recollisions.
Apart from these two steps, the bulk of our proof
is devoted to giving sharp estimates for a 
large class of Feynman graphs.

To get an idea, imagine that we expand the solution to the Schr\"odinger
equation by using the Duhamel formula repeatedly. This rewrites the solution
into a sum of elementary wave functions, each of which is characterized
by  a sequence of collisions with random obstacles.
 When we take  the expectation
of $\|\psi_t\|^2$
with respect to a Gaussian randomness,
we pair the random obstacles  by Wick's theorem and obtain 
a sum of amplitudes of Feynman graphs.
In case of a non-Gaussian randomness the higher order cumulants are
also present due to various recollision patterns
(their contribution turns out to be negligible, 
but proving this is rather involved).

If we take only the Laplacian as the free part in the expansion,
even the amplitudes of individual graphs diverge in the limit we consider.
However, this can be remedied by a
simple resummation  of all two-legged insertions
caused by the lowest order self-energy contribution
(it turns out that higher order corrections to the self-energy
do not play a role in the scaling limit we consider). 
The resummation is performed by choosing an appropriate
reference Hamiltonian $H_0$ for the expansion. 
After this rearrangement, all graphs have a finite amplitude 
in our scaling limit, and the ladder graphs give the leading 
contribution.

However, we have to estimate not only individual graphs
but the sum of all graph amplitudes,
which requires beating down the factorial growth 
of the number of graphs.
This problem has been addressed in constructive field theory.
For field theories with bosons,
the graphical expansion to infinite order diverges.
Borel summability was
proven by
cluster expansion and renormalization group methods
\cite{GK,FMRS,BY,BDH}.
In fermionic theories,
the anticommutation relations entail cancellations
which lead to analyticity in the presence of regulators
\cite{GKGN,FMRSGN,FKTCO,SW}.
Our method to control
the combinatorial growth is completely different:
it is by very sharp bounds on the individual graphs.
We give a classification of arbitrary large graphs,
based on counting the number of
vertices carrying extra oscillatory effects. The number of
these vertices is called  the {\it degree } of the graph
and it measures the improvement over the standard power counting.
For the ladder  graphs, the degree is zero, 
for the anti-ladder (i.e., complete crossing) graph it is 2.
For general graphs, the degree is roughly
the number of vertices after removing all ladder and anti-ladder
subgraphs. We thus obtain an extra $\lambda^c$ factor
(for some $c>0$) {\em
per non-(anti)ladder vertex}. This strong improvement is sufficient
to beat the growth of the combinatorics in the time scale we consider.
To our knowledge, nothing like this has been done in a
graphical expansion before.
Improved phase space estimates have been used
to prove regularity in two-dimensional many--fermion systems,
but the improvement exponent was fixed
independently of the number of vertices
\cite{FST1,FST3,FKTOL}.

For a comparison, the  unperturbed Green
 functions in the perturbation expansion
for the many--fermion systems and for the random Schr\" odinger equation are
given  by
$$
\frac 1 {ip_0+p^2-\mu },  \qquad
\frac 1 {p^2-\alpha + i \eta}\; .
$$
In the many--fermion case, $\mu>0$,  
$p_0 \in M_F = \{\frac{\pi}{\beta} (2 n + 1): n \in \bZ\}$
where $\beta \sim T^{-1}$ is the inverse temperature. 
In the random Schr\" odinger case, $\eta \sim t^{-1}$. 
Their $L^2$-properties  are different:
$$
\frac{1}{\beta} \sum_{p_0 \in M_F}
\int \rd p  \big |ip_0+p^2-\mu \big |^{-2}  \sim |\log \beta|, \quad
\int \rd p \big |p^2-\alpha + i \eta \big |^{-2}  \sim \eta^{-1}\; .
$$
Notice the divergence is more severe for the random Schr\"odinger equation case. 
In the many-fermion case, 
there is one $p_0$--summation per line of the graph; in the random Schr\" odinger case
there are just two overall $\alpha$--integrals for graphs with arbitrarily many lines.

\medskip

This paper is organized as follows. In Section~\ref{sec:tech} we
perform the self-energy renormalization, we smooth out the 
data  and restrict the problem to a finite box. The
Duhamel expansion is introduced in Section~\ref{sec:stoprule}. 
In Section~\ref{sec:mainpr} we reduce the Main Theorem
to Theorems~\ref{7.1}, \ref{thm:L2}    and \ref{thm:laddheat}.
The key result is Theorem~\ref{thm:L2} which we prove in
the rest of this  paper. The other two theorems are more technical
and they are proven in the companion paper \cite{ESY3}.
The Feynman graphs are introduced in Section~\ref{sec:gra}. In
Sections~\ref{sec:nolump} and \ref{sec:lump} we reduce all estimates
to Theorem~\ref{thm:Vsi}. This theorem   is our
 main technical bound on Feynman graphs and it is 
 proven in Section~\ref{sec:mainproof}.

Since the random potential in our model is given by
 general i.i.d.  random variables, 
the rule for taking the expectation is different
from the case of Gaussian random field used in \cite{EY}.
This produces  technical difficulties
especially for the Anderson model, where,
in addition to the usual pairing from the Wick theorem,
we have to introduce higher order partitions of the vertices,
called {\it non-trivial lumps}. Our
continuum model  avoids  this complication due to
the additional randomness of the obstacle centers.
Nevertheless, we present  the general proof  here so that the
key technical results in this paper could be applied to
the Anderson model as well.
For readers  interested only 
in the continuum case, we recommend
 to ignore the non-trivial lumps.

Universal constants and constants that depend only on the dimension $d$,
on the final time $T_0$ and
 on the Schwarz norms $\| B\|_{k,k}$ from \eqref{Bcondi}
 will be denoted by $C$ 
and their value may vary from line to
line.

\section{Preparations}\label{sec:tech}
\setcounter{equation}{0}

\subsection{Renormalization}

The purpose of this procedure is to include immediate recollisions
with the same obstacle into  the propagator itself.
This is also called the renormalization of
 "one-particle propagators" or 
 two legged subdiagrams.
Without renormalization, these graphs individually
are exponentially large (``divergent''), but
their sum is finite. Renormalization removes this
instability and the analysis
of the resulting Feynman graphs will become simpler.

The self-energy operator is given by the multiplication operator
in momentum space
\be
 \theta(p) := \Theta(e(p)), \qquad    \Theta (\alpha): = \lim_{\e \to 0+ }
 \Theta_\e (\alpha) \; , \qquad \Theta_\e(\alpha) : = \Theta_\e
(\alpha, r)
\label{eq:thetalim}
\ee
for any $r$ with $e(r)=\alpha$, 
where
\be\label{theta}
    \Theta_\e (\alpha, r): =  \int \frac  { |\wh B(q-r)|^2
\rd q} { \alpha- e(q) + i \e} \; .  
\ee
Note that by spherical symmetry of $B$ and $e(q)$,
$\Theta_\e(\alpha, r)$ depends only on the length of $r$,
therefore $\Theta_\e(\alpha)$ in (\ref{eq:thetalim}) is well defined.
Clearly $\theta(p)$ is spherically symmetric.
The existence of the limit and  related properties of $\Theta$
have been proven in \cite{EY} using that $\|\wh B^2\|_{2d,2d}<\infty$.
Here we summarize the results:
\begin{lemma}  In $d\ge 3$
the following hold:
\be
  | \Theta_\e(\alpha, r) - \Theta_\e (\alpha, r')|\leq C\big| \; |r|-|r'|\;\big|
\label{lipr}
\ee
(Eq. (3.80) in \cite{EY}) and
\be
  |\Theta_\e (\alpha, r) - \Theta_{\e'}(\alpha', r)|\leq 
 C (|\e-\e'| + |\alpha - \alpha'|) \e^{-1/2}
\label{zhold}
\ee
if $\e\ge \e' >0$ (Eq. (3.68) in \cite{EY}).
 {F}rom this latter estimate the existence
of the limit $\lim_{\e\to0+0} \Theta_\e(\alpha, r)$ follows.
Moreover, $\Theta$ is H\"older continuous
\be   
|\Theta(\alpha) - \Theta(\alpha')| \leq C|\alpha-\alpha'|^{1/2} \; .
\label{eq:holder}
\ee
\end{lemma}

{\it Proof.}  
 We have only to prove the H\"older continuity.
For any $\e$ and any $r, r'$
with $\alpha = e(r)$,  $\alpha'= e(r')$ we have
\begin{align}
   |\Theta(\alpha) - \Theta(\alpha')| 
\leq & \lim_{\e'\to0+0} |\Theta_{\e'}(\alpha, r) - \Theta_\e(\alpha, r)|
   + |\Theta_\e(\alpha, r) - \Theta_\e(\alpha, r')| \nonumber\\
 &  + |\Theta_\e (\alpha, r') - \Theta_\e(\alpha', r')|
  + \lim_{\e'\to 0+0} |\Theta_\e (\alpha', r') -\Theta_{\e'}(\alpha', r')| 
 \nonumber\\
  \leq &  \;
 C\Big(\e^{1/2} + \big|\; |r|-|r'|\;\big| + |\alpha - \alpha'|\e^{-1/2}
 \Big) \; . \nonumber
\end{align}
By optimizing $\e$ and using $e(r)=\alpha$, we obtain (\ref{eq:holder}).
$\;\;\Box$

\bigskip

We have the following estimate on $\theta(p)$ and in parallel on $\Theta(e)$:
\begin{lemma}\label{lemma:theta}
For any $d\ge 3$ there exist universal positive constants $c_1, c_2$ such that
\be
 |\theta(p)|\leq \frac{c_2 \log \langle p\rangle}{ \langle p\rangle}\;, \qquad
    |\Theta(e)|\leq \frac{c_2 \log \langle e \rangle}{
 \langle e\rangle^{1/2}} \; ,
\label{eq:thetaest}
\ee
\be
 \mbox{Im}\;\Theta(e) \leq -c_1 \min\{ |e|^{\frac{d}{2}-1}, |e|^{-1/2} \}\;,
\qquad
     \mbox{Im}\;\theta(p) \leq -c_1 \min\{ |p|^{d-2}, |p|^{-1} \} \; .
\label{eq:imthetaest}
\ee
\end{lemma}

{\it Proof of Lemma \ref{lemma:theta}.}
By performing the angular integration,
we can write $\Theta_\e(\alpha, p)$ 
with $e(p)=\alpha$ as
\begin{equation}\label{co-area}
   \Theta_\e(\alpha, p)
 =  \int_{0}^{\infty} \frac{ (2e)^{\frac{d}{2}-1}
   \rd e}{\a - e +i\e} \; S(e) \; ,
   \qquad \mbox{with}\quad
   S(e):=\int_{ S^{d-1}}
   |\wh B(\sqrt{2e}(\phi_r-\phi))|^2\rd\phi \; ,
\end{equation}
where $\phi_r$ is a fixed vector on the unit sphere $S^{d-1}$.
For small $e$ values
$$
  |S(e)| = O(1),  \qquad |\nabla S(e)| = O(e^{-1/2})\; .
$$
For large $e$ values, using the regularity of $\wh B$, 
$$
  |S(e)| = O(e^{-\frac{d-1}{2}}), \qquad
 |\nabla S(e)| = O(e^{-\frac{d}{2}}) \; .
$$
These estimates, in particular, 
 stand behind the proof that $\lim_{\e\to0+0} \Theta_\e(\alpha, p)$
is finite, since they guarantee the sufficient decay for large $e$
and the sufficient smoothness around the singularity of the denominator
in  \eqref{co-area}.
The imaginary part therefore is
$$
 \mbox{Im}\, \theta(p)=  \mbox{Im} \lim_{\e\to0+0}\Theta_\e(\alpha, p)
   = -\pi (2\alpha)^{\frac{d}{2}-1}S(\alpha)
$$
which behaves as $\sim -|p|^{d-2}$ for small $p$ and as $\sim -|p|^{-1}$
for large $p$. 
The real part of $\Theta_\e(\a, r)$ is bounded for small $\a$.
For large $\alpha$
one splits the integration
$$
   \mbox{Re}\, 
\Theta_\e(\alpha, r) = \Bigg(\int_{\alpha-1}^{\alpha+1} + 
\int_{|\alpha-e|\ge 1}\Bigg)
 \frac{ (2e)^{\frac{d}{2}-1}
   \rd e}{\a - e +i\e} \; S(e) \; .
$$
After Taylor expanding $(2e)^{\frac{d}{2}-1}S(e)$ around $\alpha\gg 1$,
the first term is bounded by 
$$
  \Bigg| (2\a)^{\frac{d}{2}-1}S(\a) \int_{\alpha -1}^{\alpha + 1} 
  \frac{\rd e}{\a -e+i\e}\Bigg| + 2\sup_{|e-\a|\le 1}
\Big|
\frac{\rd}{\rd e} [(2e)^{\frac{d}{2}-1}S(e)]\Big|
= O(\alpha^{-1/2})\; ,
$$
and the second term by
$$
   \int_{|\alpha -e|\ge 1} \frac{\rd e}{|\alpha -e| e^{1/2}} \leq 
\frac{c\log \langle \alpha \rangle }{\langle \alpha\rangle^{1/2}} \; .
$$

If we write $\Theta(e)=  {\cal R} (e)- i{\cal I}(e)$, where ${\cal R} (e)$ and
${\cal I}(e)$ are real functions,
and recall  $ Im (x+  i 0)^{-1} = - \pi \delta(x)$,  we have
\be
   {\cal I}(e)=- \mbox{Im} \,\Theta (e) = \pi\int \delta(e(q)-e) |\wh B(q-r)|^2 \rd q
\label{eq:opt} \ee 
for any $r$ satisfying $e=e(r)$. $\qquad\Box$

\bigskip

We rewrite the Hamiltonian as
$$
    H= H_0 + \wt V,
$$
where
\be\label{renH}
  H_0:=\omega(p):= e(p) +\lambda^2 \theta(p), \qquad \wt V := \lambda
    V -\lambda^2 \theta(p)\; .
\ee
We note that our renormalization is only an approximation to
the standard self-consistent renormalization  given
by the solution to the equation
\be
     \om(p) = e(p) + \lambda^2 \lim_{\e\to 0+0}
     \int \frac  { |\wh B(p-q)|^2 \rd q } { \om(p)- \om(q) + i \e} \; .
\label{eq:selfc}
\ee
Due to our truncation procedure, the definition (\ref{eq:thetalim})
is sufficient and is more convenient for us.
Since $e(p)$ is spherically symmetric, so are $\theta(p)$ and
$\om(p)$.

\bigskip

The following lemma collects some estimates on the renormalized  propagators
we shall use to prove Theorem~\ref{main}.
The proof is fairly simple and will be given in \cite{ESY3}.
We note that formula (2.8) in \cite{ESY3} erroneously
contain a factor 2 in the exponent of $\eta$, the correct
bound is \eqref{eq:3aint}. 

\begin{lemma}\label{le:opt}
Suppose that $\lambda^2 \ge \eta \ge \lambda^{2+ 4 \kappa}$ with
$\kappa \le 1/12$. Then we have,
\be
    \int \frac{|h(p-q)| \rd p}{|\alpha - \om(p)+ i\eta|}
    \leq \frac{C \| h\|_{2d,0}\, |\log\lambda| \; 
\log \langle \alpha\rangle}{\langle \alpha\rangle^{1/2}
 \langle |q| -\sqrt{2|\alpha|}\rangle } \; ,
\label{eq:logest}
\ee
and for $0\le a<1$
\begin{align}
     \int \frac{|h(p-q)| \rd p}{|\a -\om(p) + i\eta|^{2-a}}
 \leq & \frac{C_a \| h \|_{2d, 0}\,\lambda^{-2(1-a)}}{\langle 
\alpha\rangle^{a/2}
\langle |q| -\sqrt{2|\alpha|}\rangle } \; ,
\label{eq:2aint}
\\
  \int \frac{|h(p-q)| \rd p}{|\a -e(p) + i\eta|^{2-a}}
 \leq & \frac{C_a \| h \|_{2d, 0}\,\eta^{-(1-a)}}{\langle \alpha\rangle^{a/2}
\langle |q| -\sqrt{2|\alpha|}\rangle } \; .
\label{eq:3aint}
\end{align}
For $a=0$ and with $h:= \wh B^2$, the following more precise estimate holds.
There exists a  constant $C_0$, depending only
on finitely many constants $C_k$ from \eqref{Bcondi}  such that
\be
  \int \frac{ \lambda^{2}|\wh B(p-q)|^2 \; \rd p 
 }{|\alpha-\ov\om(p)-i\eta|^{2}}  \leq
    1+ C_0\lambda^{-12\kappa}\big[\lambda + |\a- \om(q)|^{1/2}\big] \, .
\label{eq:ladderint}
\ee
\end{lemma}

\subsection{Smoothing the initial data and
the potential}\label{sec:tru}

In this section we  show that 
it is sufficient to prove the Main Theorem
under the assumptions that $\wh\psi_0(p)$ is
a bounded, smooth, compactly supported function
and $\wh B(p)$ is
supported on $\{ |p|\leq \lambda^{-\delta}\}$
for any fixed $\delta>0$. 

The approximation procedure relies on  the following
$L^2$-continuity property of the Wigner transform. If a random
wave function  is decomposed as $\psi=\psi_1+\psi_2$, then
\be
   \Big| \bE\langle \wh\cO, \wh W_\psi^\e\rangle -
    \bE\langle \wh\cO, \wh W_{\psi_1}^\e\rangle\Big|
    \leq C\Big(\int \sup_v |\wh\cO(\xi, v)| \rd\xi \Big)
    \sqrt{\bE \big[ \| \psi_1 \|^2 +\|\psi_2\|^2\big]
 \cdot \bE \| \psi_2 \|^2} \; 
\label{wigcont}
\ee
by a simple Schwarz inequality. (Due to a misprint, the $\|\psi_2\|^2$
term was erroneously omitted in Section 2.1. of our earlier paper
\cite{EY}.)

\medskip

{\it Approximation of the initial data.}
 Let $\wh\psi_0\in L^2$ and let
$\wh \psi_n$ be a sequence of smooth, compactly supported
functions with $\|\wh \psi_n-\wh\psi_0\|\to 0$.
We decompose
$\wh \psi_0 = \wh\psi_n + (\wh\psi_0-\wh\psi_n)$. Then
$$ 
    \wh\psi(t) = e^{-itH}\wh\psi_n
   +  e^{-itH} (\wh\psi_0-\wh\psi_n) \; .
$$
Since
$$
    \| e^{-itH}  (\wh\psi_0-\wh\psi_n)\| = \|\wh\psi_0-\wh\psi_n\| \to 0 
$$
 as $n\to\infty$, uniformly in $t$, we see that
$$ 
   \lim_{n\to\infty} \Big| \bE\langle \wh\cO, \wh W_{\psi(t)}^\e\rangle -
    \bE\langle\wh \cO, \wh W_{\psi_n(t)}^\e\rangle\Big| =0
$$
uniformly in $t$ (and thus in $\e$),
where 
$\psi_n(t):= e^{-itH}\wh\psi_n$
is the time evolution of the approximated initial data.
This means that the approximation procedure is continuous
on the  left hand side of
(\ref{fint}).

Similarly, on the right hand side of \eqref{fint}, we
can define $f_n(T, X, e)$ to be the solution to 
(\ref{eq:heat}) with initial data $f_n(0, X, e): = 
\delta(X)\big[ |\wh\psi_n|^2\big](e)$.
Clearly $\big[  |\wh\psi_n|^2\big](e)$
converges to $\big[  |\wh \psi_0|^2\big](e)$
in $L^1(\rd e)$.
Therefore 
\be
   f_n(T, X, e) \to f(T, X, e) 
\label{fnconv}
\ee
 in $L^1(\rd X\, \rd e)$, 
uniformly in $T$. The right hand side of
(\ref{fint}) is therefore also continuous as $n\to\infty$.

We remark, that if $\wh \psi_0$ is smooth, e.g.
$|\nabla_p \wh \psi_0(p)|\leq C\langle p\rangle^{-4d}$,
 then a bounded, smooth and 
compactly supported approximant, $\wh\psi_n$, can be chosen
so that $\big[  |\wh\psi_n|^2\big](e)
\to \big[  |\wh \psi_0|^2\big](e)$ for every $e>0$ and
then the convergence in \eqref{fnconv} also holds in $L^1(\rd X)$
for any $e$. The smoothness of  $\wh \psi_0$ is 
used only at the point  when we explicitly
compute the main term of the perturbation expansion and
identify it with the Boltzmann equation, see \cite{ESY3}.

\medskip

{\it Propagation estimate.}
To verify that  a truncation is allowed for $\wh B$, we first
need  a crude  propagation estimate. 
Define the following event for any $Z>0$
$$
    \Omega_Z:=\Big\{ \om \; : \; 
\int_{|y-k|\leq 1} \rd |\mu_\om|(y)
  \leq Z \langle k\rangle \; , \forall k\in \bZ^d\;  \Big\} \; ,
$$ 
where $|\mu_\om|$ denotes the total variation of the
(random) measure $\mu_\om$.
For any fixed $k\in \bZ^d$, let 
$N_k$ be the number of Poisson points in the ball $\{ x \; : \; |x-k|\leq 1\}$.
We  compute
$$
    \bE \Big\| \int_{|y-k|\leq 1} \rd |\mu_\om|(y) \Big\|^{d+1}
   \leq \bE N_k^{d+1} |v|^{d+1} \leq C_d,
$$
using \eqref{momm}  and that $N_k$ is a Poisson random variable
with expectation $\bE N_k = \mbox{vol(unit ball)}$.
By Markov inequality
$$
   P(\Omega_Z^c) \leq \sum_{k\in \bZ^d} \frac{C_d}{Z^{d+1} \langle k \rangle^{d+1}}
  = O(Z^{-d-1})\; ,
$$
thus we have 
\be
    \lim_{Z\to\infty}{\bf P} (\Omega_Z)=1 \; .
\label{LD}
\ee
We decompose
$$ 
   \bE\langle \wh\cO, \wh{W_{\psi_t}^\e}\rangle = 
 \bE \;\big[ {\bf 1}(\Omega_Z)\langle \wh\cO, 
\wh{W_{\psi_t}^\e}\rangle \big] +  
\bE\; \big[ {\bf 1}(\Omega_Z^c)\langle \wh\cO, 
\wh{W_{\psi_t}^\e}\rangle\; \big],
$$
where ${\bf 1}(\cdot )$ is the characteristic function.
On the set $\Omega_Z^c$ we use 
\be
\Big| 
\bE \;\big[ {\bf 1}(\Omega_Z^c)\langle \wh \cO, \wh{W_{\psi_t}^\e}\rangle \big]
 \Big|
  \leq \Big(\int \sup_v |\wh\cO(\xi, v)| \rd\xi \Big)  
\|\psi_t\|^2 {\bf P} (\Omega_Z^c) \to 0
\label{vani}
\ee
as $Z\to\infty$, uniformly in $t$ (hence in $\lambda$).

For $\om\in\Om_Z$ we have
$|V_\om(x)|\leq C Z\langle x\rangle$ using the decay properties
of $B$.  Computing
the time derivative of the mean square displacement,
 we obtain
$ \partial_t (\psi_t, x^2 \psi_t) = i( \psi_t, [H, x^2]
 \psi_t)\; .$
Using  $[H, x^2]= -(\nabla\cdot x + x\cdot\nabla)$ and
 a Schwarz estimate  we have
\be
    \Big|  \partial_t( \psi_t, x^2 \psi_t)\Big|
 \leq  C (\psi_t, x^2 \psi_t)^{1/2}\big[ E  
 + \lambda (\psi_t, |V_\om|,\psi_t )\big]^{1/2}
\label{grom}
\ee
with $E:= (\psi_t, H\psi_t)= (\psi_0 , H\psi_0)$
by energy conservation. We
estimate $(\psi, |V_\om|\psi)
\leq CZ+ CZ(\psi, x^2\psi)^{1/2}$, in particular 
the energy $E$ is bounded (depending on $\psi_0$ and $Z$).
{F}rom \eqref{grom} we thus  have
\be
    (\psi_t, x^2\psi_t) \leq c_1(Z,\psi_0)t^4  + c_2(Z,\psi_0) 
\label{finspeed}
\ee
on $\Omega_Z$ with some constants $c_{1,2}(Z,\psi_0)$.

\medskip

{\it Approximation of the potential.}
We define the truncation of $B$ in Fourier space
as 
 $\wh B^\delta(p): =  \varphi( \lambda^\delta \langle p\rangle)\wh B (p)$,
where  $\varphi :\bR_+\to [0,1]$ is a fixed smooth cutoff function
with $\varphi(a) \equiv 1$ for $a\leq 1/2$ and $\varphi(a) \equiv 0$
for $a\ge 1$. 
In position space, we have
 for any $M\in \bN$,
\be
    | B(x)-B^\delta(x)|\leq \langle x\rangle^{-2d}
   \int \Big| \langle \nabla_p \rangle^{2d}
   \big[\wh B(p)[1- \varphi(\lambda^\delta \langle p\rangle)\big]\; \Big|\rd p
   \leq  C_{\delta,M}\lambda^M \langle x\rangle^{-2d} \; 
\label{bde}
\ee
by using that $B$ is in Schwarz space \eqref{Bcondi}.

Let
$$
   H^\delta : = - \frac{1}{2}\Delta_x + \lambda\int_{\bR^d} 
   B^\delta(x-y)\rd\mu_\om(y)
$$
be the Hamiltonian with the truncated potential.
Let $\psi^\delta_t: = e^{-itH^\delta} \psi_0$ be the 
evolution of the wave function under the modified Hamiltonian $H^\delta$.
On the set $\Omega_Z$ and for $t\ll \lambda^{-4}$
$$
   \partial_t \| \psi_t - \psi^\delta_t\|^2 =-2 \, \mbox{Im} \, (\psi^\delta_t,
 (H-H^\delta)\psi_t)
  \leq C_{\delta} Z \lambda^{13}
(\psi_t, \langle x^2\rangle \psi_t)^{1/2} \leq C(Z,\delta,\psi_0)\lambda^5
$$
by using  \eqref{bde} with $M=12$.
In particular, $\psi_t$ and $\psi^\delta_t$ remain
close up to time scale $t\sim \lambda^{-2-\kappa}$, $\kappa<2$.
This bound, together with the $L^2$-continuity of the
Wigner transform \eqref{wigcont} guarantees that the truncation
of $B$ does not influence the left hand side of \eqref{fint}.

As for the right hand side of \eqref{fint}, notice
that the  collision kernel, $\sigma(U,V)$,
of the momentum jump process \eqref{Lgen} is restricted to the energy surface
$e(V)=e(U)=e$. Therefore $U, V$ are bounded, depending on $e$,
so $\wh B(U-V) = \wh B^\delta (U-V)$ for these momenta, if $\lambda$
is sufficiently small. Thus the truncation of $B$ does not influence
 the right hand side of \eqref{fint}.

Armed with these results, we assume for
the rest  of the paper that $\wh \psi_0(p)$ is 
smooth, compactly supported, bounded
and $\wh B(p)$ is supported on
 $\{ |p|\leq \lambda^{-\delta}\}$ for any
 fixed $\delta>0$.
We thus extend the convention from the end of Section
\ref{sec:mainresult} 
that general constants denoted by $C$ may depend on the truncated
version of $\wh B$ and $\wh\psi_0$. The same applies to the
hidden constants in the $O(\cdot)$ and $o(\cdot)$ notations.

\subsection{Restriction to a finite box}

We will reduce the problem to a finite box of size $L$, $L\gg1$,
with periodic boundary conditions. In this way, for technical
convenience, we avoid the infinite summation in \eqref{def:muom}.
Let $\Lambda_L
:= [-L/2, L/2]^d\subset \bR^d$ be a 
finite torus and let $\Lambda_L^*:= (\bZ/L)^d$
be the dual lattice. We introduce the notation
\be
    \int_{\Lambda_L^*} f(x) \rd p : = \frac{1}{|\Lambda_L^*|}
   \sum_{p\in \Lambda_L^*} f(p) \; .
\label{riem}
\ee
The integrals $\int_{\Lambda_L}$ and $\int_{\Lambda_L^*}$
converge to their infinite volume counterparts as $L\to\infty$.
Let $(\cdot, \cdot)_L$ and $\| \cdot \|_L$ denote the  scalar
product and the norm on $L^2(\Lambda_L)$.

For any $L, M\gg 1$ 
we consider the random Schr\"odinger operator
$$
   H'=H_{L, M}':  =-\frac{1}{2}\Delta
   + \lambda V_\om'\; \qquad V_\om'(x):=
 \sum_{\gamma=1}^M  v_\gamma' \; B(x- y_\gamma') = \int_{\Lambda_L}
  B(x-y)\rd\mu_\om'\; ,
$$
with periodic boundary conditions on $\Lambda_L$ and
$\mu_\om':=\sum_{\gamma=1}^M v'_\gamma \delta_{y_\gamma'}$.
Here $\{  y_\gamma'\; : \; \gamma=1 ,\ldots, M \}$ are
i.i.d. random variables uniformly distributed on $\Lambda_L$
and $\{ v_\gamma'\; : \; \gamma=1 ,\ldots, M \}$
are i.i.d. variables distributed according to $\bP_v$
and they are independent of the $y_\gamma'$.  $M$ itself
will be random; it is chosen to be an independent Poisson
variable with expectation $|\Lambda_L|$.
The expectation with respect to the joint measure of $\{ M, y_\gamma' ,
 v_\gamma'\}$ is denoted by $\bE'$.
Sometimes we will use
the decomposition 
\be
\bE'= \bE_M \bE_y^{\otimes M} \bE_v^{\otimes M}
\label{facte}
\ee
 referring to
the expectation of $M$, $\{ y_\gamma\}$ and $\{ v_\gamma\}$ separately.
The parameter $L$ is implicit in these notations.
In particular, $\bE_y^{\otimes M}$ stands for the normalized integral
\be
       \frac{1}{|\Lambda_L|^M} \int_{\Lambda_L}
 \prod_{\gamma=1}^M \rd y_{\gamma} \; .
\label{ey}  
\ee

It is well known that the restriction of the random measure $\mu_\om$ 
(see \eqref{def:muom}) to the box $\Lambda_L$ has the same 
distribution as $\mu_\om' $.
In particular, given
a realization $\om$ of the infinite volume random measure $\mu_\om$,
we can associate to it the number of points in $\Lambda_L$ ($M=M(\om)$)
and  the operator $H'_\om= H_{L, M(\om)}$
with random measure $\mu_\om'$. We can thus realize the random
operator $H'_\om$ on the same probability space as $H_\om$.
Due to the periodic boundary 
and the nontrivial support of $B$, 
the potential of $H_\om$ and $H'_\om$ will not be the same on $\Lambda_L$,
but the difference will be negligible far away from the boundary.

Let $\chi_L$ be a smooth cutoff function, supported on $\Lambda_L$,
with $\chi_L\equiv 1$ on $\Lambda_{L/2}$ and $|\nabla \chi_L|\leq CL^{-1}$.
Let $\psi_{L}(t): = \chi_L e^{-itH}\psi_{0}'$ and let
$\psi'(t): = e^{-itH'} \psi_{0}'$ be the two dynamics
applied to the cutoff initial data $\psi_{0}':= \chi_L\psi_0$
supported on $\Lambda_L$. We also define the cutoff
observable $\cO_L:= \chi_L\cO$.
Clearly
\be
\lim_{L\to\infty}
\bE \langle \wh\cO_L, \wh W^\e_{\psi_L(t)} \rangle_L=
\bE \langle \wh\cO, \wh  W^\e_{\psi(t)} \rangle
\label{ocut}
\ee
for any $t$.
We  estimate
\be
   \partial_t \| \psi_L(t) - \psi_L'(t)\|_L^2 
  \leq C\| (H-H')\psi_L(t)\|^2_L + C\| [H,\chi_L]\psi_L(t)\|^2_L \; .
\label{Lgrom}
\ee
The second term is bounded by $CL^{-1}\| \nabla \psi_L(t)\|$
and on $\Omega_Z$ it  can be
estimated by the total energy as in \eqref{grom}. With a
propagation estimate similar to \eqref{finspeed}
but applied to the evolution
$e^{-itH}\psi_{0}'$, we easily obtain that the
right hand side of \eqref{Lgrom} vanishes as $L\to\infty$
for any $t$. On the complement set, $\om \in \Omega_Z^c$,
we use the uniform  bound \eqref{vani} and
finally let $Z\to\infty$.
In summary, we have shown the following

\begin{lemma}\label{lemma:L} Let $\psi'(t): = e^{-itH'_{L,M}} \psi_{0}'$,
where $M$ is a Poisson random variable with mean $|\Lambda_L|$, then
$$
\limsup_{L\to\infty}
   \Big| \bE \langle \wh \cO, \wh  W^\e_{\psi(t)}\rangle 
    -  \bE' \langle \wh \cO_L, \wh W^\e_{\psi'(t)}\rangle_L  \Big|= 0 \;
  \qquad \Box
$$
whenever $\int \sup_v|\wh\cO(\xi,v)|\rd \xi < \infty$. $\;\;\Box$
\end{lemma}

\section{The Duhamel expansion}\label{sec:stoprule}
\setcounter{equation}{0}

We expand the unitary kernel of $H= H_0 + \wt V$ (see \eqref{renH})
by the Duhamel formula. Due to the restriction to $\Lambda_L$,
we really work with $H_{L,M}' = H_{0}' + \wt V'$, where the 
renormalized free evolution, $H_{0}'$, is given by
$\om(p)$ in Fourier space 
 and $\wt V'= \lambda V' -\lambda^2\theta(p)$,  $p\in \Lambda^*_L$.
The prime indicates the restriction to $\Lambda_L$ and
the dependence  on $L$ and $M$. 
In this section we work on $\Lambda_L$ 
 but we will mostly omit the primes in the notation.

For any fixed integer $N\ge 1$
\begin{equation}\label{duh}
        \psi_t : = e^{-itH}\psi_0 = \sum_{n=0}^{N-1} \psi_n (t)
     + \Psi_{N}(t) \; ,
\end{equation}
with
\be
        \psi_n(t) : = (-i)^n\int_0^t [\rd s_j]_1^{n+1} \; \;
    e^{-is_{n+1}H_0}\wt V e^{-is_nH_0}\wt V\ldots
       \wt V e^{-is_1 H_0}\psi_0
\label{eq:psin}
\ee
being the fully expanded terms
and
\be
        \Psi_{N} (t): = (-i) \int_0^t \rd s \, e^{-i(t-s)H}
   \wt V \psi_{N-1}(s)
\label{eq:PsiN}
\ee
is the  non-fully expanded or error term. We used the shorthand notation
$$
    \int_0^t [\rd s_j]_1^n : = \int_0^t\ldots \int_0^t
 \Big(\prod_{j=1}^n \rd s_j\Big)
        \delta\Big( t- \sum_{j=1}^n s_j\Big) \; .
$$

Since each potential $\wt V$ in (\ref{eq:psin}), (\ref{eq:PsiN})
is a summation itself, $\wt V=
-\lambda^2\theta(p)+\sum_{\gamma=1}^M V_\gamma$,
$V_\gamma(x):= v_\gamma B(x-y_\gamma)$,
 both of these terms in (\ref{eq:psin}) and (\ref{eq:PsiN})
are actually big summations over so-called
 elementary wave functions,
which are characterized by their collision history, i.e. by a sequence
of obstacles and, occasionally, by $\theta(p)$.
Denote by
$\tGamma_n$, $n\le\infty$,  the set of sequences
\be\label{Gamman}
\tgamma = (\tgamma_1, \tgamma_2, \ldots , \tgamma_n), \qquad
\tgamma_j\in \{ 1, 2, \ldots , M\}\cup \{ \vartheta\}
\ee
and by $ W_\tgamma$ the associated  potential
$$
        W_\tgamma :=  \left\{ \begin{array}{cll} \lambda V_\tgamma & \qquad
\mbox{if} \quad & \tgamma\in \{ 1, \ldots, M\} \\
     - \lambda^2 \theta(p)  & \qquad \mbox{if} \quad & \tgamma =\vartheta
 \; . \end{array} \right.
$$
The tilde refers to the fact that the additional $\{ \vartheta\}$
symbol is also allowed. An element $\tgamma \in \{ 1, \ldots , M\}
\cup \{ \vartheta\}$
is identified with the 
potential $W_\tgamma$ and it is called {\it potential label}
if $\tgamma\in\{ 1, \ldots , M\}$,
otherwise it is a {\it $\vartheta$-label.}
 A potential label carries a factor $\lambda$,
a $\vartheta$-label carries $\lambda^2$.

For any $\tgamma\in\tGamma_n$ we define the
following fully expanded wave function with truncation
\be
    \psi_{*t, \tgamma}: = (-i)^{n-1}\int_0^t [\rd s_j]_1^{n} \; \; W_{\tgamma_n}
    e^{-is_nH_0} W_{\tgamma_{n-1}} \ldots
    e^{-is_2H_0} W_{\tgamma_1} e^{-is_1H_0} \psi_0
\label{eq:trunc}
\ee
and without truncation
\be
    \psi_{t, \tgamma}: =  (-i)^{n}\int_0^t [\rd s_j]_1^{n+1} \; \;
    e^{-is_{n+1}H_0} W_{\tgamma_n}
    e^{-is_nH_0} W_{\tgamma_{n-1}} \ldots
    e^{-is_2H_0} W_{\tgamma_1} e^{-is_1H_0} \psi_0\; .
\label{eq:exp}
\ee
In the notation the star $(*)$ will always refer to truncated functions.
Note that
$$
    \psi_{t,\tgamma} = (-i)\int_0^t \rd s \; e^{-i(t-s)H_0}
    \psi_{*s,\tgamma}\; .
$$
Each term (\ref{eq:exp})
 along the expansion procedure is characterized by its order
$n$ and by a sequence $\tgamma\in\tGamma_n$. We now identify the main
terms.

Denote by  $\Gamma_k^{nr} \subset \tGamma_k$ the set of 
{\it non-repetitive} sequences that contain only potential labels, i.e.
$$
    \Gamma_k^{nr}:=\Big\{
    \gamma = (\gamma_1, \ldots , \gamma_k) \; : \;
   \gamma_j\in\{1, \ldots, M\}, \;
 \gamma_i\neq \gamma_j \; \mbox{if} \; i\neq j\Big\} \; .
$$
Let
$$
   \psi_{t,k}^{nr}:= \sum_{\gamma\in \Gamma_k^{nr}} \psi_{t,\gamma} 
$$
denote the corresponding elementary wave functions. 

The typical
number of collisions up to time $t$ is of order $\lambda^2t$. 
To allow us for some room, we set 
\be
     K := [\lambda^{-\delta}(\lambda^2 t)]\; , 
\label{def:K}
\ee
($[ \; \cdot \; ]$ denotes integer part),
where $\delta=\delta(\kappa)>0$ is 
a small positive number to be fixed later on.
$K$ will serve as an upper threshold for the number of
collisions in the expansion.

\section{Proof of the Main Theorem}\label{sec:mainpr}
\setcounter{equation}{0}

The proof is divided into three theorems.
 The first one states that all terms other than $\psi_{t,k}^{nr}$,
$0\leq k< K$,
are negligible. 
For the precise statement we use the previous notations, in particular
we recall that 
the prime indicates the dependence on $L,M$

\begin{theorem} [$L^2$-estimate of the error terms]\label{7.1}
Let $t=O(\lambda^{-2-\kappa})$ and $K$ given by \eqref{def:K}.
If $\kappa < \kappa_0(d)$ and $\delta$ is sufficiently small (depending only
on $\kappa$),  then
$$
   \lim_{\lambda\to0}\lim_{L\to\infty}\bE' 
\Big\| \psi_t' -\sum_{k=0}^{K-1} \psi_{t,k}^{\prime \; nr} \Big\|^2_L =0 \; .
$$
In $d=3$ dimensions, one can choose $\kappa_0(3)=\frac{1}{370}$.
\end{theorem}

 The second key theorem 
gives an explicit formula
for the main terms,  $\psi_{t,k}^{\prime \; nr}$. It
really identifies the so-called ladder diagram as the
only contributing term. We introduce the notation
$$
        R_\eta(\a, v):= \frac{1}{\alpha - \om(v) + i\eta} \; , 
$$
for the renormalized propagator.

\begin{theorem} [Only the ladder diagram contributes] \label{thm:L2}
Let $\kappa<\frac{2}{33d+36}$,  $\e=\lambda^{2+\kappa/2}$,
 $t=O(\lambda^{-2-\kappa})$,
 and $K$ given by \eqref{def:K}.
 For a sufficiently small 
positive $\delta$, for
 $\eta = \lambda^{2+\kappa}$ and for  any $1\leq k < K$ we have
\be
   \lim_{L\to\infty} \bE' \|\psi_{t,k}^{\prime \; nr}\|^2_L=  V_\lambda(t, k)
    + O\Big( \lambda^{\frac{1}{3}-
(6+\frac{11}{2}d)\kappa-O(\delta)}\Big)
\label{eq:L2bound}
\ee
\be  
\lim_{L\to\infty} \langle \wh \cO_L, \bE' \wh W^\e_{\psi_{t,k}^{\prime \; nr}}
\rangle_L=  W_\lambda(t, k, \cO)
    + O\Big( \lambda^{\frac{1}{3}-
(6+\frac{11}{2}d)\kappa-O(\delta)}\Big)
\label{eq:Wbound}
\ee
as $\lambda\ll 1$. Here
\begin{align}
   V_\lambda(t, k): = & \frac{\lambda^{2k}e^{2t\eta}}{(2\pi)^2}
  \iint_{-\infty}^\infty \rd\a\rd\beta  \; e^{i(\a-\beta)t} 
 \int\Big(  \prod_{j=1}^{k+1}  \rd p_j \Big) \; 
|\wh\psi_0(p_1)|^2
\nonumber\\
&\times \prod_{j=1}^{k+1} \ov{ R_\eta(\a, p_j)} R_\eta(\beta, p_j)
  \prod_{j=1}^k |\wh B(p_{j+1}-p_j)|^2 \; 
\label{ladder} \\
W_\lambda(t, k, \cO): = & \frac{\lambda^{2k}e^{2t\eta}}{(2\pi)^2}
  \iint_{-\infty}^\infty \rd\a\rd\beta  \; e^{i(\a-\beta)t} \int \rd \xi
\int  \Big( \prod_{j=1}^{k+1} \rd v_j \;  \Big)
\wh\cO(\xi, v_{k+1})\overline{\wh W_{\psi_0}^\e}(\xi, v_1) 
  \nonumber \\
 & \times 
\prod_{j=1}^{k+1} 
  \ov{R_\eta\Big(\a, v_j +\frac{\e\xi}{2}\Big)}
   R_\eta\Big(\beta, v_j -\frac{\e\xi}{2}\Big)
  \prod_{j=1}^k |\wh B(v_j-v_{j+1})|^2 \; .
\label{9.51}
\end{align}
\end{theorem}
We   adopt the
notation $O(\delta)$ in the exponent of $\lambda$. This always
means $\mbox{(const.)}\delta$ 
with  universal, explicitly computable positive constants
that depend on $\kappa$ and that can be easily computed
 from the proof.

The formula \eqref{ladder} is the value of the so-called {\it ladder
Feynman graph} in the diagrammatic expansion of $\bE' \| \psi_{t,k}^{\prime \;
nr}\|^2$.
 We will see in Proposition \ref{prop:lump}
that this
expansion generates  $k!B_k$ terms, where $B_k$ is
the number of partitions of a set with $k$ elements
(note that $B_k$ is almost of order $k!$).
Theorem \ref{thm:L2} states that only one diagram is
relevant; the contribution of all the other Feynman graphs
 is negligible even after summation. The extension of 
\eqref{eq:L2bound}  to the Wigner transform  \eqref{eq:Wbound}
is straightforward. Theorem~\ref{thm:L2}
 is the most important step in the proof of the Main Theorem.

\medskip

The third theorem identifies the limit of $\sum_k W_\lambda(t, k,\cO)$
as $\lambda\to0$
with the solution to the heat equation. 
We note that the definition \eqref{9.51} does not apply literally
 to the free evolution term $k=0$; this term is defined
separately: 
\be
   W_\lambda(t, k=0,\cO)  : = \int \rd\xi\rd v\;
    e^{it\e v\cdot \xi}\; e^{2t\lambda^2 {\scriptsize \mbox{Im}}\,
 \theta (v)}\;
 \wh\cO(\xi, v)\ov{\wh W_0}(\e\xi, v) \; .
\label{xi0}
\ee

\begin{theorem} [The ladder diagram converges to the heat equation]
\label{thm:laddheat}
Under the conditions of Theorem~\ref{thm:L2} and setting
$t=\lambda^{-2-\kappa}T$, we have 
\be
  \lim_{\lambda\to0} \sum_{k=0}^{K-1} W_\lambda(t, k,\cO) = 
  \int \rd X  \int \rd v \; \cO(X, v)f(T, X, e(v))\; ,
\label{eq:laddheat}
\ee 
where $f$ is the solution to the heat equation \eqref{eq:heat}.
\end{theorem}

{\it Proof of the Main Theorem \ref{main} using
Theorems~\ref{7.1}, \ref{thm:L2}    and \ref{thm:laddheat}.}
We compute the expectation of the rescaled Wigner transform, 
$\bE W^\e_t = \bE W^\e_{\psi_t}$,
 tested against
a Schwarz  function  
$$
   \int  \rd X
   \int \rd v \;  \cO(X, v)  \bE  W_{t}^\e(X, v) = 
   \int \rd \xi \int \rd v \; \wh \cO(\xi, v) 
  \bE \ov{\wh W_{t}^\e}( \xi , v) 
  =\langle \, \cO,  \bE  W_{t}^\e \rangle\; .
$$
Combining
Lemma~\ref{lemma:L}, Theorem~\ref{7.1} and the finite box version of the
$L^2$-continuity of the Wigner transform (\ref{wigcont}),
it is sufficient to compute the Wigner transform of 
$\psi'(t, K): =  \sum_{k=0}^{K-1}\psi^{\prime \;  nr}_{t,k}$.
The Wigner transform $W_{\psi'(t, K)}$ 
is quadratic in $\psi'$, so it contains a
 double sum over $k$ and $k'$
$$
     W_{\psi'(t, K)} =\sum_{k,k'=0}^{K-1}\ov {\psi^{\prime \;  nr}_{t,k}}
 (\cdots)
 \psi^{\prime \;  nr}_{t,k'}(\cdots) \; .
$$
The potential labels are not repeated
within $\ov\psi$ and $\psi$. Moreover, 
 the expectation of a single potential in \eqref{eq:exp} is zero.
Thus the potential labels in the $\psi$ and $\ov\psi$ must pair,
in particular 
 taking expectation reduces 
this double sum to a single sum over $k$
$$
   \bE' \,  W_{\psi'(t, K)} =\sum_{k=0}^{K-1}
\bE'\,  W_{\psi^{\prime \;  nr}_{t,k}}\; .
$$ 
 By using \eqref{eq:Wbound}
and  \eqref{eq:laddheat} together with $K= O(\lambda^{-\kappa-\delta})$,
we obtain Theorem~\ref{main}. $\;\;\Box$

\medskip

The main result of the present paper is the proof of 
Theorem \ref{thm:L2}.
 The proofs of Theorem~\ref{7.1} and Theorem~\ref{thm:laddheat}
will be given in the companion paper 
\cite{ESY3}. 
 For the reader's convenience, we summarize below
the key ideas of the proof of  Theorem~\ref{7.1} from
 \cite{ESY3}. 

\medskip

The Duhamel expansion allows for the flexibility that
at every new term of the expansion we perform the
separation into elementary waves, $\psi_{*s,\wt\gamma}$,
and we can decide whether we want to stop (keeping the full propagator
as in \eqref{eq:PsiN}) or we 
continue to expand that term further.
This decision will depend on the collision history, $\wt\gamma$.
 In particular, not every error term will be expanded up to 
the same order $N$, in some cases we may
decide to stop the expansion earlier. 

To estimate a non-fully expanded term, 
we will use the unitarity of the full evolution,
\be
  \Big\| (-i)\int_0^t e^{-i(t-s)H} \psi_{*s,\wt\gamma} \rd s\Big\|^2
   \leq t\int_0^t \| \psi_{*s,\wt\gamma}\|^2 \rd s \; .
\label{unit}
\ee
Typically we lose a factor of $t$ by using this estimate since
the oscillatory character of the time integration
is lost. We can use this crude estimate only if the
fully expanded term, $\|  \psi_{*s,\wt\gamma} \|^2$,
is  small, i.e. if $\wt\gamma$ represents an atypical collision sequence.
Once $\wt\gamma$ is ``sufficiently'' atypical, we stop the expansion 
for that elementary wave function to reduce
the number and the complexity of the expanded terms.

There are basically two patterns how a collision history can become atypical;
either the total number of collisions exceeds the typical
number of collisions, $O(\lambda^2t)$, or there is a recollision.
This explains why only  the non-repetition terms $\psi_{t,k}^{nr}$ 
with $k\leq K$ contribute to the main term. 

A recollision is typically penalized by a factor 
$\lambda^2$ in the weak coupling
environment. This is, however, not the case for the immediate repetition of a
potential label,  
$\wt\gamma_j=\wt\gamma_{j+1}\in \{1, \ldots, M\}$. The renormalization
\eqref{renH} compensates for these terms.  Up to the highest order,
the contribution of  a sequence with an immediate repetition
cancels that of the same sequence where the
repetition is replaced by a $\theta$-label.  Technically,
all these estimates have to be combined with the key method of
the present paper (proof of Theorem~\ref{thm:L2})
to show that the sum of all $k!B_k$
repetition diagrams is sufficiently small to compensate
for the unitarity estimate \eqref{unit}.

The result of the current paper (Theorem \ref{thm:L2}) 
have been fundamentally used in
\cite{ESY3}. While that paper was already in print, we 
have improved the possible range of $\kappa$ and the 
exponent in the error bounds in 
Theorem \ref{thm:L2} (compare with Theorem
2.3 \cite{ESY3}) and,
therefore, several exponents in \cite{ESY3}
 can be improved. While these improvements are minor (and the
exponents used in \cite{ESY3} are still correct),
 we list them in the following for the convenience of the
readers. 
Instead of 
$\kappa< 2/(34d+39)$ required in Theorem 2.3 of \cite{ESY3},
the upper bound $\kappa <2/(33d+36)$ is sufficient.
The exponent
$\frac{1}{3}-\big(\frac{17}{3}d+\frac{13}{2}\big)\kappa -O(\delta)$
appearing in Theorem 2.3 of \cite{ESY3} has been
improved to $\frac{1}{3} - \big(6+\frac{11}{2}d\big)\kappa 
- O(\delta)$.
Using the improved exponent
in \eqref{eq:jointdeg} instead of $\frac{1}{3} - \big(\frac{17}{3}d
+ \frac{3}{2}\big)\kappa - O(\delta)$ in (5.18) of \cite{ESY3},
the exponent in (4.37) of Proposition 4.6 in \cite{ESY3}
can be improved to $\frac{1}{3} - \big(7 + \frac{11}{2}d\big)\kappa
-O(\delta)$. The better estimates on terms (I) and (II)
at the end of Sections 4.5 and 4.6 of \cite{ESY3} will lead 
to a somewhat better threshold $\kappa_0(d)$ for $\kappa$.
More precisely, the upper bound (4.31) 
of \cite{ESY3} is changed to  
$\kappa< (2q-48)/((33d+36)q+108+72d)$ and (4.32) of \cite{ESY3}
is changed to $\kappa<(2q-72)/((33d+36)q+108+72d)$
after correcting a typo $-(16+2d)\kappa$ to $ -(16+12d)\kappa$
in the exponent of the previous line.
In Section 4.6 of \cite{ESY3}, the bound (4.40) is
changed to $\kappa < 2/(18q+33d+96)$ and (4.41) is unchanged.
These bounds yield an explicit  $\kappa_0(d)>0$
depending on the dimension, so that Theorem 4.1 of \cite{ESY3}
holds for $\kappa < \kappa_0(d)$ (the explicit upper
bound in Theorem 4.1 was a typo, it should have been
$\kappa<\kappa_0(d)$). 
For $d=3$, explicitly  $\kappa_0(3)>\frac{1}{370}$.

\section{Pairing potential labels}\label{sec:pair}
\setcounter{equation}{0}

The wave function 
$$
    \psi_{t,k}^{\prime \; nr}=  (-i)^k\sum_{\gamma\in \Gamma_{k}^{nr}}
    \int_0^t [\rd s_j]_1^{k+1} \; \; e^{-is_{k+1}H_0'}V'_{\gamma_k}
    e^{-is_kH_0'} V'_{\gamma_{k-1}} \ldots
    e^{-is_2H_0'} V'_{\gamma_1} e^{-is_1H_0'} \psi_0'
$$
contains $k$  potential terms
with different potential labels.
Every term in
$$
    \bE' \|\psi_{t,k}^{\prime \; nr}\|^2_L = \sum_{\gamma, \gamma'}
    \bE'\; \ov{\psi_{t,\gamma}}\psi_{t,\gamma'}
$$
has $2k$ potential terms, and their expectation is
\be 
    \bE' \; \ov{ V_{\gamma_1}' V_{\gamma_2}' \ldots  V_{\gamma_k}'}
    V_{\gamma_1'}' V_{\gamma_2'}' \ldots  V_{\gamma_k'}'\; .
\label{2kex}
\ee
Since there is no repetition within $\gamma$ and $\gamma'$,
and $\bE' V_\gamma' =0$, the expectation in \eqref{2kex} is nonzero only if
there is a complete pairing between  $\gamma$ and $\gamma'$. Such  pairings
correspond to permutations on
$I_k=\{ 1,2, \ldots , k\}$. We denote by $\fS_k$ the set
of all permutations on $k$ elements.

We recall the $K$-identity from Lemma 3.1 of \cite{EY}
(with a corrected $(2\pi)^{-1}$ factor)
\be
    \int_0^t [\rd s_j]_1^{k+1} \prod_{j=1}^{k+1} e^{-is_j\om(p_j)}
    = \frac{ ie^{\eta t}}{2\pi}\int_{\bR} \rd \alpha \; e^{-i\alpha t}  \prod_{j=1}^{k+1}
    \frac{1}{\alpha - \om(p_j)+i\eta}
\label{eq:K}
\ee
for any $\eta>0$.
Therefore, we have
\begin{align}
    \bE' \|\psi_{t,k}^{\prime \; nr}\|^2_L 
 = \frac{\lambda^{2k}e^{2t\eta}}{(2\pi)^2}
    & \sum_{\sigma\in \fS_k}
    \int\!\!\int_{\Lambda_L^*}\rd\bp \rd \tbp \;
    \delta(p_{k+1}-\tp_{k+1})
\label{eq:psiM1}\\
    &\times\bE'  
\sum_{\gamma_1, \ldots ,\gamma_k=1\atop \gamma_i\neq\gamma_j}^M
     \prod_{j=1}^k 
  \overline{\wh V_{\gamma_j}(p_{j+1}-p_{j})}
     \wh V_{\gamma_j}( \tp_{\sigma(j)+1}-\tp_{\sigma(j)})
     \overline{\wh\psi_0'(p_1)}
    \wh\psi_0'(\tp_1)
\nonumber\\
&\times
\int\!\!\int_{\bR}\rd\alpha\rd\beta
    \; e^{i(\alpha-\beta)t}
    \Bigg(\prod_{j=1}^{k+1}
  \frac{1}{\alpha - \ov\om(p_j)-i\eta}\;
    \frac{1}{\beta -\om(\tp_j)+i\eta}\Bigg) \; ,
\nonumber
\end{align}
where the summation runs over all ordered $k$-tuples
$(\gamma_1, \ldots, \gamma_k)$ of $\{ 1, 2, \ldots , M\}$ with
disjoint elements.
We compute the expectation, using $m_2=1$ from \eqref{momm}  and 
the factorization of $\bE'$ from \eqref{facte}
\be
    \bE \prod_{j=1}^k  \overline{\wh V_{\gamma_j}(p_{j+1}-p_{j})}
     \wh V_{\gamma_j}( \tp_{\sigma(j)+1}-\tp_{\sigma(j)})=
   P(\sigma, \bp,\tbp) \ov{\cB(\bp)}\cB(\tbp)
\label{eq:treegraph}\ee
with
\be
       \cB(\bp): = \prod_{j=1}^k {\wh B}(p_{j+1}-p_{j})
\label{def:cB}
\ee
and
\be
   P(\sigma, \bp,\tbp) := \bE_M \bE_y^{\otimes M}
    \sum_{\gamma_1, \ldots, \gamma_k=1\atop \gamma_i\neq\gamma_j}^M
   \prod_{j=1}^k \exp\Big[ 2\pi i y_{\gamma_j}(p_{j+1}-p_{j}-
   (\tp_{\sigma(j)+1}-\tp_{\sigma(j)}))\Big]\; .
\label{def:P}
\ee
We obtain from \eqref{eq:psiM1} that
\be
    \bE' \|\psi_{t,k}^{\prime \; nr}\|^2_L = \lambda^{2k}
    \sum_{\sigma\in \fS_k}
    \int\!\!\int_{\Lambda_L^*}\rd\bp \rd \tbp \;
    \delta(p_{k+1}-\tp_{k+1}) \qquad\qquad \qquad\qquad
\label{eq:psiM}
\ee
$$
    \qquad\qquad
    \qquad\qquad\times  P(\sigma, \bp,\tbp)
     M^\circ(k,\bp,\tbp)\overline{\wh\psi_0'(p_1)}
    \wh\psi_0'(\tp_1)
$$
with
\be
    M^\circ(k, \bp,\tbp):=   \frac{e^{2t\eta} }{(2\pi)^2}
  \int\!\!\int_{\bR}\rd\alpha\rd\beta
    \; e^{i(\alpha-\beta)t}
    \Bigg(\prod_{j=1}^{k+1}
  \frac{\ov{\wh B}(p_{j+1}-p_j)}{\alpha - \ov\om(p_j)-i\eta}\;
    \frac{ \wh B( \tp_{j+1} - \tp_j)}{\beta -\om(\tp_j)+i\eta}\Bigg) \; 
\label{def:M}
\ee
(with the convention that for $j=k+1$ we set the
superfluous term $\wh B(p_{j+1}-p_j):=1$).

The expectation value in \eqref{def:P} can be easily computed
to yield a product of delta functions since the variables $y_{\gamma_j}$
are independent. The constraint $\gamma_i\neq\gamma_j$ induces
only a trivial combinatorial factor that becomes
irrelevant in the $L\to\infty$ limit.

If the obstacle centers were deterministic, i.e., 
$y_{\gamma_j}=\gamma_j$, then 
the constraint $\gamma_i\neq\gamma_j$ has more serious consequences.
This is the case for the lattice Anderson model, where the
summation in \eqref{def:P} extends to all $\gamma_j\in \bZ^d$
with $\gamma_i\neq\gamma_j$ and the momentum variables are
on the dual torus, $\Tor^d:=[-\frac{1}{2}, \frac{1}{2}]$.
Due to the constraint  $\gamma_i\neq\gamma_j$, the formula
\eqref{def:P} is not a simple
product of delta functions
and we have to  use a connected graph expansion 
that is well-known from statistical physics.

Let $\cA_n$ be the set of partitions of $I_k:=
\{ 1,2, \ldots, k\}$,
i.e. $\bA =\{ A_\mu \; : \; \mu\in I(\bA) \} \in\cA_k$
if $\cup_{\mu\in I(\bA)} A_\mu =I_k$
 and the elements of $\bA$ are
disjoint and non-empty.  The sets in the partition are labelled
by the index set $I(\bA)$ and let $m(\bA)=|I(\bA)|$
denote the number of elements in
$\bA$. The elements of the partition $\bA$ will 
be called {\it lumps}. A lump is {\it trivial} if it
has only one element. The trivial partition,
where every lump is trivial, is denoted by $\bA_0$.

\begin{lemma}\label{lemma:con}
(i) [Continuum model] For any fixed $L$, $k$ and $M$, $k\leq M$, and any fixed 
momenta $q_j\in \Lambda_L^*$,
\be
    \bE_y^{\otimes M}
\sum_{\gamma_1, \ldots ,\gamma_k=1\atop \gamma_i\neq\gamma_j}^M
    \prod_{j=1}^k \exp\Big[2\pi i q_jy_{\gamma_j}\Big] = 
    \frac{M!}{|\Lambda_L|^k(M-k)!}\sum_{\bA \in\cA_k}
    \prod_{\nu\in I(\bA)} c(|A_\nu|)\delta\Big(\sum_{\ell \in A_\nu}
     q_\ell\Big)
\label{eq:conngr}
\ee
with $c(1)=1$, $c(n)=0$ for any $n\ge 2$.

(ii) [Lattice model] For any fixed $k$ and momenta $q_j\in \Tor^d$,
\be
   \sum_{\gamma_1, \ldots, \gamma_k\in \bZ^d\atop \gamma_i\neq \gamma_j}
 \prod_{j=1}^k \exp\Big[2\pi i q_j\gamma_j\Big] = 
\sum_{\bA \in\cA_k}
    \prod_{\nu\in I(\bA)} c(|A_\nu|)\delta\Big(\sum_{\ell \in A_\nu}
     q_\ell\Big)
\label{eq:conngr1}
\ee
with
$$
     c(n): = \sum_{\Gamma\subset K_n\atop\Gamma \; connected}
     (-1)^{|\Gamma|}
$$
where $K_n$ denotes the complete graph on $n$ vertices
and $|\Gamma|$ denotes the number of edges in
the subgraph $\Gamma$.
The following estimate holds for $n\ge 2$
\be
     |c(n)|\leq n^{n-2} \; .
\label{eq:ajest}\ee
\end{lemma}

 {\it Remark 1.}  
Recall that $M$ is a Poisson random variable with
expectation $|\Lambda_L|$.
Therefore, apart from the prefactor that converges to 1
almost surely as $L\to \infty$,
the right hand side of \eqref{eq:conngr} is simply $\prod_{j=1}^k
\delta(q_j)$. With an obvious  choice of $c(n)$ we
write it in the same form as \eqref{eq:conngr1}. In this way
the continuum and lattice models can be treated simultaneously.
The explicit form of $c(n)$ will not be needed.
The arguments in the sequel will use only the bound \eqref{eq:ajest}
that is valid for both choices of $c(n)$.

{\it Remark 2.}
 Analogous formulas hold if the natural
index set $I_k=\{1 , 2, \ldots , k\}$
is replaced by an arbitrary finite set $S$. In this case, the
summation on the right hand side of \eqref{eq:conngr}--\eqref{eq:conngr1}
 is over all partitions of $S$.
The set of these partitions is denoted by $\cA(S)$.

\bigskip

{\it Proof of Lemma \ref{lemma:con}.} 
Part (i) is straightforward from the definition of $\bE_y$.
For part ii) we use the connected graph expansion
$$
    \prod_{i\neq j=1}^k (1-\delta_{\gamma_i,\gamma_j})
    =\sum_{\bA\in\cA_k} \prod_{\nu\in I(\bA)} \delta_c(A_\nu) \; ,
$$
where
$$
    \delta_c(A) = c(|A|) \prod_{\ell,\ell'\in A} 
\delta_{\gamma_\ell,\gamma_{\ell'}}
$$
is the Ursell coefficients of the hard-core lattice gas (see: e.g. \cite{S}).
Therefore
\bey
   \mbox{L.h.s of }\; \eqref{eq:conngr1}
    &=& \sum_{\gamma_1, \ldots, \gamma_k\in \bZ^d}
    \sum_{\bA\in\cA_k} \prod_{\nu\in I(\bA)}\Bigg(
    e^{2\pi i\sum_{\ell\in A_\nu} q_\ell \gamma_\ell}
    c(|A_\nu|)\prod_{\ell,\ell'\in A_\nu} \delta_{\gamma_\ell,\gamma_{\ell'}}
    \Bigg)\nonumber\\
    &=&\sum_{\bA \in\cA_k}
    \prod_{\nu\in I(\bA)} \Big[ c(|A_\nu|)
    \delta(\sum_{\ell \in A_\nu}
     q_\ell)\Big] \; .  \qquad\Box
\nonumber
\eey
We will use the identity \eqref{eq:conngr} to express $P$ 
in \eqref{def:P}
as a linear combination of products
of delta functions of the momenta and insert 
it into \eqref{eq:psiM}. After the limit $L\to\infty$, 
each term in the summation  $\sum_\sigma\sum_\bA$
 will be expressed by a Feynman graph. The precise definitions
will be given
in the next Section.

\section{Graphical representation}\label{sec:gra}
\setcounter{equation}{0}

 Traditionally,
the Feynman graphs consist of interaction vertices 
and particle lines among them. In case of Gaussian random
potentials, the interaction vertices are paired
according to the Wick theorem \cite{EY}. For non-Gaussian
randomness, the non-vanishing  higher order cumulants
correspond to joining several vertices \cite{Ch}. 
In our case, the appearance of the non-trivial
subsets are due to selecting the non-repetition sequences.
This requires us to define Feynman graphs in a more general
setup than usual. In this section we  introduce
the necessary graphical representation in full generality
and we will 
define the value of a Feynman graph, $V^\circ(\bA, \sigma)$, with
 permutation $\sigma$ and partition $\bA$ in \eqref{def:VsigmaA}. 
The final result of this section is given in Proposition \ref{prop:lump}
at the end.

\subsection{Circle graphs and their values}\label{sec:genest}

We start with an oriented circle graph with two 
distinguished vertices, denoted by $0$, $0^*$. The number
of vertices is $N$.
The vertex set is $\cV$, the set of oriented edges  is $\cL(\cV)$.
For $v\in \cV$ we use the notation $v-1$ and  $v+1$ for
the vertex right before and after $v$ in the circular ordering.
We also denote $e_{v-}=(v-1,v)$ and $e_{v+}=(v,v+1)$ 
the edge right before and after
the vertex $v$, respectively. In particular  $e_{(v+1)-} = e_{v+}$.
For each $e\in\cL(\cV)$, we introduce a momentum $w_e$ and
a real number $\alpha_e$ associated to this edge. The
collection of all momenta is denoted by $\bw=\{ w_e \; :\; e\in \cL(\cV)\}$
and $\rd\bw = \otimes_e \rd w_e$ is the Lebesgue measure.
We sometimes use the notation $v\sim e$ to indicate
that an edge $e$ is adjacent to a  vertex $v$.

Let $\bP = \{ P_\mu \; : \; \mu \in I\}$ be a
 partition of the set $\cV \setminus\{0, 0^*\}$
$$
        \cV\setminus\{0, 0^*\} = \bigcup_{\mu\in I}P_\mu \; ,
$$
(all $P_\mu$ nonempty and pairwise disjoint)
where $I=I(\bP)$ is the index set to label the sets in
the partition. Let $m(\bP):= |I(\bP)|$. The sets $P_\mu$ are called
$\bP$-{\it lumps} or just {\it lumps}.  If two elements  
$v,v'\in\cV \setminus\{0, 0^*\}$ belong to the same lump
within a partition $\bP$, we denote it by $v\equiv v' \; (mod \; \bP)$.
We  assign a variable, $u_\mu\in \bR^d$, $\mu \in I(\bP)$,
to each lump. We call them {\it auxiliary momenta};
they will be needed for a technical reason.
We  always assume that the
auxiliary momenta add up to 0
\be
    \sum_{\mu\in I(\bP)} u_\mu =0\; .
\label{sumu}
\ee

 The set of all partitions of the vertex
set $\cV \setminus\{0, 0^*\}$  is denoted by $\cP_\cV$.
For any $P\subset\cV$, we
let
$$
L_+(P) : = \{ (v,v+1)\in\cL (\cV)\; : \; v+1\not\in P, \; v\in P\}
$$
denote the set of  edges that go out of $P$, with respect to the
orientation of the circle graph, and similarly
 $L_-(P)$ denote the set of  edges that go into $P$. 
We set $L(P):= L_+(P) \cup L_-(P)$.

For any $\xi\in\bR^d$ we define the following product of delta functions
 \be
   \Delta(\bP, \bw, \bu): = 
\delta \Big(  \xi+\sum_{e\in L_\pm( \{ 0^*\} )} \pm w_e\Big)
\prod_{\mu\in I(\bP)} \delta\Big( \sum_{e\in L_\pm(P_\mu)}
    \pm w_e - u_\mu\Big) \; ,
\label{def:Delta}
\ee
where $\bu: = \{ u_\mu\; : \; \mu\in I(\bP)\} \in \bR^d$ is
a set of auxiliary momenta.
The sign $\pm$ indicates that momenta $w_e$ is added or subtracted
depending whether the edge $e$ is outgoing or incoming, respectively.
The function $\Delta(\cdots)=\Delta_\xi(\cdots)$
 depends on $\xi$, but we will mostly omit
this fact from the notation. All estimates will be uniform in $\xi$.

Summing up all arguments of these delta functions and
using (\ref{sumu}) we see that these delta functions
force the two momenta corresponding to the
two edges adjacent to $0$ to differ by $\xi$: $w_e-w_{e'}=\xi$
for $e\in L_+(\{ 0\})$, $e'\in L_-(\{ 0\})$.

\medskip

As a motivation for these definitions, we mention that
the lumps  naturally arise from the connected graph
formula (Lemma \ref{lemma:con}).  According to this formula,
the Kirchoff Law must be satisfied for all lumps,
i.e. the incoming and outgoing momenta must sum up to zero.
This fact would be described by the delta functions
\eqref{def:Delta} with all $u_\mu=0$. In  certain
recollision terms, however, the non-repetition 
condition leading to Lemma \ref{lemma:con}
 is not fully satisfied and the Kirchoff Law breaks down
for a few lumps.
 The  nontrivial auxiliary momenta will 
bookkeep this deviation from the Kirchoff Law
 (see \cite{ESY3} for more details).
Finally, the shift by $\xi$ at the vertex $0^*$ in \eqref{def:Delta}
 will be used when
computing the Wigner transform in Fourier representation
\eqref{FW}.

\medskip

For each subset $\cG\subset \cV\setminus \{ 0, 0^*\}$,
 we define
\be
   \cN_{\cG}(\bw): = \prod_{e\sim 0}|\wh \psi_0(w_e )|
 \prod_{v\in \cV \setminus\{ 0, 0^*\}\setminus\cG} 
|\wh B(w_{e_{v-}} - w_{e_{v+}})|  \prod_{v\in \cG} 
\langle w_{e_{v-}} - w_{e_{v+}} \rangle^{-2d} \; .
\label{def:cN}
\ee
In our application,
the subset $\cG$ collects those vertices, where the original
potential decay $|\wh B(w_{e_{v-}} - w_{e_{v+}})|$ could not
be explicitly kept along the estimates
 and this will happen
only at a few places; the size of $\cG$ will be at most 8.
 For the purpose
of this paper, i.e. for the proof 
of Theorem \ref{thm:L2},
 we will need only $\cG=\emptyset$, but
for the analysis of the repetition terms in 
 \cite{ESY3} we need the more general definition.

Due to the support properties of $\wh B$ and $\wh \psi_0$, we will
see that all intermediate momenta $w_e$  satisfy
$|w_e|\leq  N\lambda^{-\delta}$. The maximal number of
vertices in our graphs will be $N\leq 2K+2 
= O(\lambda^{-\kappa-\delta})$, therefore all intermediate
momenta will be smaller than $\zeta:=\lambda^{-\kappa-3\delta}$.
This justifies to define the restricted Lebesgue measures 
\be
  \rd\mu(w): = {\bf 1}(|w|\leq \zeta) \rd w\; ,
  \quad
  \zeta:= \lambda^{-\kappa-3\delta}, 
 \qquad \rd \mu(\bw) : = \otimes_e \rd \mu (w_e) \; .
\label{def:mu}
\ee
Moreover, each auxiliary momenta, $u_\mu$,
 will always be a sum (or difference)
of different $w_e$ momenta (see \eqref{def:Delta}), therefore each
of them always satisfies $|u_\mu|\leq O(\lambda^{-2\kappa-4\delta})$.
We will often take the supremum of all possible auxiliary momenta
and $\sup_\bu$ is always considered subject to this bound.

With these notations, we define,
 for any $\bP\in \cP_\cV$ and $g=0,1,2,\ldots$, the 
{\bf $E$-value of the partition}
\be
    E_g(\bP, \bu,\balpha): =\lambda^{N-2} \sup_{\cG \; : \; |\cG|\leq g}
    \int \rd\mu(\bw)\prod_{e\in \cL(\cV)}   \frac{1}{|\alpha_e- \om(w_e)+i\eta|}
   \; \Delta(\bP, \bw, \bu) \cN_{\cG}(\bw)  \; .
\label{def:E}
\ee
The prefactor $\lambda^{N-2}$ is due to the fact that in
the applications all but the two distinguished  
vertices, $\{ 0, 0^*\}$, will carry a factor $\lambda$. 
The $E$-value depends also on the parameters $\lambda, \eta$, 
but we will not specify them in the notation.
In the applications,
the regularization $\eta$ will be mostly chosen 
as $\eta=\lambda^{2+\kappa}$.

We will also need a slight modification of these definitions, indicated
by a lower star in the notation:
\be
    E_{*g}(\bP, \bu,\balpha): =\lambda^{N-2}\sup_{\cG \; : \; |\cG|\leq g}
    \int \rd\mu(\bw)\prod_{e\in \cL(\cV)\atop
      e\not\in L( \{ 0^*\} )}   \frac{1}{|\alpha_e- \om(w_e)+i\eta|}
    \; \Delta(\bP, \bw, \bu) \cN_\cG(\bw)\; .
\label{def:E*}
\ee
The only difference is that the denominators carrying the momenta associated
to edges that are adjacent to $0^*$ are not present in $E_{*g}$.
We call $E_{*g}$ the {\bf truncation} of $E_{g}$.
We will see that Feynman diagrams arising
from the perturbation expansion can naturally be
estimated by quantities of the form (\ref{def:E}) or (\ref{def:E*}).

\bigskip
\subsection{Feynman graphs}\label{sec:FG}

We apply this general setup to the following situation
that we will call {\bf Feynman graph}. Every quantity
in our perturbation expansion will be expressed by
values of Feynman graphs that are defined below.

For experts we mention that our Feynman graphs differ 
from those that one typically
  obtains after averaging over a Gaussian disorder. 
In the latter, potential lines never appear as 
external lines but only as pairing lines, 
and one can identify
vertices connected by pairing lines 
so that the graph becomes four--valent.
In our case, the graph is still trivalent 
and has external potential lines,
with a corresponding dependence on momentum variables $\bf u$.  
Also, averaging the disorder will not simply pair up lines 
but can also join more than two potential lines.
which correspond to the higher moments.

Consider the cyclically ordered set $\cV_{n,n'}: =
\{0, 1, 2, \ldots, n, 0^*, \tilde n', \wt{n'-1},
 \ldots ,\tilde 1 \}$ and view this as the
 vertex set of an oriented circle graph on $N=n+n'+2$
 vertices. We set $I_n:=\{ 1, 2, \ldots n\}$
and $\wt I_{n'} :=\{ \wt 1, \wt 2, \ldots, \wt n'\}$
and the vertex set can be identified with $\cV_{n,n'}= I_n\cup \wt I_{n'}
\cup\{ 0, 0^*\}$.

The set of edges $\cL(\cV_{n,n'})$ is 
partitioned into $\cL(\cV_{n,n'}) = \cL\cup
\wt\cL$ such that $\cL$ contains the edges between $I_n\cup \{0,0^*\}$
and $\wt\cL$ contains the edges between $\wt I_{n'}\cup \{0,0^*\}$.

Let $\cP_{n,n'}$ be the set of all partitions $\bP$
on the set $I_n \cup \wt I_{n'}$. The lumps of a partition
containing only one vertex will be called {\it single lumps.}
The vertices $0$ and $0^*$ are not part of the partitions
hence they will not be considered single lumps.
Let $G=G(\bP)$ be the set of edges that go into  a single lump 
 and let $g(\bP):= |G(\bP)|$ be its cardinality.
In case of $n=n'$, we will use the shorter 
notation $\cV_n=\cV_{n,n}$, $\cP_n=\cP_{n,n}$
etc. The Feynman graphs arising from  the non-repetition
terms will always have $n=n'$ and no single lumps, $g(\bP)=0$,
 but the more
general definition  will be needed for the repetition
terms in \cite{ESY3}. 
We remark that even in \cite{ESY3} we will always have
\be
|n-n'|\leq g(\bP) \leq 4, \qquad n,n'\leq K\; .
\label{nn'}
\ee

We also introduce a function $Q$ that will represent 
the momentum dependence of the observable. In our estimates, we 
 will always bound $Q$ in supremum norm;
no decay or smoothness will be necessary. We will need extra
conditions on the observable only 
to evaluate the ladder in the proof of Theorem~\ref{thm:laddheat} 
(see \cite{ESY3} for details).
Since $Q$ will always appear linearly in our formulae,
we can assume, for convenience, that $\|Q\|_\infty\leq 1$.
General $Q$ can be accommodated by a multiplicative factor
$\|Q\|_\infty$ in the final estimate but it will not be
carried along the proofs.

We define the following function to collect all potential terms:
\begin{align}
   \cM(\bw) :=  & \prod_{e \in \cL\cap G}[- \ov{\theta(w_e)}]
 \prod_{e \in \wt\cL\cap G}[- \theta(w_e)] \; 
\prod_{e \in \cL\setminus G\atop e\not\sim 0^*}  \ov{\wh B(w_e- w_{e+1})} \;
\prod_{e \in \wt\cL\setminus G\atop e\not\sim 0}  \wh B(w_e- w_{e+1}) \;  
\label{def:cM}  
\\
& \times\ov{\wh\psi(w_{e_{0+}})}\wh\psi(w_{e_{0-}}) Q\Big[ \frac{1}{2}
  (w_{e_{0^*-}} +w_{e_{0^*+}}) \, \Big] \nonumber
\end{align}
with $\bw: =\{ w_e\; : \; e\in \cL\cup\wt\cL\}$ and recalling that
for any $e\in \cL(\cV)$,
 the edge $e+1$ denotes the edge succeeding $e$ in
the circular ordering.

The delta function $\Delta (\bP, \bw, \bu\equiv 0)$ 
ensures that the two momenta adjacent to each single lump coincide.
This holds even for $\xi\neq 0$, recall that $0, 0^*$ 
are not considered lumps.
Therefore the distribution
$\cM(\bw)\Delta(\bP, \bw, \bu\equiv 0)$ is supported on the regime with
$|w_e|\leq \zeta$ for all momenta $w_e$,
 thanks to the support properties of $\wh\psi_0$,
$\wh B$ and to the control on the number of
terms, $n,n'\leq K$ (see Section \ref{sec:genest}).
 In particular 
\be
    \cM(\bw)\Delta(\bP, \bw, \bu\equiv 0)
 \rd \bw = \cM(\bw)\Delta(\bP, \bw, \bu\equiv 0)\rd\mu(\bw) \; .
\label{suppM}
\ee
Using the boundedness of $\theta$
 and $|\wh\psi_0(w)|\leq C\langle w
\rangle^{-10d}$,  we easily obtain
\be
  | \cM(\bw)|\Delta(\bP, \bw, \bu\equiv 0)\rd\bw \leq
 C^{g(\bP)} 
\cN_{\cG}(\bw)\Delta(\bP, \bw, \bu\equiv 0)\rd\mu(\bw)\;  ,
\label{NleqM}
\ee
where $\cG$ is the set of single lumps and  $g(\bP)=|\cG|$, since
the delta function  also guarantees
that there is no additional decay at the vertices $v\in \cG$
in $\cN_\cG(\bw)$ (the last product in \eqref{def:cN} is a constant).

Let $\a, \beta\in\bR$, $\bP\in \cP_{n,n'}$ and
\be
   V(\bP,\a,\beta): =\lambda^{n+n'+g(\bP)}
\int \rd\bw
    \prod_{e\in\cL} \frac{1}{\alpha- \ov\om(w_e) - i\eta} \prod_{e\in\wt\cL}
        \frac{1}{\beta- {\om}(w_e) + i\eta}
\label{def:Vlong}
\ee
$$
    \times  \Delta(\bP, \bw, \bu\equiv 0) \cM (\bw)\; .
$$
Thanks to \eqref{suppM}, the integration measure could be
changed to $\rd\mu(\bw)$.
The truncated version, $V_{*}(\bP,\a,\beta)$,
is defined analogously
but the $\alpha$ and $\beta$ denominators that
 correspond to $e \in L( \{ 0^*\} )$ are removed.

We set $Y:=\lambda^{-100}$ and define   
\be
       V_{(*)}(\bP): = \frac{e^{2t\eta}}{(2\pi)^2}\iint_{-Y}^Y
        \rd\alpha\rd\beta \; e^{it(\alpha-\beta)}  V_{(*)}(\bP,\a,\beta)
\label{def:Vshort}
\ee
and
\be
        E_{(*)g}(\bP,\bu):= \frac{e^{2t\eta}}{(2\pi)^2}
        \iint_{-Y}^Y \rd\alpha\rd\beta \;  E_{(*)g}(\bP,\bu,\balpha) \; ,
\label{def:Eshort}
\ee
where $\balpha$ in $ E_{(*)g}(\bP,\bu,\balpha)$
is defined as $\alpha_e = \alpha$ for $e\in\cL$ and
 $\alpha_e:=\beta$ for $e\in\wt\cL$. The notation $(*)$ 
indicates the same formulas with and without truncation.
We will call these numbers 
the {\it $V$-value} and
{\it $E$-value of the partition $\bP$}, or sometimes, 
of the corresponding Feynman graph.
Strictly speaking, the $V$- and the $E$-  values
 depend on  $\xi$ through $\Delta=\Delta_\xi$.
When this dependence is important, we will make it explicit in the notation,
e.g. $V=V_\xi$. 
The $V$-value depends on the choice of  $Q$ as well. When necessary,
the notations $V_\xi(\bP; Q)$ will indicate
this fact.

Clearly, by using (\ref{NleqM}),
\be
        \big| V_{(*)}(\bP)\big|\leq (C \lambda)^{g(\bP)}\;
        E_{(*)g}(\bP,\bu\equiv 0) \; 
\label{eq:VleqE}
\ee
with $g=g(\bP)$.
We will use the notation $E_{(*)g}(\bP):=E_{(*)g}(\bP, \bu \equiv 0)$.

As we will see in  \eqref{eq:K}, for the graphical representation
of the Duhamel expansion we will really need 
\be
 V_{(*)}^\circ(\bP):= \frac{e^{2t\eta}}{(2\pi)^2}\iint_\bR \rd\alpha\rd\beta \;
     e^{it(\a-\beta)}V_{(*)}(\bP,\alpha,\beta)\; ,
\label{def:circ}
\ee
i.e. a version of
 $V_{(*)}(\bP)$ with unrestricted $\rd\alpha\, \rd\beta$ integrations.
 (The circle superscript in $V^\circ$  will refer
to the unrestricted version of $V$). However,
 the difference
between the restricted and unrestricted $V$-values are
negligible even when we sum them up for all partitions:

\begin{lemma}\label{lemma:VV} Assuming that $\eta\ge \lambda^{2+4\kappa}$ and
\eqref{nn'},
we have
\be
   \sum_{\bP\in \cP_{n,n'}}  \Big| V_{(*)}(\bP)-  V_{(*)}^\circ(\bP)\Big| 
  = O(\lambda^{5(n+n')}) \; .
\label{def:EMout}
\ee
The same result holds if $V_{(*)}(\bP)$ were defined
by restricting the $\alpha,\beta$-integral to
any domain that contains
 $[-Y, Y]\times [-Y, Y]$.
 \end{lemma}

{\it Proof.} First we consider the case $n,n'\ge 1$.
To estimate the difference, we consider
the integration domain where either $|\a|\ge Y$ or $|\beta|\ge Y$.
We assume, for definiteness, that $|\a|\ge Y$,
and we estimate all $\a$ denominators trivially,
$$
    \frac{1}{|\a -\ov{\om} (w_e)-i\eta|} \leq \frac{C}{\langle \a\rangle} \; ,
$$
by using that 
$|\om (w_e)|\leq \frac{1}{2} w_e^2 + O(\lambda^2) \leq \frac{1}{2}Y
+ O(\lambda^2)$ on the support of $\rd \mu(w_e)$.
Then we estimate all but the last 
 $\beta$-denominators  in \eqref{def:Vlong} trivially 
 by $\eta^{-1} \langle \beta /Y \rangle^{-1}$.
Thus all $w_e$ integrations are trivial except the last one
where we use  \eqref{eq:logest}.
Thanks to the bounds $|\wh \psi(w)|, |\wh B(w)|
\leq C\langle w\rangle^{-10d}$, one easily obtains that
$$
   |V(\bP,  \a, \beta)| \leq 
  \frac{(C\lambda)^{n+n'+g(\bP)} |\log\lambda| \; \
\log \langle\beta\rangle}{ \langle\a
\rangle^{n+1} \eta^{n'} \langle \beta/Y \rangle^{n'} 
 \langle \beta \rangle^{1/2}} \; .
$$
Therefore, we have 
$$
   \int_\bR \rd \beta \int_{\{|\a|\ge Y\}}  \rd \a \; \;
 |V(\bP,  \a, \beta)| \leq 
 \frac{(C\lambda)^{n+n'+g(\bP)} |\log\lambda| }{ \eta^{n'} 
 Y^{n-\frac{1}{2} -2\delta}} = O(\lambda^{6(n+n')})
$$
by using \eqref{nn'}.   
Similar bound holds for the truncated values,  $V_{*}$.
Thus
\be
  \Big| V_{(*)}(\bP)-  V_{(*)}^\circ(\bP)\Big|  = 
O(\lambda^{6(n+n')}) \; .
\label{def:EMout1}
\ee
Since the total number of partitions, $|\cP_{n,n'}|$,
 is smaller than $(n+n')^{n+n'}$
and in our applications $n,n'\leq K \ll \lambda^{-\kappa-2\delta}$,
we see that the restriction of the $\alpha,\beta$-integral to
any domain that contains
 $[-Y, Y]\times [-Y, Y]$
has a  negligible effect of order $O(\lambda^{5(n+n')})$
 even after summing up all $V$-values.

If the condition  $n,n'\ge 1$ is not  satisfied,
 say the number of   $\alpha$-denominators  is one ($n=0$),
then we will not introduce the
auxiliary variable $\a$ as in \eqref{eq:K},
because the $\int\rd\alpha$
integral would be logarithmically divergent after taking
the absolute value. In this case,  we use the  definition
$$
    V^\circ(\bP): = \frac{e^{t\eta}}{2\pi}\int_\bR \rd\beta \; e^{-it\beta} 
    V(\bP, \beta)
$$
with
$$
   V^\circ(\bP,\beta): =\lambda^{n+n'+g(\bP)}
\int \Big(\prod_{e\in \cL\cup\wt \cL} \rd w_e\Big)
     \prod_{e\in\wt\cL}
        \frac{1}{\beta- {\om}(w_e) + i\eta}
$$
$$
   \times e^{it \ov{\om}(w)}\Delta(\bP, \bw, \bu\equiv 0) \cM(\bw) \; ,
$$
directly, instead of $V^\circ$ given in \eqref{def:circ}.
Similar modifications hold for the other cases ($n'=0$, $n\ge1$ and
$n'=n=0$)  as well.
In particular, in our expansion (including all the cases in \cite{ESY3})
only a few such graphs may appear due to $|n-n'|\leq 2$.
The estimates leading to \eqref{def:EMout}
in these cases are similar but much easier than
in the general case  and they are left to the reader
(the same estimates were covered in \cite{EY} as well,
 without the renormalization of 
the dispersion relation).  $\;\;\Box$

\medskip

Sometimes we will use a numerical labelling of the edges
(see Fig.~\ref{fig:relab}).
In this case, we label the edge between $(j-1, j)$
 by $e_j$, the edge between $(\wt j, \wt{j-1})$
by $e_{\tilde j}$. At the special vertices $0, 0^*$ we
denote the edges as follows: $e_{n+1}:=(n, 0^*)$,
$e_{\wt{n'+1}}:=(0^*, \tilde n')$, $e_1=(0,1)$
 and $e_{\wt 1}:=(\wt 1, 0)$.
Therefore the
edge set $\cL=\cL(\cV_{n,n'})$ is identified with the index
set $I_{n+1} \cup \wt I_{n'+1}$ and
we set 
\be
p_j: = w_{e_j}, \qquad\tp_j: = w_{e_{\wt j}}.
\label{iden}
\ee
These two notations will sometimes  be used in parallel:
the $\bp$, $\tbp$ notation is preferred when  distinction is
needed between momenta on $\cL$ and $\wt\cL$ edges and
the $\bw$ notation is used when no such distinction
is necessary. Note that   we always
have
\be
    p_1-\tp_1=\xi \; .
\label{peqp}
\ee

\bef\bec
\epsfig{file=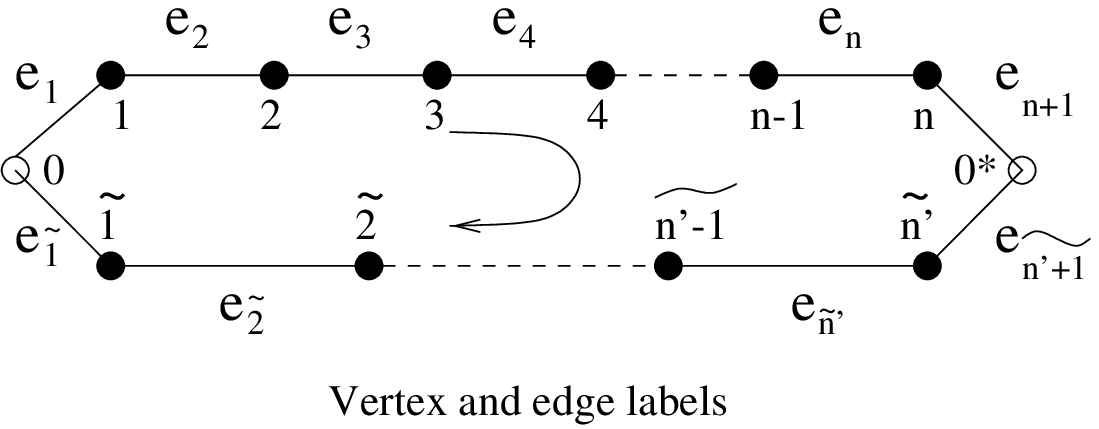,scale=.8}
\eec
\caption{Vertex and edge labels}\label{fig:relab}
\eef

\subsection{Non-repetition Feynman graphs}

A partition $\bP\in \cP_n$ of $I_n\cup\wt I_n$ is called {\bf even} if
for any $P_\mu\in \bP$ we have $|P_\mu\cap I_n|=|P_\mu\cap
\wt I_n|$. In particular, in an even
partition there are no single lumps,
 $G(\bP)=\emptyset$.

Let $\fS_n$ be the set of permutations on $I_n$ and
let $id$ be the identity permutation.
Note that $\bA\in \cA_n$ and $\sigma\in \fS_n$,
uniquely determine an even partition in $\bP(\bA, \sigma) \in \cP_n$,
by $I(\bP):=I(\bA)$  and $P_\mu : = A_\mu \cup \sigma(A_\mu)$.

Conversely, given an even partition $\bP\in \cP_n$,
we can define its projection onto $I_n$, $\bA:=\pi(\bP)\in \cA_n$,
 by $I(\bA):=I(\bP)$ and $A_\mu: = P_\mu\cap I_n$.
We let
$$
        \fS_n(\bP): = \{ \sigma\in \fS_n\; : \; \bP( \pi(\bP),\sigma)= \bP\}
$$
be the set of permutations that are {\bf compatible} with
a given even partition $\bP$. In other words, $\sigma\in \fS_n(\bP)$
if for each $i\in I_n$  the pair $(i, \sigma(i))$ belongs
to the same $\bP$-lump.
Clearly
\be
        |\fS_n(\bP)|= \prod_{\mu\in I(\bP)} \Big( \frac{|P_\mu|}{2}\Big)!
        = \prod_{\mu\in I(\pi(\bA))} |A_\mu|\, ! \; .
\label{eq:Pk}
\ee
We will use the notation
\be
V_{(*)}(\bA, \sigma, Q): = V_{(*)}(\bP(\bA, \sigma); Q)
\label{def:VsigmaA}
\ee
and similarly for $E_{(*)g}$ and $V^\circ_{(*)}$. 
In the proofs, $Q$ will be omitted. We also introduce
\be
       c(\bA): =\prod_{\nu\in I(\bA)} c(|A_\nu|) \; .  
\ee
With these notations we can state the representation
of the non-repetition terms as a summation over
Feynman diagrams:

\begin{proposition}\label{prop:lump} With $Q\equiv 1$ and  $\xi=0$ we have
\be
   \lim_{L\to\infty} \bE' \|\psi_{t,k}^{\prime \;  nr}\|^2_L
    = \sum_{\sigma\in\fS_k}
    \sum_{\bA\in\cA_k}c(\bA) V_{\xi=0}^\circ(\bA, \sigma, Q\equiv 1)
\label{eq:psicM}
\ee
and with $Q_\xi(v): = \wh\cO(\xi, v)$  we have
\be
   \lim_{L\to\infty} \bE' \langle \wh \cO_L, 
\wh W^\e_{\psi_{t,k}^{\prime \;  nr}}\rangle_L
    = \sum_{\sigma\in\fS_k}
    \sum_{\bA\in\cA_k} c(\bA) \int \rd \xi \; 
V_{\e\xi}^\circ(\bA, \sigma, Q_\xi)   \; .
\label{eq:WcM}
\ee
\end{proposition}

{\it Proof of Proposition \ref{prop:lump}.}
We insert (\ref{def:P}) and \eqref{eq:conngr}
into \eqref{eq:psiM} and we take the limit $L\to\infty$. 
 We use that 
$$
\bE_M  \Big[ \frac{M!}{|\Lambda_L|^k (M-k)!}\Big]\to1
$$
 for any fixed $k$.
We also replace every Riemann sum \eqref{riem} with integrals 
and we use $\wh \psi_{0}'
\to\wh\psi_0 $. By recalling (\ref{peqp}) and 
by choosing $Q\equiv 1$ in the definition \eqref{def:cM},
we obtain \eqref{eq:psicM}.  The proof of \eqref{eq:WcM}
is identical.
$\;\;\Box$.

\medskip

{\it Proof of Theorem~\ref{thm:L2}.} We will prove only 
\eqref{eq:L2bound}; the proof of \eqref{eq:Wbound} is analogous.
Starting with \eqref{eq:psicM}, we
notice that the graph with  the trivial partition $\bA_0$ 
and with the identity permutation on $I_k$ 
gives the main term in Theorem~\ref{thm:L2}
since
$$
   V_\lambda(t, k) = V^\circ_{\xi=0}(\bA_0, id) \; .
$$
This graph is  called the {\bf ladder graph}
(Fig.~\ref{fig:ladd}). All other graphs will be negligible.

\bef\bec
\epsfig{file=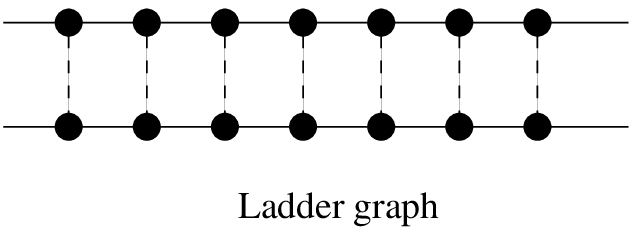, scale=.8}
\eec
\caption{Ladder graph}\label{fig:ladd}
\eef

We first replace $V^\circ(\cdots)$ with $V(\cdots)$; the error
is negligible by Lemma \ref{lemma:VV}. 
In Section \ref{sec:nolump} we then estimate $ V(\bA, \sigma)$ 
for the trivial partition  $\bA=\bA_0$, where every lump has one element.
The result is Proposition \ref{prop:Vsi}.
 In this case we set $V(\sigma):=V(\bA_0, \sigma)$.
In Section \ref{sec:lump} we treat the general case $\bA\neq\bA_0$.
The final result of this section is Proposition \ref{prop:sumup}.
The proof of both propositions rely on Theorem \ref{thm:Vsi}
that is the core of our method. Its proof is given separately
in Section \ref{sec:mainproof}.
Finally, the proof of Theorem~\ref{thm:L2}
follows from Proposition \ref{prop:sumup},
together with (\ref{eq:psicM}), \eqref{def:EMout}
and \eqref{eq:ajest}.  $\;\;\Box$

\medskip

We remark that the $E$- and $V$-values of the partitions depend on
the parameters $\lambda, t, \xi, \zeta, k$ and $g$; a fact
that is not explicitly
included in the notation. In Sections \ref{sec:nolump},
\ref{sec:lump} and
\ref{sec:mainproof}
 we will always assume the following
relations
\be
    \eta=\lambda^{2+\kappa}, \quad
 t= \lambda^{-2-\kappa} T,\quad T\in [0,T_0],\quad
   K=[\lambda^{-\delta}(\lambda^2t)],
 \quad k< K,\quad \zeta=\lambda^{-\kappa
  -3\delta}, \quad g\leq 8
\label{param}
\ee
with a sufficiently small positive $\delta>0$  that is
independent of $\lambda$ but depends on $\kappa$.
All estimates will be uniform in $\xi$ and in $T\in [0,T_0]$.
We mention that for the proof of
Theorem~\ref{thm:L2} we need only $g=0$, but
the more general case is used in \cite{ESY3}.

\section{Estimates on Feynman graphs 
without nontrivial lumps}\label{sec:nolump}
\setcounter{equation}{0}

We use the letters $p_j$, $\tp_j$,
$j\in I_{k+1}$ for the momenta variables (see the convention
at the end of Section \ref{sec:FG}) and $I(\bA_0)= I_k$
for the index set of the trivial partition. In the following
sections we always assume $Q\equiv 1$.

We introduce the restricted version of $M^\circ$ (see \eqref{def:M}) as
\be
    M(k, \bp,\tbp):=  \frac{e^{2t\eta} }{(2\pi)^2}
 \int\!\!\int_{-Y}^Y\rd\alpha\rd\beta
    \; e^{i(\alpha-\beta)t}
    \Bigg(\prod_{j=1}^{k+1}
  \frac{\ov{\wh B}(p_{j+1}-p_j)}{\alpha - \ov\om(p_j)-i\eta}\;
    \frac{ \wh B( \tp_{j+1} - \tp_j)}{\beta -\om(\tp_j)+i\eta}\Bigg) \; 
\label{def:Mre}
\ee
and  we also define the trivial estimate of
 $M$ as
\be
    N(k, \bp,\tbp):=\frac{e^{2t\eta}}{(2\pi)^2}
   \int\!\!\int_{-Y}^Y\rd\alpha\rd\beta
    \Bigg(\prod_{j=1}^{k+1}
  \frac{|\wh B(p_{j+1}-p_j)|  }{|\alpha - \ov\om(p_j)-i\eta|}\;
   \frac{ |\wh B( \tp_{j+1} - \tp_j)| }{|\beta -\om(\tp_j)+i\eta|}\Bigg)\; .
\label{def:N}
\ee
The truncated versions of these quantities, denoted by
 $M_{*}(k, \bp, \tbp)$ and $N_{*}(k, \bp, \tbp)$, are
defined by removing the
  $(k+1)$-th $\alpha$ and $\beta$ denominators from
the definitions (\ref{def:Mre}), (\ref{def:N}) but keeping
all numerators and all other denominators.

{F}rom  the definitions
(\ref{def:Vshort}), \eqref{def:VsigmaA} and  $V(\sigma)= V(\bA_0, \sigma)
=V(\bA_0,\sigma, Q\equiv 1)$,
 we obtain
\be
  V_{(*)}(\sigma)=\lambda^{2k}\int\!\!\int \rd\bp \rd\tbp
\; M_{(*)}(k, \bp,\tbp) \Delta_\xi(\sigma, \bp,\tbp, \bu\equiv 0)  
\ov{\wh\psi_0}(p_{1})\wh\psi_0(\tp_1) \; ,
\label{eq:VsiA}
\ee
\be
  E_{(*)}(\sigma,\bu)=\lambda^{2k}\int\!\!\int \rd\bp \rd\tbp
  \; N_{(*)}(k, \bp,\tbp)\Delta_\xi(\sigma, \bp,\tbp, \bu)
\ov{\wh\psi_0}(p_{1})\wh\psi_0(\tp_1)\; 
\label{eq:VsiAest}
\ee
with
$$
  \Delta_\xi(\sigma, \bp,\tbp,\bu):= \delta(\tp_{k+1}-p_{k+1}+\xi)
\prod_{\ell=1}^k \delta\Big(\; p_{\ell+1}-p_{\ell}
    -(\tp_{\sigma(\ell)+1}-\tp_{\sigma(\ell)}) - u_\ell \; \Big) \; .
$$
Clearly $|V_{(*)}(\sigma)|\leq  E_{(*)g=0}(\sigma,\bu\equiv 0)$ for any $\xi$
(see (\ref{eq:VleqE})).

\medskip

We introduce a convenient  notation.
For any  $(k+1)\times (k+1)$ matrix $M$ and for any vector
of momenta $\bp=(p_1, \ldots p_{k+1})$, we let $M\bp$
denote the following $(k+1)$-vector
of momenta
\be
   M\bp : = \Big( \sum_{j=1}^{k+1} M_{1j} p_j, \;  
 \sum_{j=1}^{k+1} M_{2j} p_j, \ldots \Big) \; .
\label{def:Mp}
\ee
Furthermore, we introduce the vector $\bv=(v_1, \ldots, v_{k+1})$
as
\be
    v_\ell: = \xi + u_1 + u_2 + \ldots + u_{\ell-1}, \quad
 \mbox{for all} \;\; \ell =1, 2,\ldots, k+1 \; .
\label{def:v}
\ee
Note that $v_{k+1} = \xi$ by \eqref{sumu}.

Given a permutation $\sigma\in \fS_k$, we define a $(k+1)\times (k+1)$
matrix $M=M(\sigma)$ as follows
\be
       M_{ij}(\sigma): = \left\{ \begin{array}{cll} 1 & \qquad \mbox{if}
    \quad & \tsi(j-1) < i \leq \tsi(j)\\
       -1 & \qquad \mbox{if} \quad & \tsi(j) <i \leq \tsi(j-1)\\
       0 & \qquad \mbox{otherwise,} \quad & \;  \end{array} \right.
    \label{def:Mmat}
\ee
where, by definition, $\tsi$ is the {\bf extension} of $\sigma$ to
 a permutation of $\{ 0, 1, \ldots, k+1\}$
by $\tsi(0):=0$ and $\tsi(k+1):=k+1$. In particular $[M\bp]_1= p_1$,
$[M\bp]_{k+1}= p_{k+1}$.

It is easy to check that
\be
     \Delta_\xi(\sigma, \bp,\tbp, \bu)=
    \prod_{j=1}^{k+1}\delta\Big( \; \tp_j - [M\bp]_j +[M\bv]_j 
    \Big)\; ,
\label{eq:deltas}
\ee
in other words, the matrix $M$ encodes the dependence of the $\tp$-momenta on
the $p$-momenta and the $v$-momenta. This rule is
 transparent in the graphical representation
of the Feynman graph: the momentum $p_j$ appears in
those $\tp_i$'s which fall into its "domain of dependence",
i.e. the section between the image of the two endpoints of
$p_j$, and the sign depends on the ordering of these images
(Fig.~\ref{fig:domdep}).

\bef\bec
\epsfig{file=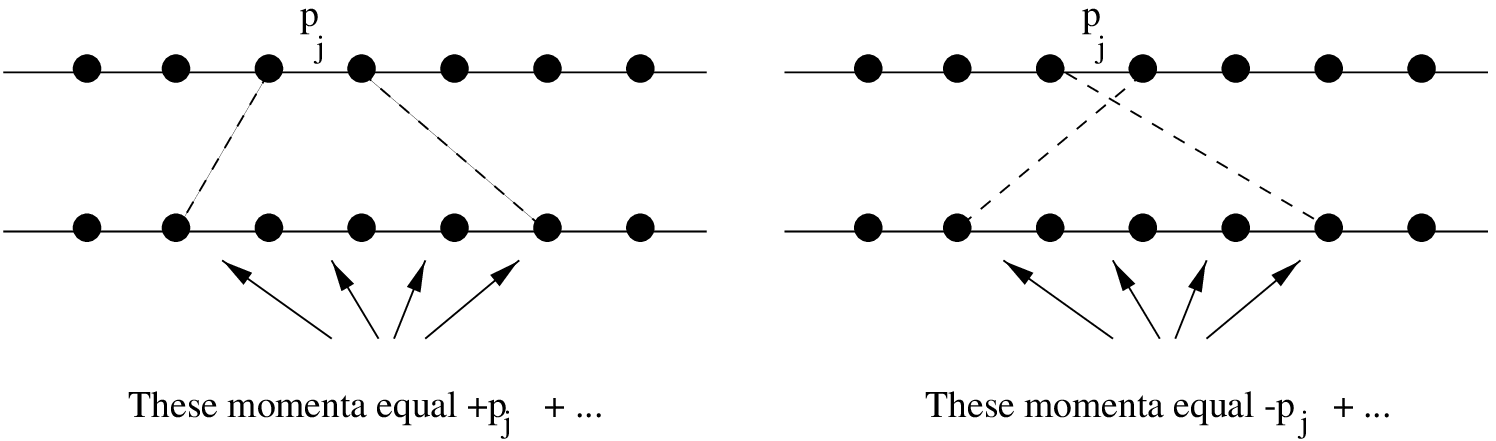, scale=0.8}
\eec
\caption{Domain of dependencies of the momenta}\label{fig:domdep}
\eef

\begin{definition}
A matrix $M$ with entries $0, +1$ or $-1$ is called
{\bf tower matrix} if in each column the non-zero entries
are consecutive and  identical. The collection
of these consecutive 1 or $-1$ entries are
called the {\bf tower} of that column.
\end{definition}

By construction,
the matrix $M(\sigma)$ is a tower matrix.

\begin{proposition}\label{prop:unimod}
For any permutation $\sigma\in\fS_k$ the matrix $M(\s)$ is

(i) invertible;

(ii) totally unimodular, i.e. each subdeterminant is $0$ or $\pm 1$.
\end{proposition}

{\it Proof.} The invertibility follows from the fact that $\bp$ and $\tbp$
play a symmetric role in (\ref{eq:deltas}) if $\bv\equiv 0$, $\xi=0$, 
in particular $M(\sigma)^{-1}= M(\sigma^{-1})$.
It is easy to prove by
induction on the size of the matrix 
that any tower matrix is totally unimodular. $\;\;\Box$.

\medskip

The following definition is crucial. It establishes the necessary concepts
to measure the complexity of  a permutation.

\begin{definition}[Valley, peak, slope and ladder]\label{def:slope}
Given a permutation $\sigma\in \fS_k$ let $\tsi$ be its extension. A point
 $(j, \sigma(j))$, $j\in I_k=\{1,2, \ldots, k\}$,
on the  graph of $\sigma$ is called {\bf peak} 
if $ \sigma(j) < \min\{ \tsi(j-1), \tsi(j+1)\}$,
it is called {\bf valley} if $\sigma(j) > \max\{ \tsi(j-1),\tsi(j+1)\}$.
Furthermore, 
if $\sigma(j)-1 \in \{ \tsi(j-1), \tsi(j+1)\}$ and $(j,  \sigma(j))$ is not
a valley, then the point $(j, \sigma(j))$, $j\in I_k$, is called {\bf ladder}.
Finally, a point  $(j, \sigma(j))$, $j\in I_k$, on the graph
of  $\sigma$ is called {\bf slope} if it is not  a peak, valley or ladder.

Let $I=\{ 1, 2, \ldots, k+1\}$ denote the set of row indices of $M$.
This set is partitioned
into five disjoint subsets, $I=I_p\cup I_v \cup I_\ell\cup I_s\cup I_{last}$,
 such that $I_{last}:= \{ k+1\}$ is the last index, and
 $i\in I_p, I_v, I_\ell$ or $I_s$ depending
on whether $(\sigma^{-1}(i),i)$ is a peak, valley, ladder or slope, respectively.
The cardinalities of these sets are denoted by $p:=|I_p|$, $v:=|I_v|$,
$\ell:= |I_\ell|$ and $s:= |I_s|$, and, if necessary, we indicate the dependence
on $\sigma$ as $p=p(\sigma)$, etc.
We define the {\bf degree} of the permutation $\sigma$ as
\be
{\rm deg} (\sigma): = k-\ell(\sigma)\; .
\label{def:d}
\ee

A maximal collection of
consecutive ladder indices, $i+1, \ldots, i+b \in I_\ell$,
 is called a {\bf ladder} of length $b$. The index $i$ 
is called the {\bf top index of a ladder}. 
The {\bf bottom index of a ladder}  is defined to be $i+b$ or
$i+b+1$, depending on whether
 $|\tsi^{-1}(i+b+1)-\sigma^{-1}(i+b)|\neq 1$ or
$|\tsi^{-1}(i+b+1)-\sigma^{-1}(i+b)| = 1$, respectively.
The set of bottom and top indices are 
denoted by $I_b, I_t$ and $I_t\subset \{ 0, 1, \ldots, k-1\}$,
$I_b\subset \{1,2, \ldots , k+1\}$.
Note that the top index of a ladder never belongs to $I_\ell$
and it may be 0. The bottom index is either a ladder, a valley or $k+1$.
 \end{definition}

{\it Remarks:} (i) The terminology of peak, valley, slope, ladder
comes from the graph of the permutation $\tsi$ 
drawn in a coordinate system where the axis of the dependent variable,
$\sigma(j)$, is oriented downward (see Fig.~\ref{fig:slope6}).
It immediately follows from the definition of the extension $\tsi$
that the number of peaks and valleys are the same, $p(\sigma)= v(\sigma)$.

(ii) The nonzero
entries in the matrix $M(\sigma)$
follow the same geometric pattern as the graph: each downward
segment of the graph corresponds to a column with
a few consecutive 1's, upward segments correspond to
columns with $(-1)$'s. On Fig.~\ref{fig:slope6} we also
pictured the towers of $M(\sigma)$ drawn inside the graph of $\sigma$.

(iii) Because our choice of orientation of the vertical axis 
follows the convention of labelling rows of a matrix,
a peak is a local minimum of  $j\to \sigma(j)$. 
We fix the convention that  the notions ``higher'' or ``lower'' 
for objects related to the vertical axis (e.g. row indices)
always refer to the graphical picture.
In particular the ``bottom'' or the ``lowest element'' of a tower is 
located in the row with the highest index.

Also, a point on the graph of the function $j\to \sigma(j)$ is traditionally
denoted by $(j,\sigma(j))$, where the first coordinate
$j$ runs  on the horizontal axis,
while in the labelling  of the  $(i,j)$--matrix element $M_{ij}$
of a matrix $M$ the first coordinate $i$ labels rows, i.e.
it runs vertically. 
To avoid confusion, we will always specify whether
a double index $(i,j)$ refers to a point on the graph of $\sigma$
or a matrix element.

(iv) We note that for the special  case of the identity permutation
$\sigma = id$ we have $I_p=I_s=I_v=\emptyset$,
and $I_\ell=\{1, 2, \ldots , k\}$. In particular, 
${\rm deg}(id)=0$
and ${\rm deg}(\sigma)\ge2$ for any other permutation $\sigma\neq id$.

\bef\bec
\epsfig{file=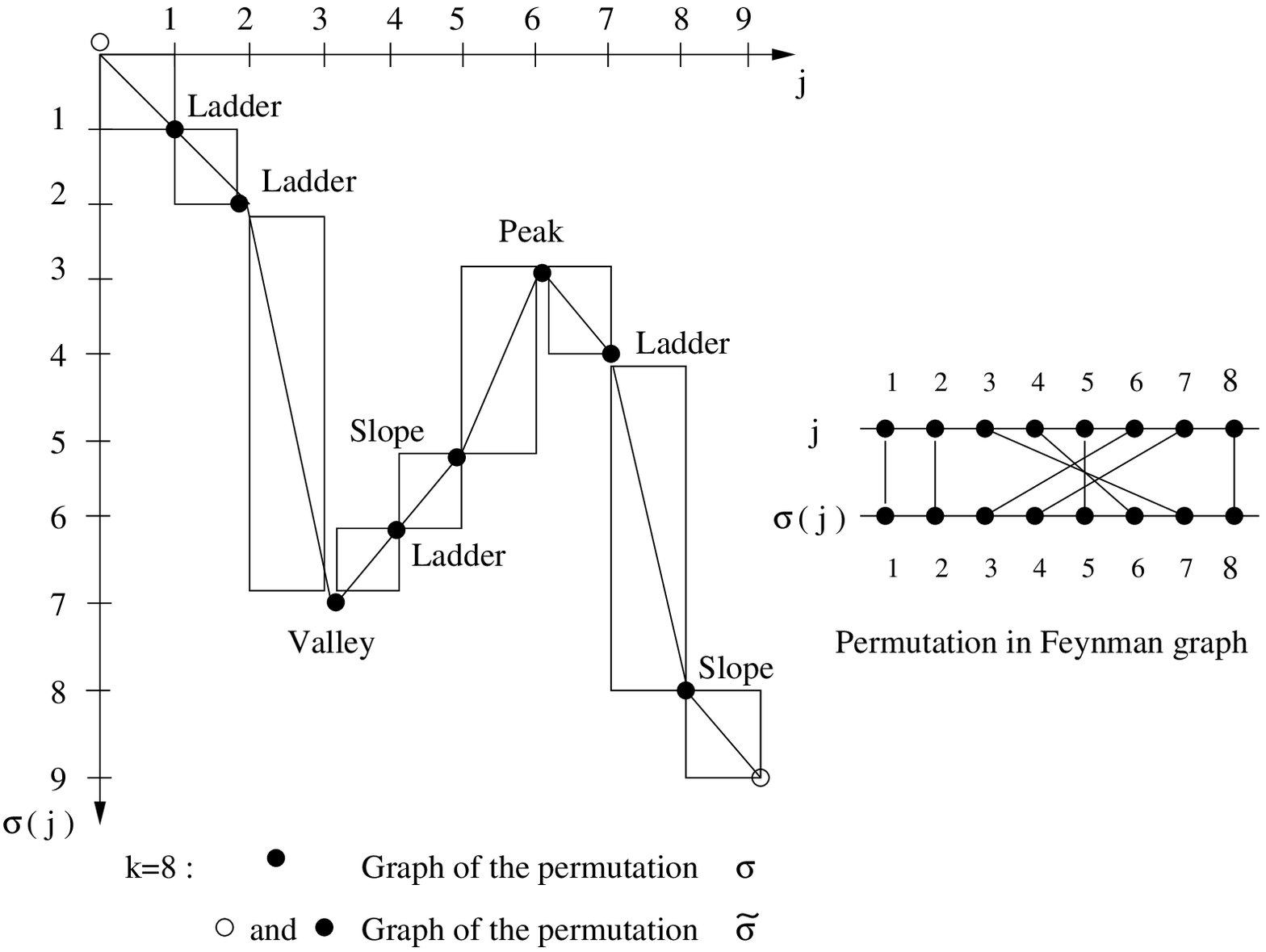, scale=0.9}
\eec
\caption{Graph of a permutation with the towers}\label{fig:slope6}
\eef

An example is shown on Fig.~\ref{fig:slope6} with $k=8$.
The matrix corresponding to the permutation on this figure is the following
(zero entries are left empty)
\be
    M(\sigma):= \begin{pmatrix}
  1 & &  &  & &  &  & &\cr
&   1 & &  &  & &  &  & \cr
& & 1&  &  & &  &  & \cr
&& 1&  &  &-1 & 1 &  & \cr
& & 1&  &  &-1 &  & 1 & \cr
& & 1&  & -1 & &  & 1 & \cr
& & 1& -1 &  & &  & 1 & \cr
& & &  &  & &  & 1 & \cr
&  & &  &  & &  &  & 1 
\end{pmatrix}
\begin{array}{cl}
1 & \ell \cr
2 & \ell \cr
3 & p \cr
4 & \ell \cr
5 & s \cr
6 & \ell \cr
7 & v \cr
8 & s \cr
9 &  (last)
\end{array}  
\label{Mex}
\ee
The numbers on the right indicate the column indices and the
letters show whether it is peak/slope/valley/ladder or last. 
In this case $I_p= \{ 3 \}$, $I_v=\{ 7\}$, $I_s = \{5,8\}$.
$I_\ell=\{ 1,2, 4,6\}$, $I_{last}=\{ 9\}$, $I_t=\{0, 3, 5 \}$, $I_b=\{ 2, 4, 7\}$ 
and ${\rm deg}(\s)=4$. 
There are three ladders, two of them of length one and one is of length two.

\bigskip

Now we are ready  to estimate $|V(\sigma)|\leq 
E_{g=0}(\sigma, \bu\equiv 0)$.
 The following theorem shows that the degree of
the permutation, ${\rm deg}(\sigma)$, measures the size of $V(\sigma)$.
 The proof is
the key step in our 
method and it is given in Section \ref{sec:mainproof}.

\begin{theorem}\label{thm:Vsi} Assume \eqref{param} with
$\kappa < \frac{2}{6+9d}$ and let $\sigma\in\fS_k$.
 Then the $E$-value of the graph of the trivial partition
with permutation $\sigma$ is estimated by
\be
    \sup_\bu E_{(*)g}(\sigma,\bu) \leq
    C \Big(\lambda^{\frac{1}{3}-(1+\frac{3}{2}d)\kappa-O(\delta)} 
\Big)^{\de(\sigma)}|\log\lambda|^2
\label{eq:Vsi}
\ee
if $\lambda\ll 1$.
\end{theorem}

This theorem is complemented by the following combinatorial lemma.

\begin{lemma}\label{lemma:comb}
Let $k \leq K=O(\lambda^{-\kappa-\delta})$, $D\ge0$ integer, and let
 $\gamma>\kappa+\delta$ be fixed.
 Then
\be
    \sum_{\sigma\in\fS_k\atop \dee(\sigma)\ge D} \lambda^{\gamma 
  \,\de(\sigma)} \leq
    O\Big( \lambda^{D(\gamma-\kappa-\delta)}\Big)
\label{eq:lambdasum}
\ee
for $\lambda\ll 1$.
\end{lemma}

Since ${\rm deg}(\sigma)\ge 2$ if $\sigma\neq id$,
from Theorem \ref{thm:Vsi}, Lemma \ref{lemma:comb}, $g(\bP)=0$ and 
the estimate $|V(\sigma)|\leq E_{g=0}(\sigma, \bu\equiv 0)$ we
immediately obtain:

\begin{proposition}\label{prop:Vsi} Assuming \eqref{param} with
$\kappa < \frac{2}{12+9d}$ we have
\be
    \sum_{\sigma\in \fS_k\atop\sigma\neq id}  |V(\sigma)| \leq
    O\Big(\lambda^{\frac{2}{3}-(4+3d)\kappa-O(\delta)}\Big)   
 \;\qquad
\label{eq:Vsisum}
\ee
 for $\lambda\ll 1$.
$\Box$
\end{proposition}

{\it Proof of Lemma \ref{lemma:comb}.} Notice that $\ell(\sigma)=k$ only if
$\sigma=id$, for all other permutations $\ell(\sigma)\leq k-2$.
We shall prove that, for any $\ell$,
\be
    \# \{ \sigma\in \fS_k\; : \; \ell(\sigma)=\ell\} \leq 2(2k)^{k-\ell} \; .
\label{eq:permnumb}
\ee
Then  (\ref{eq:lambdasum}) follows  by  recalling $k-\ell(\sigma)
={\rm deg}(\sigma)$
and $k\leq K =O(\lambda^{-\kappa-\delta})$ and by
summing up the geometric series
$$
 \sum_{\sigma\in \fS_k\atop \dee(\sigma)\ge D} \lambda^{\gamma \de(\sigma)}
 \leq 2\sum_{m=D}^{k} \big[ 2\lambda^{(\gamma-\kappa-\delta)}\big]^m \; .
$$

\medskip

To prove \eqref{eq:permnumb},  let $\sigma$ be a permutation with $m$
ladders of size $b_1, \ldots,  b_m$ such that $\sum_j b_j= \ell$ and $b_j \ge 1$.
If we remove these ladder indices, we have $k-\ell$ indices in
 $\{1, 2, \ldots , k\}\setminus I_\ell=\{ i_1, i_2,  \ldots,  i_{k-\ell}\}$. 
The permutation $\sigma$ induces a unique permutation $\sigma^*\in
\fS_{k-\ell}$
on the indices of this set by $\sigma^*(j) <\sigma^*(j')$
iff $\sigma(i_j)< \sigma(i_{j'})$.

Let $I^*:=\{ 0, i_1, i_2,  \ldots,  i_{k-\ell}\}$ and set 
$\nu: =|I^*|=k-\ell +1$.
Clearly, the top of any ladder belongs to $I^*$ and each element of $I^*$
can be the top of at most one
ladder.
We assign the length $b_j$ of the ladder to its top and
for simplicity, we
assign  the value zero to any other element of $I^*$.
Thus we obtain numbers
$b_1, \ldots,  b_{\nu}$ with $\sum_{j=1}^{\nu} b_j= \ell$
and $b_j \ge0$.

If the permutation $\sigma^*$ and the
 numbers $b_1, \ldots,  b_{\nu}$ are given,
then we have $2^{\nu}$ ways to reconstruct the original permutation
$\sigma$. To see this, first notice
that within $I^*$
the ladder-tops are identified by the condition $b_j>0$ and the
corresponding ladder in $\sigma$ can emanate 
either ``to the right'' or ``to the left'' down from its top 
on the graph of $\s$. Once this choice is made, the permutation
$\sigma$ can be uniquely reconstructed from $\sigma^*$ by
inserting ladders of given length starting from their tops.
Therefore, the number of permutations $\sigma\in \fS_k$
 with $\ell$ ladder indices is bounded by
$$
2^{\nu} (k-\ell)! \times \#\Big\{ (b_1, b_2, \ldots, b_{\nu})
\; : \; \sum_{j=1}^{\nu} b_j= \ell, \; b_j \ge0 \Big\}
$$
$$
\leq 2^{\nu} (k-\ell)! { \nu-1+\ell\choose \nu-1}
\leq 2(2k)^{k-\ell} \;.
$$
This completes the proof of  Lemma \ref{lemma:comb}.
 $\Box$

\section{Estimates on Feynman graphs with nontrivial lumps}\label{sec:lump}
\setcounter{equation}{0}

In this section we estimate $V(\bA, \sigma)$ 
for a general partition $\bA$.
  We start with a definition.

\begin{definition}\label{def:deg}
(i) Let $\bA\in \cA_k$.  Set $a_\nu:=|A_\nu|$,
$\nu\in I(\bA)$, to be the size of the $\nu$-th lump.  Let
$$
        S(\bA): = \bigcup_{\nu\in I(\bA)\atop a_\nu \ge 2} A_\nu
$$
be the union of nontrivial lumps. The cardinality of this set,
$s(\bA): = |S(\bA)|$, is called the {\bf degree of the partition} $\bA$.

(ii) Let  $\bA\in \cA_k$ and $\s\in \fS_k$. The number
\be
   q(\bA, \sigma):=\max\Big\{ {\rm deg}(\sigma), \frac{1}{2} s(\bA)\Big\}
\label{def:q}
\ee
is called the {\bf joint degree} of the pair $(\sigma, \bA)$
of the permutation $\s$  and partition $\bA$.
\end{definition}

The goal is the following generalization
of Proposition \ref{prop:Vsi} 
that includes summations over non-trivial lumps.

\begin{proposition}\label{prop:sumup} We assume \eqref{param}.
Let $D\ge 0$ and $s\ge 2$ be
 given integers, let $q:= \max\{ D,
\frac{1}{2}s\}$. For any $\kappa< \frac{2}{33d+36}$ 
we have
\be
   \Xi(k,D,s):= \sum_{\sigma\in \fS_k\atop \dee (\sigma)\ge D}
   \sum_{\bA\in \cA_k\atop s(\bA)\ge s}
   \sup_{\bu, g\leq 8} E_{(*)g}
   (\bA, \sigma,\bu)|c(\bA)|\leq C \Big( \lambda^{ \frac{1}{3}-
  (6+\frac{11}{2}d)
   \kappa -O(\delta)}\Big)^q |\log \lambda|^2\; .    
\label{eq:sumup}
\ee
\end{proposition}

We recall that
for the continuum model non-trivial lumps do
not contribute since $c(\bA)=0$ unless each $A_\nu$ is trivial, $|A_\nu|=1$.
Thus \eqref{eq:sumup} can be proved 
 directly from 
Theorem \ref{thm:Vsi}, Lemma \ref{lemma:comb} and Lemma 
\ref{lemma:breakup} below  with a somewhat better exponent.
We will give a proof for the general case that is valid for
both the discrete and continuum models.

\medskip

{\it Proof of Proposition \ref{prop:sumup}.}
The following lemma shows that any even partition $\bP\in \cP_k$
can be generated by a permutation with high degree, depending
on the size of nontrivial lumps. The proof will be given later
at the end of this section.

\begin{lemma}\label{lemma:ds}
For any even partition $\bP\in \cP_k$
there exists a compatible permutation $\wh\sigma=\wh\sigma(\bP)\in \fS_k(\bP)$
such that
\be
     {\rm deg}(\wh\sigma)\ge \frac{1}{2} s(\pi(\bP)) \; .
\label{eq:ds}
\ee
\end{lemma}

\begin{corollary}\label{cor:jointdeg}
Given $\sigma\in \fS_k$ and $\bA\in \cA_k$,  we have, for $\kappa
 < \frac{2}{6+33d}$
\be
     \sup_{ \bu} E_{(*)g}
   (\bA, \sigma,\bu) \leq C|\log \lambda|^2
   \Big(\lambda^{\frac{1}{3}-(1+\frac{11}{2}d)\kappa-O(\delta)} 
     \Big)^{q(\bA, \sigma)}  \; .
\label{eq:jointdeg}
\ee
\end{corollary}

{\it Proof of Corollary \ref{cor:jointdeg}.} We define
a permutation $\s^*:=\s^*(\bA, \s)$ as $\s^*:=\s$ if
 ${\rm deg}(\sigma)\ge \frac{1}{2} s(\bA)$,
and $\sigma^* := \wh\s(\bP(\bA, \s))$ otherwise.
By Lemma \ref{lemma:ds} we have  ${\rm deg}(\sigma^*)= q(\bA, \s)$.
Clearly $\bP(\bA, \sigma) = \bP(\bA, \sigma^*)$, in particular,
$E_{(*)g}(\bA, \sigma,\bu) = E_{(*)g}(\bA, \sigma^*,\bu)$.

We wish to estimate  the value of an arbitrary partition $\bA$
by that of the trivial partition $\bA_0$. We can artificially break
up lumps and using the auxiliary momenta to account for
the additional Kirchoff rules.  
We describe 
this procedure
in full generality for any circle graph.
We will call it Operation I because further similar operations
will be introduced in the companion paper \cite{ESY3}.

\medskip

{\it Operation I: Breaking up lumps}

\medskip

Consider a circle graph on $N$ vertices (Section \ref{sec:genest}).
Given a partition of the set $\cV\setminus \{ 0, 0^*\}$, 
$\bP=\{P_\mu\; :\; \mu \in I(\bP)\}
\in \cP_\cV$, we define a new partition $\bP^*$
by breaking up one of the lumps into two smaller nonempty lumps.
 Let $P_\nu = P_{\nu'} \cup P_{\nu''}$ with
$P_{\nu'}\cap P_{\nu''}=\emptyset$ and $\bP^*=\{ P_{\nu'},  P_{\nu''},
 P_\mu  \;  : \; \mu\in I(\bP)\setminus \{ \nu\}\}$.
In particular $I(\bP^*) = I(\bP)\cup \{ \nu', \nu''\} \setminus \{ \nu\}$
and $m(\bP^*)= m(\bP)+1$.

\begin{lemma}\label{lemma:breakup}
With the notation above, we have
$$
     E_{(*)g}( \bP, \bu, \balpha) \leq 
\int_{|r|\leq N\zeta}
 \rd r \; E_{(*)g}
    (\bP^*, \bu^*(r,\nu),
     \balpha)\; ,
$$
where the new set of momenta $ \bu^*=\bu^*(r,\nu)$ is given
by $u^*_\mu:=u_\mu$, $\mu\in I(\bP)\setminus \{\nu\}$
and $u^*_{\nu'}= u_\nu-r$, $u^*_{\nu''} = r$. In our estimates
we will always have $N\leq 2K$ and then
$$
    \sup_\bu E_{(*)g}( \bP, \bu, \balpha) \leq \Lambda
    \sup_\bu E_{(*)g}
    (\bP^*,\bu,
     \balpha)\; 
$$
with $\Lambda:= [CK\zeta]^d = O(\lambda^{-2d\kappa -O(\delta)})$
(see \eqref{def:K} and \eqref{def:mu}).
\end{lemma}

{\it Proof of Lemma \ref{lemma:breakup}.} 
The break-up of the lump $P_\nu$ corresponds
to
\be
\delta\Big( \sum_{e\in L_\pm(P_\nu)}
    \pm w_e - u_\nu\Big) =
    \int \rd r \; \delta\Big( \sum_{e\in L_\pm(P_{\nu'})}
    \pm w_e  - u_\nu +r\Big)\delta\Big(\sum_{e\in L_\pm(P_{\nu''})}
    \pm w_e -r\Big) \; .
\label{eq:breakdelta}
\ee
Note that $L(P_\nu)\subset L(P_{\nu'})\cup L(P_{\nu''})$
and for any edge $e\in (L(P_{\nu'})\cup L(P_{\nu''}))\setminus
L(P_\nu)$, we inserted an extra $w_e-w_e$ in the left hand side of
(\ref{eq:breakdelta}). Note that the property 
(\ref{sumu}) on the sum of auxiliary momenta is preserved.
The integration in (\ref{eq:breakdelta}) can be restricted to
$|r|\leq N\zeta$ since  $|w_e|\leq \zeta$
for all $e$.  $\;\;\Box$

\medskip

We return to the proof of Corollary \ref{cor:jointdeg}.
We apply   the break-up operation until all lumps become trivial (singlets).
This requires  not more than $s(\bA)$ steps,  
and by using Lemma \ref{lemma:breakup}, we
obtain 
$$
     \sup_\bu E_{(*)g}(\bA, \sigma^*,\bu) \leq 
   \Lambda^{s(\bA)}
     \sup_\bu E_{(*)g}(\bA_0, \sigma^*,\bu)\;
$$
with $\Lambda := [CK\zeta]^d$.
Then (\ref{eq:jointdeg}) immediately follows from 
$s(\bA)\leq 2 q(\bA, \sigma)$ and from
Theorem \ref{thm:Vsi}. This completes the proof of 
Corollary \ref{cor:jointdeg}.
$\;\;\Box$

\bigskip

{\it Proof of Lemma \ref{lemma:ds}.}  The inequality
\eqref{eq:ds} is trivial if the partition has no nontrivial lump,
i.e. $s(\pi(\bP))=0$, so we can assume $s(\pi(\bp))\neq 0$.
For any $\sigma\in \fS_k$ we define the
set of {\bf internal ladder indices} as
$$
  I^*_\ell=  I^*_\ell(\sigma):= \{ i\in I_\ell\;  : \;
    |\tsi^{-1}(i-1)-\tsi^{-1}(i)|=|\tsi^{-1}(i+1)-\tsi^{-1}(i)|=1\}
$$
where $\tsi$ is the extension of $\sigma$ as before.
The indices $i-1, i+1$ are called the {\bf protectors}
of the internal ladder index $i\in  I^*_\ell$. They ensure that
the index $i$ is neither the bottom nor the top  index
of the ladder.

We first claim that for any $\sigma\in\fS_k$, $\sigma\neq id$, we have
\be
       k-|I_\ell^*(\sigma)| \leq 2{\rm deg}(\sigma)
\label{ineq1}
\ee
To see this inequality, we first recall
the definition of a ladder and its bottom and top indices from 
Definition \ref{def:slope}.
Since every ladder has a unique bottom and top index, that are not internal
ladder indices, we see that  the sets $I_\ell^*$, $I_b$ and $I_t$ are disjoint
subsets of $\{0, 1, \ldots, k+1\}$.
Since $I_v, I_\ell^*\subset \{1, 2, \ldots ,k\}$,
 $I_v\cap I_\ell^*= I_v\cap I_t=\emptyset$
and  $k+1\not\in I_t$, we obtain
$|I_\ell^*| + |I_t| + |I_b|+ |I_v\setminus I_b| \leq k+ 1+\gamma $, where 
$\gamma$ is the characteristic function of the event
  $k+1\in I_b$, i.e. $\gamma:= |\{ k+1\}\cap I_b|$.
Moreover, $|I_t|=|I_b|$, thus $|I_\ell^*| + 2|I_b| + 
|I_v\setminus I_b|\leq k+1+ \gamma$.
Notice that $I_\ell \subset I_\ell^* \cup ( I_b\setminus I_v)$,
and $k+1\not\in I_\ell, I_v$, therefore
$\ell \leq |I_\ell^*| + |I_b\setminus I_v| -\gamma = |I_\ell^*| + |I_b|  - 
|I_v\cap I_b|-\gamma$. From the last two inequalities
and from $\mbox{deg}(\sigma)= k-\ell$, we obtain 
$$
  k-|I_\ell^*|\leq 2\mbox{deg}(\sigma) +1-
\gamma-|I_v\setminus I_b| -2|I_v\cap I_b|
 \leq  2\mbox{deg}(\sigma) +1-|I_v| \; .
$$
Since $\sigma\neq id$, we have $|I_v|\ge 1$ and which proves \eqref{ineq1}.

Next we will show that there exists a compatible permutation $\wh\sigma \in \fS_k(\bP)$,
\be
     I_\ell^*(\wh\sigma) \cap \wh\s(S(\bA)) = \emptyset\; ,
\label{emptyint}
\ee
where we set $\bA:=\pi(\bP)$ for simplicity.
Since $S(\bA)\neq \emptyset$, $I_\ell^*(\wh \sigma) \neq \{1, 2,\ldots, k\}$,
i.e. $\wh\sigma\neq id$. Combining  (\ref{ineq1}) with \eqref{emptyint} and 
with the fact that both $I_\ell^*$ and $ \wh\s(S(\bA)) $ are
subsets of $ \{1, 2, \ldots, k\}$,  
we  obtain (\ref{eq:ds}).

To construct $\wh\sigma$ satisfying (\ref{emptyint}), we apply a greedy
algorithm. Since $\bP$ is even, $\fS_k(\bP)$ is nonempty and we pick
a $\sigma_0 \in \fS_k(\bP)$. If (\ref{emptyint}) is not satisfied
for $\sigma_0$, then some internal ladder index $i$ is
in the image of a nontrivial lump $A\in\bA$; $i=\s_0(i')$, $i'\in A$.
 Let $j'\in A$ be another element of
this lump. Flip the permutation $\sigma_0$ on these two elements,
i.e. define a new permutation $\sigma_1$ such that $\sigma_1(i'):= \sigma_0(j')=j$,
$\sigma_1(j'):= \sigma_0(i')=i$, and $\sigma_1(r):= \sigma_0(r)$ for
any $r\neq i,j$.
 Clearly $\sigma_1\in \fS_k(\bP)$,
$\s_1(s(\bA))=\s_0(s(\bA))$.
We claim that
\be
    |I_\ell^*(\sigma_1) \cap \s_1(S(\bA))| < |I_\ell^*(\sigma_0) \cap \s_0(S(\bA))| \;,
\label{eq:decr}
\ee
i.e. the total number of internal ladder indices in nontrivial lumps
decreased. Continuing this flipping process for $\sigma_1$ etc., we obtain
a permutation $\wh\sigma$ satisfying (\ref{emptyint}).

To see (\ref{eq:decr}) we note that after the flip $i$ is not
an internal ladder index any more. This is clear if $j\neq i-1, i+1$;
in that case the points  $(\tsi^{-1}(i-1), i-1)$ and $(\tsi^{-1}(i+1), i+1)$
have not changed and they would uniquely fix the location
of an internal ladder index in between. The preimage of the index $i$ has moved
out from this position, $\s_1^{-1}(i)\neq \s_0^{-1}(i)$. The  index $j$ however would
not become internal ladder since $\s_1^{-1}(j)=i'$ is between $\s_1^{-1}(i-1)$
and $\s_1^{-1}(i+1)$, but $j$ is not between $i-1$ and $i+1$.
It is easy to see that the fixed points $(\tsi^{-1}(i-1), i-1)$ and $(\tsi^{-1}(i+1), i+1)$
also prevent any other indices from becoming an internal ladder index
after the flip. This could only be possible if due to
the new point $(\tsi^{-1}_1(j), j)= (\tsi^{-1}_0(i),j)$, one of
the neighbors of $j$, say $j+1$, would become an internal ladder index. It is easy
to see that then $j+1$ must be equal to $i-1$
and the other protector of the new internal ladder index $j+1$
must be $i$. In this case $i-1$ was already an internal ladder index before
the flip as well, so no new internal ladder was created.

A similar but simpler argument shows that if $j=i-1$ or $j=i+1$,
the number of internal ladder indices also decreases.
This completes the 
 proof of Lemma \ref{lemma:ds}
 $\;\;\Box$

\bigskip

We continue the
proof of Proposition \ref{prop:sumup}.
Given $\sigma \in \fS_k$ and $\bA\in \cA_k$,
we recall the permutation  $\sigma^*=\sigma^*(\bA, \sigma)$
defined in the proof of Corollary \ref{cor:jointdeg}.
We also  note that $s(\bA)\leq 2 \, {\rm deg}(\sigma^*)$. Hence
$$
   \Xi(k,D,s) \leq \sum_{\bA\in \cA_k\atop s(\bA)\ge s}|c(\bA)|
   \sum_{ \sigma^* \in \fS_k\atop \dee(\sigma^*)\ge q}
   \sum_{\sigma \in \fS_k\atop \sigma^*(\bA, \sigma)=\sigma^*}
   \sup_\bu E_{(*)g}(\bA, \sigma,\bu) \; .
$$
Note that $\s\in \fS_k(\bP(\bA, \s^*))$, so by (\ref{eq:Pk}) the
summation over $\sigma$ contributes by a factor of at most
 $\prod_\nu a_\nu!$ and we obtain
$$
    \Xi(k,D,s)\leq \sum_{\sigma^* \in \fS_k\atop \dee(\sigma^*)\ge q}
   \sum_{\bA\in \cA_k\atop  2\leq s(\bA)\leq 2 \dee(\sigma^*)}
   \Big(\prod_{\nu\in I(\bA)} a_\nu^{a_\nu-2} a_\nu!\Big)
\sup_\bu   E_{(*)g}(\bA, \sigma^*,\bu) \; .
$$
 We also used the estimate (\ref{eq:ajest}).
By using (\ref{eq:jointdeg}) and ${\rm deg}(\sigma^*)= q(\bA,\s)$, we obtain
$$
   \Xi(k, D,s)
 \leq  \sum_{\sigma^* \in \fS_k\atop \dee(\sigma^*)\ge q}
\Big(C\lambda^{\frac{1}{3}-(1+\frac{11}{2}d)\kappa-O(\delta)}
\Big)^{\de(\sigma^*)}
   \sum_{\bA\in \cA_k\atop s\leq s(\bA)\leq 2 \dee(\sigma^*)}
   \Big(\prod_{\nu\in I(\bA)} a_\nu^{a_\nu-2} a_\nu!\Big)\; .
$$
We introduce the notation
$$
    \sum^* f(a_\nu) : = \sum_{\nu\in I(\bA)\atop a_\nu\ge 2} f(a_\nu)\; , 
\qquad
     \prod^* f(a_\nu) : = \prod_{\nu\in I(\bA)\atop a_\nu\ge 2} f(a_\nu)\;.
$$

First we fix the  sizes of the nontrivial lumps $a_\nu\ge 2$.
 Given these sizes,
the number of $\bA$ partitions is bounded by
$$
 {k \choose a_1}{k-a_1\choose a_2}{k-a_1-a_2\choose a_3} \ldots
\leq \frac{k!}{(k-\sum^* a_\nu)!\prod^* a_\nu!}\leq
\frac{k^{\sum^* a_\nu}}{\prod^* a_\nu!} \; .
$$
Recalling $s(\bA)= \sum^* a_\nu$, and that $s(\bA)\leq 2\, {\rm deg}(\s^*)$,
we have
\be
 \Xi(k,D,s)
 \leq \sum_{\sigma^* \in \fS_k\atop  \dee(\sigma^*)\ge q}
   (Ck^2\lambda^\tau)^{\de(\sigma^*)}
   \sum_{a_\nu \; : \; \sum^* a_\nu \leq 2\de(\sigma^*)}
 \Big(\prod^* a_\nu^{a_\nu-2} \Big)
\label{eq:inter}
\ee
with $\tau< \frac{1}{3}-(1+\frac{11}{2}d)\kappa$.
We  use the bound $a^{a-2}\leq C^{a-1}(a-1)!$.
To estimate the summation over $a_\nu$'s we use the following inequality.
For any fixed $m$, $H$ we have
\be
     \sum^{\#}
 \Big(\prod^* (a_\nu-1)! \Big) \leq (H-1)!
\label{fact}
\ee
where the summation $\#$ is over all sequences $(a_1, a_2, \ldots , a_m)$
of positive integers at least 2, whose sum is  $H$.
The proof of (\ref{fact}) is easily obtained by induction on $m$ from
$$
     \sum_{a=2}^{H-2} (a-1)! (H-a-1)! \leq \sum_{a=2}^{H-2} (H-2)! < (H-1)! \; .
$$
Summing (\ref{fact}) over all $H\leq 2 {\rm deg}(\s^*)$ 
and $m\leq H/2$, we obtain
the bound
$$
\sum_{a_\nu \; : \; \sum^* a_\nu \leq 2 {\rm  deg}(\sigma^*)}
 \Big(\prod^* a_\nu^{a_\nu-2} \Big)\leq
 2 \, \big[ 2 {\rm deg}(\s^*)\big]!\leq (Ck)^{2 \de(\s^*)}
$$
for the $a_\nu$ summation in
(\ref{eq:inter}) since $ {\rm deg}(\sigma^*)\leq k+1$ by definition.

In summary, we  obtain from (\ref{eq:inter})
$$
\Xi(k,D,s ) \leq \sum_{\sigma \in \fS_k\atop \dee(\sigma)\ge q}
   (Ck^4\lambda^\tau)^{\de(\sigma)}
$$
Recalling that $k = O(\lambda^{-\kappa-\delta})$,
 we can apply Lemma \ref{lemma:comb}
with $\gamma = \tau -4(\kappa +\delta)$ as long as $\gamma >\kappa+\delta$.
For sufficiently small positive $\delta$
this gives the condition $\kappa <\frac{2}{33d+36}$ 
in Proposition \ref{prop:sumup}
and the estimate (\ref{eq:sumup}). $\;\;\Box$

\section{ Proof of Theorem \ref{thm:Vsi}}\label{sec:mainproof}
\setcounter{equation}{0}

For any $(k+1)\times (k+1)$ matrix $M$ we set 
\begin{align}
    E(M): = &\sup_{\tbu} 
    \lambda^{2k}\iint_{-Y}^{Y} \rd\alpha \rd\beta
 \int\rd \mu(\bp) \, |\cB(\bp)| \, 
    | \cB ( M\bp+\tbu)|
 |\wh\psi(p_1)||\wh\psi(p_1+\tu_1)|
\nonumber
\\
  &   \times \prod_{j=1}^{k+1}
    \frac{1}{|\alpha-\ov\om(p_j)-i\eta|}
    \frac{1}{|\beta -\om([M\bp+\tbu]_j )+i\eta|}\; .
\label{def:EM}
\end{align}
The key step in the proof of Theorem \ref{thm:Vsi} is
the following Lemma:

\begin{lemma}\label{lemma:ems} 
Suppose \eqref{param},
assume  that $\kappa< \frac{2}{6+9d}$
and $\delta \leq \delta(\kappa)$ is sufficiently small.
Then for any $\sigma\in \fS_k$ we have
\be
E(M(\sigma))\leq  C \Big(\lambda^{\frac{1}{3}-\big(1+\frac{3}{2}d\big)
\kappa
  - O(\delta)}\Big)^{\de(\sigma)} |\log\eta|^2\; ,
\label{eq:VsiM}
\ee
where the matrix $M(\sigma)$ was defined in \eqref{def:Mmat}
and the degree of the permutation, ${\rm deg}(\sigma)$, was
defined in \eqref{def:d}.
\end{lemma}

{F}rom  (\ref{eq:VsiAest}), (\ref{eq:deltas}), clearly
\be
      \sup_\bu E_{g=0}(\sigma, \bu)\leq
     \frac{e^{2t\eta}}{(2\pi)^2} E(M(\sigma))
\label{EE}
\ee
after integrating out all $\tp_j$ variables in (\ref{eq:VsiAest})
and by using $\tp_1= p_1-\xi$.
The estimate \eqref{eq:VsiM}
will  then complete the proof of Theorem \ref{thm:Vsi} for $g=0$.

The proof of  Theorem \ref{thm:Vsi}
 for other (but finitely many) $g$ values follows exactly
in the same way. 
 This requires to  slightly redefine $\cB(\bp)$
(see \eqref{def:cB}) in the definition of $E(M)$
by allowing the factor
 $\langle p_{j+1} - p_j\rangle^{-2d}$
instead of $\wh B(p_{j+1}-p_j)$ at a few places exactly 
as in the definition of $\cN_\cG$ (see \eqref{def:cN}).
  As we will
see along  the proof of \eqref{eq:VsiM}, this change 
will require using the less precise bound  \eqref{eq:2aint} with $a=0$ and
$h(p-q) = \langle p-q \rangle^{-2d}$ 
instead of the  more accurate estimate \eqref{eq:ladderint} at most
$g$ times. Each time  we lose a constant factor compared with
the proof for $g=0$. Since $g\leq 8$, this results only
in a constant factor. 
Finally, the proof for the truncated $E$-values requires
to define a truncated  version of $E(M)$, where the
last product in \eqref{def:EM} runs only up to $j=k$,
i.e. the last $\a$ and $\beta$ denominators are not present.
It will be clear from the proof of Lemma \ref{lemma:ems}
that the same bound holds for the truncated version of $E(M)$
as well.
$\;\;\Box$

\subsection{Pedagogical detour}

The size of the multiple integral in (\ref{def:EM}) heavily depends on
the structure of $M=M(\sigma)$.
 Before we go into the algorithm to evaluate this
multiple integral, we present two calculations, that
introduce  the techniques that we are going to use in the actual proof.
The second calculation also provides the bound (\ref{eq:VsiM}), hence
 (\ref{eq:Vsi}),
for the case of the trivial permutation ${\rm deg}(\sigma)=0$.

\subsubsection{Method I. Pointwise bound}\label{sec:MethodI}

The most straightforward bound on (\ref{def:EM}) estimates all but one of
the $\beta$-denominators by $L^\infty$ norm
\be
    \sup_{\beta, \bp,\tbu, j}
    \frac{1}{|\beta -\om([M\bp+\tbu]_j )+i\eta|}\; 
\leq \eta^{-1}\; .
\label{eq:Linftybound}
\ee
It would be possible to estimate this denominator by
$C\lambda^{-2}$ apart from a neighborhood of zero
using (\ref{eq:imthetaest}) and  treat the $|p|\sim0$ regime
 separately. Here we choose the simplest argument and we do not optimize for
the best possible exponent $\kappa$.

We use (\ref{eq:Linftybound}) $k$ times 
 to obtain
\begin{align}
    E(M)\leq &
    \Big(\frac{C\lambda^2}{\eta}\Big)^k\sup_\tbu
    \int\frac{\rd\mu( p_{k+1})}{\langle p_{k+1}+\tu_{k+1}\rangle^{2d}}
    \int\!\!\int_{-Y}^{Y} 
   \frac{\rd\alpha \rd\beta}{|\alpha-\ov\om(p_{k+1})-i\eta|
     |\beta-\om(p_{k+1} +\tu_{k+1})+i\eta|} \nonumber\\
    &\times
    \int\ldots \int \rd p_1 \ldots \rd p_k \,
     \prod_{j=1}^{k}
     \frac{|\wh B(p_{j+1}-p_j)|}{|\alpha-\ov\om(p_j)-i\eta|}\; .\nonumber
\end{align}
Here we used the following bound for any $\bq=(q_1, \ldots , q_{k+1})$:
\be
      |\cB (\bq)||\wh\psi_0(q_1)| \leq C^k\langle q_{k+1}\rangle^{-2d}
\label{telesc}
\ee
to obtain the decay in $\tp_{k+1}=p_{k+1}+\tu_{k+1}$.
We integrate out $p_1, p_2, \ldots , p_k$ using (\ref{eq:logest}), then we
perform the $\rd \alpha, \rd\beta$  integrals and finally $\rd p_{k+1}$
to obtain
\be
    E(M)\leq (C|\log\lambda|)^{k+2}(\lambda^2\eta^{-1})^k \; .
\label{eq:linftybound}
\ee

The estimate (\ref{eq:linftybound}) is
 off by a factor $(\lambda^2\eta^{-1})^k=(\lambda^{-\kappa})^k$
because we did not use the stronger
estimate mentioned after (\ref{eq:Linftybound}).
We also collected many logarithmic factors and the constant is not optimal.
We note that in the typical term $k\sim \lambda^2 t\sim
 \lambda^{-\kappa}\gg 1$,
so even an error $C^k$ may not be
affordable.
To improve this estimate,
 for a typical matrix $M$, we will not use the pointwise bound
\eqref{eq:Linftybound}
 for all $\beta$-denominators. We will carefully
 select those $\beta$-denominators whose singularities cannot overlap with
other singularities, hence they can be integrated out at a $|\log \eta|$
expense instead of $\eta^{-1}$.

Before we explain this algorithm, we show another method to estimate $E(M)$.
It practically estimates $E(M)$ by $E(I)$, i.e. by the ladder
graph, that can be computed more precisely.
The same calculation will be important when evaluating embedded ladder graphs.

\subsubsection{Method II. Successive integration scheme
for ladder graphs}\label{sec:MethodII}

We separate all but one
  $\alpha$ and $\beta$ denominator by
a Schwarz inequality.
We obtain
\begin{align}
    E(M) \leq &\;
    \lambda^{2k}\sup_\tbu\int
    \rd \mu(\bp)
\int\!\!\int_{-Y}^{Y}
    \frac{\rd\alpha\rd\beta}{ |\alpha-\ov\om(p_1)-i\eta|
      |\beta - \om(p_1+\tu_1 ) +
    i\eta|}\nonumber\\
    &\times 
    \Bigg[ |\wh\psi_0(p_1)|^2 \prod_{j=2}^{k+1}
    \frac{|\wh B(p_{j}-p_{j-1})|^2}{|\alpha-\ov\om(p_j)-i\eta|^2} +
    |\wh \psi_0(q_1)|^2
 \prod_{j=2}^{k+1}
  \frac{ |\wh B(q_{j}-q_{j-1})|^2 }{|\beta -\om(
   q_j)+i\eta|^2}\Bigg] \nonumber
\end{align}
with the shorthand notation $\bq:=M\bp+\tbu$. 
Because $M$ is an invertible matrix with determinant $\pm 1$
(Proposition \ref{prop:unimod}),
the contributions of the two terms in the square bracket are identical
up to exchange of $\alpha $ and $\beta$. 
To estimate the first term, we use iteratively
\eqref{eq:ladderint} and \eqref{eq:2aint} (with $a=1/2$)
to integrate out $ p_{k+1},  p_k, \ldots ,  p_2$
(in this order):
\be
  \lambda^2 \int 
\frac{|\wh B(p_{k+1}-p_k)|^2}{|\a -\ov\om(p_{k+1})-i\eta|^2}\; \rd p_{k+1}
   \leq \Big[1+C_0\lambda^{-12\kappa} (\lambda+ |\a-\om(p_k)|^{1/2})\Big]\; ,
\label{ladddint1}
\ee 
\begin{align}\label{ladddint2}
  \lambda^2 \int 
   &\frac{|\wh B(p_{k}-p_{k-1})|^2}{|\a -\ov\om(p_k)-i\eta|^2} 
\Big[1+C_0\lambda^{-12\kappa} (\lambda+ |\a-\om(p_k)|^{1/2})\Big]
\rd p_k \cr
   &\leq (1+C\lambda^{1-12\kappa}) 
\Big[1+C_0\lambda^{-12\kappa} (\lambda+ |\a-\om(p_{k-1})|^{1/2})\Big]
\end{align}
etc, 
with $C:= C_0(1+ C_{1/2}\| \hat B^2\|_{2d, 0})$.
 In the last step we use only \eqref{eq:2aint} once for $a=0$ and
once for $a=1/2$:
\be
  \lambda^2 \int 
  \frac{|\wh B(p_{2}-p_{1})|^2}{|\a -\ov\om(p_2)-i\eta|^2} 
\Big[1+C_0\lambda^{-12\kappa} (\lambda+ |\a-\om(p_2)|^{1/2})\Big]
\; \rd p_2
   \leq C \; .
\label{ladddint3}
\ee
Then we integrate $\rd\a\,\rd\beta$ and finally $\rd p_1$ to obtain
\be  
   E(M)\leq C(1+C\lambda^{1-12\kappa})^k |\log \lambda|^2 \leq C|\log\lambda|^2
\label{eq:trivest}
\ee
by using $k\leq K\ll \lambda^{-1+12\kappa}$ as $\kappa<1/13$.

We note that this method also gives 
 a robust bound for the truncated $E$-value,
since the truncation means that Lemma \ref{le:opt} is used only $k-1$ times.
Summarizing, we have proved

\begin{lemma}\label{lemma:Etrunc} We assume \eqref{param} and $\kappa<1/13$.
Then
\begin{align}
   \sup_{\s\in \fS_k} \sup_\bu E ( \s, \bu) & \leq C|\log \lambda|^2 \; 
\label{eq:ladlogweak}
\\
   \sup_{\s\in \fS_k}
  \sup_\bu E_{*} (\s, \bu) & \leq C\lambda^2 |\log \lambda|^2\; . \qquad \Box
\label{eq:Etrunc}
\end{align}
\end{lemma}

\subsection{Choice of the integration variables}

Before we start the proof of Lemma \ref{lemma:ems},
 we explain the main
idea. We use a combination of Methods I and II.
We will assume in the sequel that $\sigma\neq id$. The lemma for the
trivial  $\sigma=id$ case has been proven in \eqref{eq:trivest}.

Note that each factor in the integrand in (\ref{def:EM})
is almost singular on a set of codimension one of the
form $\{ \alpha = \mbox{Re} \;\om (p_j)\}$
or $\{ \beta = \mbox{Re}\; \om ( [M \bp+\tbu]_j)\}$
in the high dimensional space of integration, $(\bR^d)^{k+1}\times \bR^2$.
 The singularities are
regularized by $\eta$ and $\mbox{Im} \; \om$ and the two regularizations
always have the same sign.
The matrix $M$ may enhance the strength of these singularities
by forcing these ``almost singularity'' sets to overlap.
 For example, in the ladder
diagram with $\tbu\equiv 0$, we have $\tbp\equiv\bp$, hence
the singularity sets are pairwise identical if $\a=\beta$.
Therefore  singularities of quadratic type
necessarily occur. It is expected that this is the only mechanism that creates relevant
overlaps of singularities. Hence, ideally,
one would integrate out the ladder momenta with the precise
bound (\ref{eq:ladderint}). This would remove all denominators with
indices $j\in I_\ell$
and the remaining integral should be of $O(\lambda^{2(k-\ell)})$
with possible logarithmic corrections, i.e. one should gain a
factor $\lambda^2$ from each non-ladder index.

However, the singularity sets of the remaining denominators may still
intersect on  higher codimensional manifolds. 
For example, since
$$
  \sup_{\a, \beta} \int \frac{1}{|\a - \bar\om (p+q)-i\eta|} \; 
\frac{1}{|\beta - \om (p)+i\eta|} \; \rd \mu(p) \ge \frac{C}{|q|+\eta},
$$
the integration of these two propagators
develops an almost-singularity at $q=0$. We will call such factors
{\it point-singularities}, although they are regularized at a very
short scale $\eta$. After integration, one point singularity
is harmless as $(|q|+\eta)^{-1}\in L^1_{loc}(\rd q)$, however,
several point singularities may accumulate along
the procedure whose simultaneous integration may lead to further 
divergences.

These enhancements
of singularities are expected to have  contributions of lower order, but their
estimate is not easy. Note that every integration variable $p_j$ may appear
in many denominators. 
This interdependence renders the effective
 $L^1$-estimate of each
integral practically impossible. A Schwarz inequality (Method II)
can remove all correlations between denominators, but the resulting
$L^2$-estimate is of the same order as the main (ladder) term
and we would not gain anything from the higher degree of the permutation $\s$.

The idea is to estimate
 many, but not all $\beta$-denominators in (\ref{def:EM})
in the trivial way (\ref{eq:Linftybound}). These denominators
are chosen in such a way that the remaining ones
can be successively integrated out without ever computing
an integrand with more than two propagators and without collecting more
than at most one point-singularity factor.
This choice of integration variables
are determined from $M=M(\sigma)$  by an algorithmic procedure that we
describe now.

We first recall that $M$ is a tower matrix, and
we define $b(j)$ and $t(j)$ to be the bottom and
the top of the tower in the $j$-th column, i.e.
$$
   b(j): = \max \{ i\; : \; M_{ij} \neq 0\}, \qquad
    t(j): = \min \{ i\; : \; M_{ij} \neq 0\} \; .
$$
We also recall the concepts 
from Definition \ref{def:slope}.
Any non-peak index, $i\in I\setminus I_p$, clearly has
the property that $i$  is the bottom of the tower of some column of $M$,
 i.e. there exists $j$ such that $b(j)=i$.
 This column index can be either 
 $j=\tsi^{-1}(i)$ or $j=\tsi^{-1}(i)+1$ or both;
ambiguity occurs only if $i\in I_v$ is a valley index.
For any ladder or slope or last index,
$i\in I_\ell\cup I_s\cup I_{last}$, we therefore define $c(i)$
to be the unique 
column index $j$ such that $b(j)=i$.
For $i\in I_{last}$, i.e. $i=k+1$,
we always have $c(k+1)=k+1$.
If $i\in I_v$, we define $c(i)$ 
to be index of the column whose tower is  higher, i.e.
$c(i): = \tsi^{-1}(i)$ if $t(\tsi^{-1}(i))<t(\tsi^{-1}(i)+1) $ and
 $c(i): = \tsi^{-1}(i)+1$
otherwise. The other column index will denoted by $\wt c(i)$;
this concept is defined only for $i\in I_v$.
  The $(i, c(i))$, $i\in I\setminus I_p$,
 elements of the matrix $M$ are called
{\bf pivot} elements; these will be used for
determining the integration variables.
The $c(i)$-column will be called the pivot column of $i$
and the corresponding tower in this column is called the
pivot tower of $i$. The definition guarantees that
the pivot tower of an index $i\in I_s\cup I_v$ has
at least length two.

\begin{definition}\label{def:cov}
 Let $h_1< h_2< \ldots $ denote the elements of
$I_v\cup I_s$ in increasing order. A slope index $h_\mu\in I_s$,
is called {\bf covered} if $t(c(h_{\mu+1})) \leq  h_\mu$, i.e. 
if the top of the tower of the next pivot element of a valley or a slope 
 is higher than $h_\mu$.
The remaining slope indices are called {\bf uncovered}.
The set of covered and uncovered slope indices will be denoted
by $I_{cs}$ and $I_{us}$, respectively.
\end{definition}

Fig.~\ref{fig:cover} shows three examples. In the
first two cases, the index $h_\mu$ is
covered  by slope and by a valley. The last case is an uncovered slope
index. The shaded boxes indicate the pivot elements.

\bef\bec
\epsfig{file=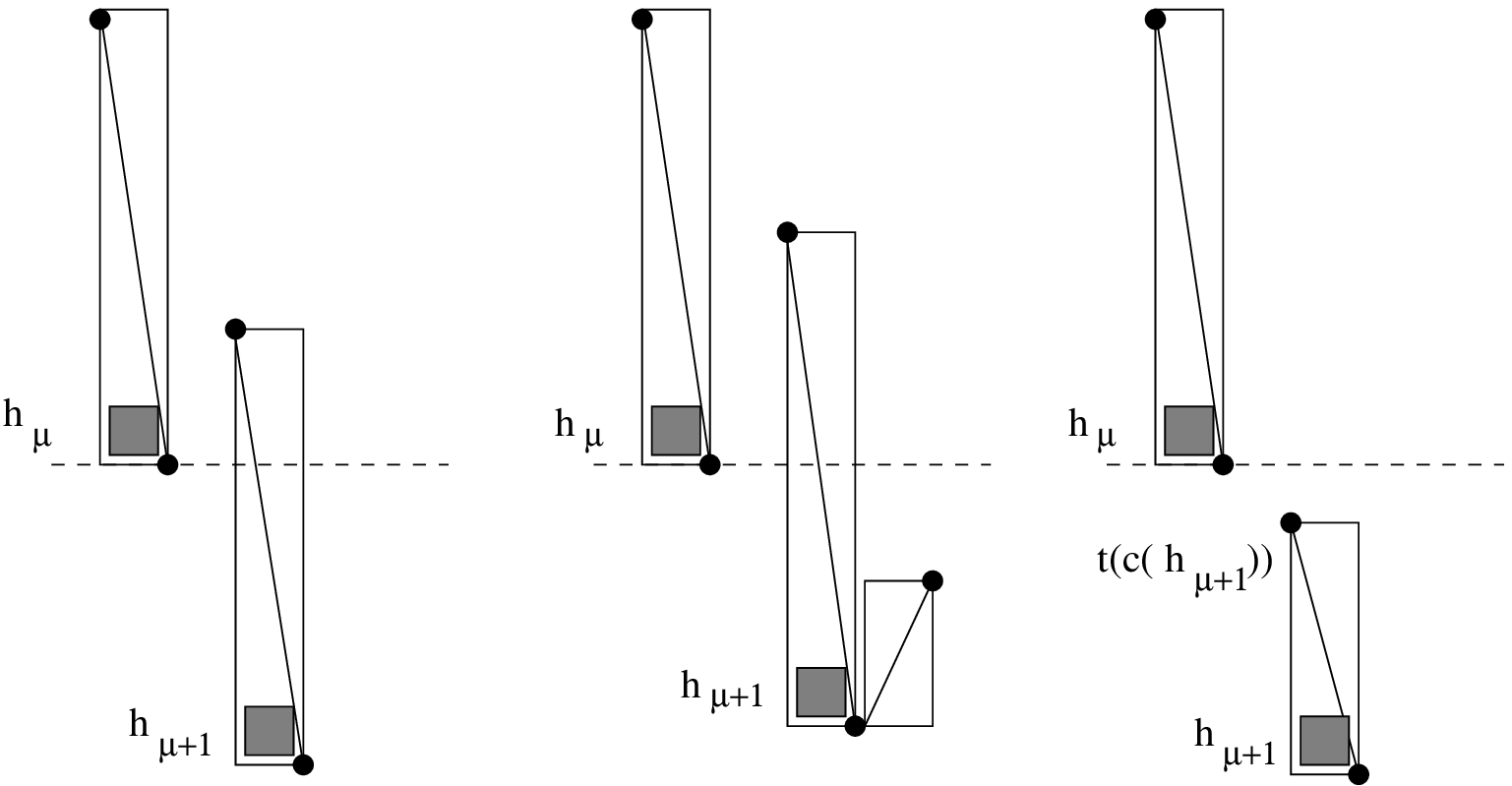, scale=0.9}
\eec
\caption{Two examples for a covered slope index $h_\mu$ and an uncovered
index}
\label{fig:cover}
\eef

Apart from the ladder momenta, $p_{c(i)}$, $i\in I_\ell$, 
we will integrate out the variables $p_{c(h_1)}, p_{c(h_2)}, \ldots$,
in this order. We will show that
at every integration step only one $\alpha$-denominator,
namely $|\a -\bar \om (p_{c(h_\mu)}) -i\eta|^{-1}$, and one
$\beta$-denominator, namely $|\beta - \om (\tp_{h_\mu})+i\eta|^{-1}$,
will contain the integration variable $p_{c(h_\mu)}$, so
we will never have to estimate integrals with more than two
propagators.

The significance of the covered slope indices is that
their integration step yields a point singularity
that will be integrated out immediately at the next
non-ladder integration step.  Therefore covered slope indices
do not give rise to accumulation of point singularities.

The covered slope indices, together with the valley indices and the
last index will be used to gain a $\lambda^{const}$ factor in
the integration procedure. They will not accumulate point singularities.
 The uncovered slope indices
and the peak indices will be treated by the trivial $L^\infty$ bound
 (\ref{eq:Linftybound}) and no gain is obtained from them.
That is, we prefer giving up the potential gain of a $\lambda^{const}$ factor 
from the uncovered slope indices to dealing with 
the accumulated point singularities. 
Finally, the ladder indices will be treated
 by Method II with the Schwarz inequality.

\begin{lemma}\label{lm:num}
The number of uncovered slope indices is at most $v$, 
where $v=|I_v|$ is the number of valleys. 
\end{lemma}

This lemma ensures that treating the uncovered slope indices 
by the trivial bound
is affordable (at the expense of a worse $\kappa$), since
each valley will result in a gain of a $\lambda^{const}$ factor.

\bigskip

{\it Proof of Lemma \ref{lm:num}.} 
First we state two claims from which we easily deduce 
the Lemma, then we prove these claims. Recall
that the sequence $h_1< h_2 < \ldots $ lists all elements of
 $I_v\cup I_s$.

\medskip

{\it Claim 1.} Let  $h_\nu\in I_{s}$ be the first slope index
in the sequence $h_1< h_2 < \ldots $. Then
there is a peak index $h$ such that $h<h_\nu$.

\medskip

{\it Claim 2.} Let $h_\mu\in I_{us}$ be an uncovered slope index and suppose
that it is not the last entry in the sequence $h_1< h_2 < \ldots$,
i.e. there is a next element, $h_{\mu+1}\in I_s\cup I_v$.
Then there is an intermediate index, $h_{\mu} < h < h_{\mu+1}$,
that is a peak index.

\medskip

By Claim 2,  one can assign a distinct peak index
to every uncovered slope index $h_\mu$ (but the last one),
namely the next peak index. Moreover, by Claim 1, there
is an additional peak index smaller than any uncovered slope index.
Therefore the number of uncovered slope indices is at most
the number of peaks. The statement the lemma
then follows, since
 the number of peaks and valleys is the same, $p=v$.

\medskip

{\it Proof of Claim 1.}  Suppose that the statement is false.
 First note that $h_\nu\ge 2$ since index 1 cannot
be slope.  Then all indices $1, 2, \ldots h_\nu-1$
 must be either ladders 
or valleys. But for any valley index $j$ there is obviously
a peak index $j'<j$, since the graph of the permutation
goes down from (0,0) to $(k+1, k+1)$, so following
the two branches of the graph emanating upwards starting from a valley
at least one of them must run into a peak that lies higher.
Therefore all  indices $1, 2, \ldots h_\nu-1$ must be ladders
and they form a ladder, with the index 0 being the
top of this ladder. In particular $t(c(h))=h-1$, $h=1,2,\ldots h_\nu-1$
(recall that $t(c(h))$ denotes the index of the top
row in the pivot tower of $h$), so every index $0,1,2,\ldots h_\nu-2$
is the top of a column of length one. Moreover, none of these
indices can be the top of any other column, since an index
which is the top of two different columns must be a peak index.
But in that case there is no room for $t(c(h_\nu))$, the top 
index of the pivot column of $h_\nu$: since $h_\nu$
was a slope, $t(c(h_\nu))\le h_\nu-2$.

 \medskip

{\it Proof of Claim 2.} 
To see the claim, suppose that none of the
indices $h$, strictly between $h_{\mu}$ and $h_{\mu+1}$, is
a peak. These indices cannot be slopes or valleys
either since  $h_{\mu}$ and $h_{\mu+1}$ are
two consecutive elements of $I_s\cup I_v$
and $h$ cannot be the last index, $h\not\in I_{last}$, 
since  $h< h_{\mu+1} <k+1$. Therefore all indices $h$
with $h_{\mu} < h < h_{\mu+1}$ are ladders.

Now we look at $h^*=t(c(h_{\mu+1}))$, i.e.
the index of the top row in the pivot tower 
of $h_{\mu+1}$ (see the last picture on Fig.~\ref{fig:cover}).
Since $h_\mu$ was uncovered, $h_\mu < h^*$.
Note that the pivot tower of $h_{\mu+1}\in I_s\cup I_v$
 has length at least two, therefore $h^*< h_{\mu+1}$,
and thus $h^*\in I_\ell$. Since $h^*$ was also the top
of a tower with length at least two, $h^*-1$ must be a peak index
since it is a row index right above the top of two towers
(namely the towers in the columns $c(h^*)$ and $c(h_{\mu+1})$).
Clearly $h_\mu \leq h^*-1 < h_{\mu+1}$, but $h_\mu\neq h^*-1$
since $h_\mu \in I_s$ and $h^*-1\in I_p$. Thus $h^*-1$ is a
peak index strictly between $h_\mu$ and $h_{\mu+1}$, which
is a contradiction. $\;\;\Box$

\subsection{Integration procedure}

Our goal is to estimate (\ref{def:EM}) for $M=M(\sigma)$
when $\sigma\neq id$.
 We start with defining
\be
   \tri q \tri : = \eta+\min\{ |q|, 1\}, \qquad q\in \bR^d \; .
\label{tri}
\ee
This is not a norm, but it satisfies the triangle inequality,
$\tri p + q\tri \leq \tri p \tri + \tri q \tri$.
For any index set $I'\subset I=\{ 1,  2,\ldots,
k+1\}$  we define the function
 $\cU_{I'}$ as the product of those potential terms, $|\wh B|$,
in \eqref{def:EM}
that depend only on momenta $\{p_j\; : \; j\in I'\}$. More precisely,
$$
       \cU_{I'}(\bp,\tbu):= \Bigg| \prod^*_j \wh B(p_{j+1} - p_j)
        \prod^*_j \wh B\Big( [M\bp+\tbu]_{j+1}  - [M\bp+\tbu]_j \Big)
   \;\Bigg| \; ,
$$
where the star indicates a product on a restricted index set.
The first product is taken over all indices $j$ for which $j, j+1 \in I'$.
The second product  is taken over  those $j$'s, for which
$M_{j+1, b} = M_{j, b}$ for all $b\not\in I'$.

For  any $|I'|\times(k+1)$ 
matrix $M$ and any vector $\fb=(b_1, b_2, \ldots , b_{k+1})
\in \bR^{k+1}$ we define
\begin{align}
    E(I', M, \fb): = \lambda^{2k} \sup_{\tbu,v} &
  \iint_{-Y}^{Y} \rd\alpha \rd\beta 
  \sup_{p_j\; : \; j\not\in I'} 
  \int \prod_{j\in I'}
\rd \mu(p_j)  \;  \cU_{I'}(\bp,\tbu) 
 \frac{1}{\tri \fb\cdot \bp+ v  \tri} \;  
\nonumber \\
&\times\Bigg(\prod_{j\in I'}
    \frac{1}{|\alpha-\ov\om(p_j)-i\eta|}\frac{1}{|\beta -\om\big(
    [M\bp+\tbu]_j \big)+i\eta|}\Bigg) \; .
\label{eq:EE}
\end{align}
Here $ v\in \bR^d$ is
an additional  dummy momentum and $\fb\cdot\bp$ is defined 
analogously to \eqref{def:Mp} as
$$
   \fb \cdot \bp : = b_1p_1 + b_2p_2 + \ldots + b_{k+1} p_{k+1} \; .
$$
We will also use the notation $E(I', M, \emptyset)$ defined
exactly by the same formula as \eqref{eq:EE} without the
factor $\tri \fb\cdot\bp + v\tri^{-1}$ in the integrand
and without the supremum over $v$.
With a slight abuse of notation we will refer to this case
as chosing the ``empty vector'' $\fb=\emptyset$.

\medskip

For each $h\in I$ we will define an index set $I^{(h)}$,
a matrix $M^{(h)}$, and a vector $\fb^{(h)}$ such that
$E(I^{(h)}, M^{(h)}, \fb^{(h)})$ is the 
remaining integral 
after the $h$-th integration step.
Each step consists of eliminating one $\beta$-denominator;
in the $h$-th step we will eliminate 
$|\beta - \om([M\bp+\tbu]_h)+i\eta|^{-1}$ and sometimes
one or two additional $\a$-denominators. The
$\a$-denominators will always be eliminated by integrating
out the corresponding $p_j$-variable. The order of integrations
will be such that the integrand in $E(I^{(h)}, M^{(h)}, \fb^{(h)})$ 
does not depend on any $p_j\not\in I^{(h)}$. 
In particular, the supremum over all $p_j\not\in I^{(h)}$
in the definition \eqref{eq:EE} will be superfluous in the integrals
appearing in our iteration.

The elimination will be done differently, depending on
the type of the index $h$. If $h\in I_p$, then the trivial
$L^\infty$ bound \eqref{eq:Linftybound} will be used.
If $h$ is an uncovered slope index, then again
the trivial bound will be used, and, additionally,
the variable $p_{c(h)}$ will be integrated out.
If $h$ is a slope or ladder index, then
variable  $p_{c(h)}$ will be integrated out.
If $h$ is a valley then we integrate
out both variables, $p_{c(h)}$ and $p_{\wt c(h)}$.
Finally, in the last step, $h = k+1$, we will integrate
out the $\beta$ variable.
We will bookkeep this integration
by removing the corresponding $h$ rows from $M$ and we will say in
short that these rows have been integrated out. 
At the end we will have to integrate out the remaining
$p_j$-variables, but then only  one $\a$-denominator
will be present so the integration can be done easily.

Point singularities are created by integrating out
a covered slope index, but there is never more than one
singularity present. An existing point singularity
will not be touched by  the ladder integrations and
it will be changed to another point singularity
when a covered slope index is integrated out.
In all other integration steps an existing point singularity
will be removed.

\bigskip

Now we define $M^{(h)}$, $I^{(h)}$ and $\fb^{(h)}$.
Let $M^{(0)}: = M=M(\sigma)$, $I^{(0)}:= I = \{ 1, 2, \ldots, k+1\}$
and $\fb^{(0)}=\emptyset$. In particular, note that
\be
    E(M)\leq \| \wh\psi_0 \|_\infty^2 \; E(I^{(0)}, M^{(0)}, \fb^{(0)})\; .
\label{e0}
\ee
For any $h\ge 1$, 
define $M^{(h)}$ to be the matrix $M=M(\sigma)$  with
the first $h$ rows, with indices $i=1,2, \ldots, h$, removed.
For $h\ge 1$, let
$$
    I^{(h)}:= I\setminus \Bigg( 
\big\{ c(h')\; : \; h'\leq h, h'\in I\setminus I_p\big\} \cup 
\big \{ \wt c(h')\; : \; h'\leq h, h'\in I_v\big\} \Bigg)
$$
be the index set of columns that have not been integrated out
up to the $h$-th step.

The vectors $\fb^{(h)}=(b^{(h)}_1, \ldots, b^{(h)}_{k+1})$
are defined as follows. For any given $h\in I$, let
$h_\mu$ be the largest element
in the listing $h_1<h_2< \ldots$ of $I_v\cup I_s$ (see Definition
\ref{def:cov}) such that $h_\mu \leq h$. If there is no such $h_\mu$
or $h_\mu\not\in I_{cs}$, then
we set $\fb^{(h)}=\emptyset$.
If $h_\mu \in I_{cs}$, then 
\be
   b^{(h)}_{j} : = \left\{ 
\begin{array}{ll} M_{h_\mu, j} & \mbox{if $j\neq c(h)$} \cr
0 &   \mbox{if $j=c(h)$.}
\end{array}
\right.
\label{def:b}
\ee

We collect  a few information about $\fb^{(h)}$
that directly follow from the definition and
from the structure of $M$.

\begin{itemize}

\item[(i)]

Suppose that $\fb^{(h)}\neq\emptyset$, and $h_\mu$  as above
(i.e. $h_\mu \leq h$ and it is the largest element with
this property).  Since $h_\mu\in I_{cs}$, i.e., $t(c(h_{\mu+1}))\le h_\mu$,
we therefore  have
\be
  b^{(h)}_{c(h_{\mu+1})}=\pm 1, \qquad  h_\mu \leq h < h_{\mu+1}, \quad
  h_\mu\in I_{cs}.
\label{bpm}
\ee

\item[(ii)]  For any fixed $\mu$,
$\fb^{(h)}$ is constant as $h$ runs through
the interval  $\{ h_\mu, \ldots, h_{\mu+1}-1\}$.

\item[(iii)] 
Consider the  pivot column $c(i)$ of a ladder index $i\in I_\ell$.
Since the only non-zero entry in the $c(i)$ column of $M$ is
$M_{i, c(i)}$, we have that $M_{h_\mu, c(i)}=0$ for all $\mu$. But then
we see that  the entries corresponding to any ladder pivots
are zero in the $\fb^{(h)}$ vectors:
\be
  b^{(h)}_{c(i)}=0\quad  \mbox{for any $i\in I_\ell$ and any
$h$ such that $\fb^{(h)}\neq\emptyset$.}
\label{bc}
\ee
\end{itemize}

To describe the integration, we explain how each case,
depending on the index type of $h$, is estimated.
We also introduce the short notation
$$
    E(h) : =E(I^{(h)}, M^{(h)}, \fb^{(h)}) \; , \quad h=0,1,\ldots, k+1.
$$

\medskip

\underline{{\it Case 1:} $h\in I_p$. } 
In this case no integration is done,
$I^{(h)}= I^{(h-1)}$ and  $\fb^{(h)}= \fb^{(h-1)}$.
 We estimate the denominator
  $|\beta - \om( [M\bp+\tbu]_h )+i\eta|^{-1}$
 trivially as in (\ref{eq:Linftybound}) and we obtain
\be
    E(h-1)\leq \eta^{-1}
    E(h)\;, \qquad  h\in I_p\; .
\label{eq:ind1}
\ee

\medskip

\underline{{\it Case 2:} $h\in I_\ell $.}
  Let $h, h+1, \ldots,  h+\tau-1 \in I_\ell$
be a ladder, i.e. a maximal sequence
of consecutive ladder indices, i.e.  $h+\tau\not\in I_\ell$
with some $\tau\ge 1$. We will integrate out the whole ladder
in this step, so we can assume that $h-1$ is the top of a ladder, 
$h-1\in I_t$. 
By definition of the ladder, 
 the pivot indices, $c(h), c(h+1), \ldots, c(h+\tau-1)$
are consecutive numbers, which, for definiteness, are assumed to be
increasing (the other case is similar).
Let $c=c(h)$.
Because $h, \ldots, h+\tau-1$ are all consecutive
ladder indices,
$[M\bp]_i = \bp_i + w$ for all $c\leq i \leq c+\tau-1$,
with a vector $w =w(\bp)$ that does not depend
on $\{ p_j \; : \; c\leq j \leq c+\tau-1\}$.
Note that $\fb$ remains unchanged as $k$ runs through the ladder indices:
 $\fb^{(k)} =\fb^{(h-1)}$ for all $k=h, h+1, \ldots , h+\tau -1$.
We claim
\be
    E(h-1) \leq 
C\lambda^{-2\tau}  \,
    E(h+\tau-1)\;, \qquad h-1\in I_t\;
\label{eq:ind2}
\ee
with a constant $C$ independent of $\tau$.
This inequality entails  integrating
 out the variables $p_c, p_{c+1}, \ldots , p_{c+\tau-1}$
(in this order) to remove the rows $h, h+1, \ldots h+\tau-1$.
This requires estimating the following integral
\begin{align}
  \cI:= \sup_{\a, \beta, w}
  \int & \prod_{j=c}^{c+\tau-1}  \frac{ \rd\mu(p_j)  }
{|\a - \ov\om(p_j)-i\eta|
    |\beta - \om (p_j + w + \tu_j)+i\eta|}  
\label{laddd} \\
&
 \times \prod_{j=c}^{c+\tau-1} |\wh B(p_{j+1}-p_j)|
   |\wh B( p_{j+1}+ \tu_{j+1}- p_j-\tu_j)|
\nonumber 
\end{align}
because the possible point singularity present in 
$E(I^{(h-1)}, M^{(h-1)}, \fb^{(h-1)})$  does not contain
any of the integration variable by \eqref{bc}.
Due to the ladder structure it is  easy to see that $\cU_{I^{(h-1)}}$
indeed contains all the $\wh B$ factors shown in \eqref{laddd}.

 After a Schwarz inequality we use an iterative integration
scheme similar to \eqref{ladddint1}--\eqref{ladddint3} 
to estimate the $\tau$-fold multiple integral
\begin{align}
 \int &\frac{|\wh B(p_{c+\tau}-p_{c+\tau-1})|^2d\mu(p_{c+\tau-1})}
{|\a -\ov\om (p_{c+\tau-1})-i\eta|^2} \int
\frac{|\wh B(p_{c+\tau-1}-p_{c+\tau-2})|^2d\mu(p_{c+\tau-2})}
{|\a -\ov\om (p_{c+\tau-2})-i\eta|^2} \ldots \nonumber\\
&\ldots \int \frac{|\wh B(p_{c+2}-p_{c+1})|^2d\mu(p_{c+1})}
{|\a -\ov\om (p_{c+1})-i\eta|^2} 
\int \frac{|\wh B(p_{c+1}-p_c)|^2d\mu(p_c)}
{|\a -\ov\om (p_c)-i\eta|^2}  \nonumber\\
& \qquad \leq \big[ \lambda^{-2}(1+C\lambda^{1-12\kappa})\big]^{\tau-1}
\nonumber\\
& \qquad\quad\times  \int 
\frac{|\wh B(p_{c+\tau}-p_{c+\tau-1})|^2d\mu(p_{c+\tau-1})}
{|\a -\ov\om (p_{c+\tau-1})-i\eta|^2}\Big[ 1 + C_0\lambda^{-12\kappa}
(\lambda +|\a -\om(p_{c+\tau-1})|^{1/2})\Big] \nonumber \\
& \qquad  \leq C\big[ \lambda^{-2}(1+C\lambda^{1-12\kappa})\big]^{\tau-1} 
\leq C \lambda^{-2(\tau-1)}\; . \nonumber
\end{align}
In the last integration we use \eqref{eq:2aint} with $a=0$ and $a=1/2$
and we also used that  $\tau\leq K\ll \lambda^{-(1-12\kappa)}$.
This gives
\be
   \cI \leq C\lambda^{-2\tau}  \; .
\label{ci}
\ee

\medskip

\underline{{\it Case 3:} $h\in I_{us}$.}  We have $I^{(h)}=
I^{(h-1)}\setminus \{ c(h)\}$ and $\fb^{(h)}=\emptyset$.
 We start proceeding as in Case 1
by first estimating the  denominator
  $|\beta - \om( [M\bp+\tbu]_h )+i\eta|^{-1}$
 trivially as in (\ref{eq:Linftybound}). Then observe
that the variable $p_{c(h)}$ appears only in one propagator,
namely in $|\a - \ov\om(p_{c(h)})-i\eta|^{-1}$.
Moreover, if $\fb^{(h-1)}\neq \emptyset$, then
$p_{c(h)}$ appears also in the point singularity,
since $b^{(h-1)}_{c(h)}=\pm 1$ by \eqref{bpm}.

Before we integrate out $p_{c(h)}$, we estimate all $\wh B$ factors
that contain this variable by supremum norm.
The number of such factors is at most four so  they can
be bounded by a constant factor $C$. Then,
to integrate out $p_{c(h)}$,
we either use the bound (with $p=p_{c(h)}$)
\be
  \sup_{|\a|\leq Y}  \int \frac{\rd\mu(p)}{|\a -\ov\om(p)-i\eta|}
  \leq C\zeta^{d-2}|\log\eta|
\label{nosing}
\ee
if $\fb^{(h-1)}=\emptyset$, or we use the bound
\be
  \sup_{|\a|\leq Y}\sup_r  \int \frac{\rd\mu(p)}{|\a -\ov\om(p)-i\eta|}
 \; \frac{1}{\tri p -r\tri} \leq C\zeta^{d-2}|\log\eta|
\label{1sing}
\ee
if $\fb^{(h-1)}\neq\emptyset$
(recall $\zeta=\lambda^{-\kappa-3\delta}$ from \eqref{param}). The bound \eqref{1sing}
 follows immediately from \eqref{resex} and \eqref{1dee},
the bound \eqref{nosing} is weaker since $\sup_p\tri p \tri \leq 1+\eta$
(the bound \eqref{nosing} can be improved to $C|\log\eta|$ but
we will not need this fact).
In summary, we have proved that
\be
    E(h-1) \leq C\zeta^{d-2}\eta^{-1}|\log\eta| \,
    E(h) \; , \qquad h\in I_{us}\; .
\label{eq:ind3}
\ee

\bigskip

\underline{{\it Case 4:} $h\in I_{cs}$.}  In this step we will
have $I^{(h)} = I^{(h-1)}\setminus \{ c(h)\}$ and $\fb^{(h)}$
will be the $h$-th row of $M$ but with the pivot element
$(h, c(h))$ changed to zero, i.e., $b^{(h)}_h =0$.
Since index $h$ is among the listing $h_1<h_2< \ldots $
of the elements $I_v\cup I_s$ (see Definition \ref{def:cov}), let
$h=h_\mu$ and let $h':= h_{\mu+1}$ be the next element of $I_s\cup I_v$.
 Since $h$ is a covered slope index,  we have
$t(c(h'))\leq h$ and thus we know that
the $(h, c(h'))$ matrix element of $M$ is $\pm 1$. 
Therefore the variable
$p_{c(h')}$  appears non-trivially in $\wt p_h=[M\bp + \tbu]_h$:
\be
   \wt p_h = [M\bp + \tbu]_h = \pm p_{c(h)} \pm p_{c(h')} \pm \ldots
\label{wpt}
\ee
After estimating those $\wh B$ factors by supremum norm that contain $p_{c(h)}$,
we will integrate
out $p_{c(h)}$
 by using the following
lemma  proved in the Appendix:

\begin{lemma}\label{lemma:newrow} Let $\lambda^3\leq \eta\leq\lambda^2$,
 and $|q|\leq C\lambda^{-1}$.
Then we have
\be
    \sup_{\a,\beta, r} \int
    \frac{\rd\mu(p)}{|\alpha - \ov\om(p)-i\eta|\; |\beta - \om(p+q)+i\eta|}
     \frac{1}{\tri p-r\tri }  \;
    \leq \frac{C \eta^{-1/2}\zeta^{d-3}
 |\log\eta|^2}{\tri q\tri }\; ,
\label{eq:dp}
\ee
Without point singularity 
 we have the following improved bound
\be
    \sup_{\alpha, \beta} \int
    \frac{ \rd \mu(p)}{|\alpha - \ov\om(p)-i\eta|\; |\beta - \om(p+q)+i\eta|} 
    \leq \frac{C \zeta^{d-3}
 |\log\eta|^2}{\tri q\tri }\; . 
\label{eq:dp5}
\ee
\end{lemma}
\medskip

The first estimate \eqref{eq:dp} is used if $\fb^{(h-1)}\neq\emptyset$,
i.e. if there was a nontrivial point singularity present.
By \eqref{bpm}, this nontrivial point singularity must originate
from the previous covered slope index, in particular
$h_{\mu-1}\in I_{cs}$ and $b^{(h-1)}_{c(h)}=\pm 1$, i.e. 
 the point singularity contains the variable
$p_{c(h)}$. The new point singularity, obtained on the
right hand side of \eqref{eq:dp} as $\tri q \tri^{-1}$,
 is exactly the linear combination  appearing in \eqref{wpt}
without the integration variable, $p_{c(h)}$. Therefore
the $p$-dependence of this point singularity is given
by $\fb^{(h)}$ (up to an overall sign that is irrelevant
since $\tri \cdot \tri$ is symmetric).
If $\fb^{(h-1)}=\emptyset$, then we can use the second estimate
\eqref{eq:dp5} and otherwise we follow the same argument 
as in the $\fb^{(h-1)}\neq\emptyset$ case.

The net effect of the elimination in Case 4 is
\be
    E(h-1) \leq C\eta^{-1/2}
   \zeta^{d-3}|\log\eta|^2 \, E(h)\; , \qquad h\in I_{cs} .
\label{eq:ind4}
\ee

\bigskip

\underline{{\it Case 5:} $h\in I_{v}$.}  In this step we will
have $I^{(h)} = I^{(h-1)}\setminus \{ c(h), \wt c(h)\}$
and  $\fb^{(h)}=\emptyset$. Recall that $c(h)$ and $\wt c(h)$
are the two columns between which the valley is located,
i.e. $\{ c(h), \wt c(h)\} = \{ \tsi^{-1}(h), \tsi^{-1}(h)+1\}$.
We integrate out both $p_{c(h)}$ and $p_{\wt c(h)}$
after estimating the $\wh B$ factors that are involved by their supremum norm.
This eliminates two $\alpha$-denominators and 
the $|\beta -\om ([M\bp+\tbu]_h)+i\eta|^{-1}$ denominator.
It follows from the definition of the valley index that
$$
[M\bp+\tbu]_h = \pm p_{c(h)}\mp p_{\wt c(h)} \pm \ldots
$$
i.e. the $\beta$ denominator contains both integration variables.
If $\fb^{(h-1)}=\emptyset$, i.e. no point singularity is
present, then
the prototype of this integral can be estimated as follows:
$$
    \sup_{|\alpha|, |\beta|\leq Y} \int\!\int
    \frac{\rd\mu(p)\rd\mu(p')}{|\alpha - \ov\om(p)-i\eta|\; 
|\a -\ov\om(p')-i\eta|\;
   |\beta - \om(p-p'+q)+i\eta|}\leq C\zeta^{2d-5}|\log\eta|^3
$$
assuming $|q|\leq C\lambda^{-1}$. This estimate follows from 
\eqref{eq:dp5} and \eqref{1sing}.

If a point singularity is present, $\fb^{(h-1)}\neq \emptyset$,
we know that $b^{(h-1)}_{c(h)}=\pm 1$ by \eqref{bpm},
therefore at least one of the integration variables
appear in it. Depending on whether $\wt c(h)$ also
appears or not, we have one of the following two
prototype estimates:
$$
    \sup_{|\alpha|, |\beta|\leq Y} \sup_r \int\!\int
    \frac{\rd\mu(p)\rd\mu(p')}{|\alpha - \ov\om(p)-i\eta|\; 
|\a -\ov\om(p')-i\eta|\;
   |\beta - \om(p-p'+q)+i\eta|}\; \frac{1}{\tri p - r\tri}
$$
\be
\leq C\eta^{-1/2}\zeta^{2d-5}|\log\eta|^3
\label{onlyr}
\ee
or
$$
    \sup_{|\alpha|, |\beta|\leq Y} \sup_r\int\!\int
    \frac{\rd\mu(p)\rd\mu(p')}{|\alpha - \ov\om(p)-i\eta|\; 
|\a -\ov\om(p')-i\eta|\;
   |\beta - \om(p-p'+q)+i\eta|}\; 
\frac{1}{\tri p \pm p'- r\tri}
$$
\be
\leq C\eta^{-1/2}\zeta^{2d-5}|\log\eta|^3\; ,
\label{twor}
\ee
assuming $ |q|\leq C\lambda^{-1}$.
Both inequalities are obtained by first applying
\eqref{eq:dp} for the $\rd p$  integral, then 
applying \eqref{1sing} for the $\rd p'$ integral.

In summary, the effect of the elimination in Case 5 is
\be
    E(h-1) \leq C\eta^{-1/2}
   \zeta^{2d-5}|\log\eta|^3 \,
    E(h)\; , \qquad h\in I_v.
\label{eq:ind5}
\ee

\bigskip

We eliminate the row indices one by one in increasing
order according to Cases 1--5.
The total estimate is
\eqn
  E(0)&\leq & \big(\eta^{-1}\big)^{|I_p|}
 \lambda^{-2\ell} C^{|I_t|} 
(C\zeta^{d-2}\eta^{-1}|\log\eta|)^{|I_{us}|}
(C\eta^{-1/2}\zeta^{d-3}|\log\eta|^2)^{|I_{cs}|}\nonumber\\
&&\times (C\eta^{-1/2}\zeta^{2d-5}|\log \eta|^3)^{|I_v|}
E(k+1)\; ,
\label{E0}
\eeqn
where we recall that $I_t$ is the set of 
indices that are tops of a ladder, in particular,
Case 2 is applied $|I_t|$ times.
Since the top of the ladder is not a ladder index and not a valley, 
so it is a peak, slope or maybe 0,  thus
$$
    |I_t|\leq |I_p|+ |I_s| +1 = v+s+1\leq 2v+s
$$
taking into account that $p=v$ and $v\ge 1$ since $\sigma\neq id$.
{F}rom \eqref{E0} and from $|I_{cs}|+|I_{us}|=|I_s|=s$
it then follows that
\be
  E(0)\leq C^{v+s+1} \lambda^{-2\ell}\eta^{-\frac{1}{2}(3v+s+|I_{us}|)}
   \zeta^{d(2v+s)} E(k+1)\;
\label{E1}
\ee
(we also used that $|\log\eta|^3\ll \zeta$). 
Note that the exponent of the constant is comparable with $\mbox{deg}(\sigma)=
k-\ell = 2v+s$. A factor $C^\ell$ would not be 
affordable if $\ell\gg \mbox{deg}(\sigma)$;
this is why the ladder integration had to be done essentially with
the precise constant by using \eqref{eq:ladderint}.

Finally, when estimating $E(k+1)$, only the
 last row of $M$, $h=k+1\in I_{last}$ is present.
Since $k+1$ is 
bigger than the largest element of $I_s\cup I_v$, which
by Definition \ref{def:cov} cannot be a covered slope index,
we see from the definition of $\fb^{(h)}$ that $\fb^{(h)}=\emptyset$,
i.e. there is no point singularity.
Moreover, we have already integrated out $|I_\ell|+ |I_s| + 2 |I_v|
= \ell + s + 2v$ variables. Since $p=v$ and $I=I_\ell \cup I_s \cup
I_v\cup I_p\cup I_{last}$ is a partition, we see that
at the last step there is only one integration 
variable left, namely $p_{k+1}$.
Considering that the $(k+1)$-th row of $M$ contains only
one nontrivial entry and thus $[M\bp+\tbu]_{k+1}=p_{k+1} + \wt u_{k+1}$,
 the integral from \eqref{eq:EE} is simplified to
\be
   E(k+1)=E(I^{(k+1)}, M^{(k+1)}, \fb^{(k+1)})
\label{k+1}
\ee
$$
  =\lambda^{2k}\sup_{\wt u}\iint_{-Y}^Y \rd \a\rd \beta
  \int \frac{\rd \mu(p_{k+1})}{|\a - \ov\om(p_{k+1})-i\eta|}
  \; \frac{1}{|\beta - \om(p_{k+1} + \wt u)+i\eta|} \leq C
\lambda^{2k}\zeta^d|\log\eta|^2
\;.
$$

We now combine the
inequalities \eqref{e0}, \eqref{E1}
and \eqref{k+1}, we
arrive at
$$
    E(M)\leq C^{v+s+1}\lambda^{2(k-\ell)} \eta^{-\frac{1}{2}(3v+s+|I_{us}|)}
    \zeta^{d(2v+s+1)} |\log\eta|^2
$$
where we recall that the general constant $C$ may depend
on $\wh \psi_0$. 
Using the choice of the parameters from \eqref{param}
and the fact that $\mbox{deg}(\sigma)= k-\ell = 2v+s\ge 2$, we obtain
a total $\lambda$-power
$$
    2(2v+s)-\big(1+\frac{\kappa}{2}\big)(3v+s+|I_{us}|) -(\kappa+3\delta)
 d(2v+s+1)
$$
$$
    \ge v+|I_{cs}| - 
\kappa\big(1+\frac{3}{2}d\big)(2v+s) -(2v+s)O(\delta)
$$
where we used Lemma \ref{lm:num} and $v\ge 1$
to estimate $\frac{1}{2}(3v+s+|I_{us}|)\leq 2v+s$ and
 $2v+s+1\leq \frac{3}{2}(2v+s)$.
Using Lemma \ref{lm:num} again, 
 we see that $v+|I_{cs}|\ge 
\frac{1}{3}(2v+s)$, i.e.
$$
    E(M)\leq C \Big(\lambda^{\frac{1}{3}-\big(1+\frac{3}{2}d\big)
\kappa  - O(\delta)}\Big)^{\de(\sigma)} |\log\eta|^2
$$
if the exponent is positive,  i.e. if $\kappa< \frac{2}{6+9d}$
and $\delta \leq \delta(\kappa)$ is sufficiently small.
This completes the proof of Lemma \ref{lemma:ems}.
 $\;\;\;\Box$

\appendix

\section{Proof of Lemma \ref{lemma:newrow}.}\label{secnewroww}
\setcounter{equation}{0}

We can replace $\om(p)$ with $e(p)$ in \eqref{eq:dp}
by using 
 a straightforward resolvent expansion:
\be
  \frac{1}{|\a - \om(p) +i\eta|} \leq \frac{1}{|\a -\lambda^2 \Theta(\a)-
 e(p) +i\eta|}\Bigg[ 1+ \frac{C\lambda^2 |\a - e(p)|^{1/2}}
   {|\a - \om(p) +i\eta|}\Bigg]
  \leq \frac{C}{|\wt\a - e(p) +i\eta|}
\label{resex}
\ee
with $\wt\a = \a -\lambda^2 \, \mbox{Re} \, \Theta(\a)$.
We used the boundedness and the
H\"older continuity of $\Theta$ \eqref{eq:holder}.
Therefore the proof of Lemma \ref{lemma:newrow} is reduced to

\begin{lemma}\label{lemma:main}
For any $|q|\leq C\lambda^{-1}$
\be
  I_1:=\int \frac{\rd \mu( p)}{|\a - e(p) +i\eta|\,
  |\beta - e(p+q) +i\eta|}\, 
  \leq  \frac{C\zeta^{d-3}|\log\eta|^2}
   {\tri q \tri }
\label{withoutp}
\ee
\be
  I_2:=\int \frac{\rd \mu( p)}{|\a - e(p) +i\eta|\,
  |\beta - e(p+q) +i\eta|}\, 
  \frac{1}{\tri p-r\tri }\leq  \frac{C\eta^{-1/2}\zeta^{d-3}|\log\eta|^2}
   {\tri q \tri }
\label{withp}
\ee
uniformly in $r,\alpha, \beta$.
\end{lemma}

\medskip

{\it Proof of Lemma \ref{lemma:main}.}
The bound on $I_1$ follows from a direct calculation and $|q|\leq
C \lambda^{-1}$
\be
   I_1 \leq  \int_0^\zeta \frac{u^{d-1}\rd u}{ |\a - u^2/2 + i\eta|}
  \int_{-1}^1 \frac{\rd c}{\big|2\beta - (u^2+q^2) -  2|q|uc\big|+\eta} 
  \leq \frac{C\zeta^{d-3}|\log\eta|^2}{|q|} \;. 
\label{smallpr}
\ee
If $\eta \leq |q|$, then we use $\tri q \tri \leq |q|$ to obtain
\eqref{withoutp}. If $|q|\leq \eta$, then we use Schwarz inequality
to separate the denominators and use 
\eqref{eq:3aint}
to conclude the proof of \eqref{withoutp}.

 To prove \eqref{withp},
 we first establish the
following bound uniformly in $\a$:
\be
   J:= \int \frac{\rd \mu(p)}{|\a - e(p) +i\eta|}\frac{1}{\tri p-r\tri }
  \leq  C\zeta^{d-2}|\log\eta| \;
\label{1dee}
\ee
that follows  
 by a direct calculation 
$$
    J\leq  C \int_0^\zeta \frac{u^{d-1}\rd u}{|\a - u^2/2 + i\eta|} \Bigg[ 1+
  \int_{-1}^1 \frac{\rd c}{ \big| u^2 + r^2 -2|r|uc \, \big|^{1/2} }\Bigg]
$$
$$
  \leq C(\zeta^{d-3}+\zeta^{d-2})\int_0^{\zeta^2}
 \frac{\rd v}{|\a - v +i\eta|}\leq C\zeta^{d-2}|\log \eta|
$$
with $u=|p|$, $v=u^2/2$ using $\big| u^2 + r^2 -2|r|uc \, \big|
\ge |u|^2|1-c^2|$ for $|c|\leq 1$.

We
can  assume that $|q|\ge \eta$, otherwise
we can estimate the $\beta$-denominator in \eqref{withp} 
trivially by $\eta^{-1}$
and we can conclude with \eqref{1dee}.
We  then distinguish two regimes. If $\tri p-r \tri \ge 
\eta^{1/2}$, then we
estimate $\tri p-r\tri$ trivially and we use \eqref{withoutp}.

Now let $\tri p-r\tri \leq\eta^{1/2}$. We split this regime
to two subregimes, $|q|\leq 2|p|$ and $|q|\leq 2|p+q|$, the union of
whose clearly cover all values of $p$.

In the regime, where 
$|q|\leq 2|p|$, we estimate the square
root of the $\beta$-denominator trivially and use a Schwarz inequality
to separate the remaining $\beta$ denominator from the point singularity.
The corresponding contribution can be estimated by
$$
    C\eta^{-1/2} \int \frac{{\bf 1}( |q|\leq 2|p|)}{|\a - e(p) +i\eta|}
   \Bigg[ \frac{1}{ |\beta - e(p+q)+i\eta|}
   + \frac{{\bf 1}(\tri p - r\tri \leq
   \eta^{1/2})}{\tri p-r\tri^2} \Bigg]  \, \rd\mu(p) \; .
$$
The first term was already estimated in \eqref{withoutp}. The second
term is bounded by the co-area formula by
$$
   C\eta^{-1/2}  \int_{(|q|/2)^2}^{\zeta^2}
   \frac{ J_a \; \rd a }{|\a - a +i\eta| \, |a|^{1/2}}\; ,
   \qquad \mbox{with} \quad
  J_a:=\int_{\Sigma_a}  \frac{{\bf 1}(\tri p - r\tri \leq
   \eta^{1/2})}{\tri p-r\tri^2} \;  \rd\nu( p) \; ,
$$
where $\Sigma_a:=\{ p \; : \; e(p)=a\}$ and $\rd\nu(p)$
being the surface measure.  Clearly $J_a\leq |\log\eta|$.
and we obtain the estimate $C\eta^{-1/2}|\log\eta|^2/|q|$.

In the regime where $|q|\leq 2|p+q|$, we shift $p\to p+q$, $r\to r-q$ 
and interchange the role
of the $\alpha$ and $\beta$ denominators in the above proof.
This completes the proof of Lemma \ref{lemma:main}. $\;\;\Box$.


\begin{thebibliography}{AA}



\bibitem{AM}
M. Aizenman and S. Molchanov, {\sl Localization at large disorder and at
extreme energies: an elementary derivation}, Commun.
 Math. Phys. {\bf 157},  245--278  (1993)


\bibitem{ASW} M. Aizenman, R Sims, S. Warzel, {\sl Absolutely continuous
spectra of quantum tree graphs with weak disorder.}
Commun. Math. Phys. {\bf 264} no. 2, 371-389 (2006)

\bibitem{A}
P. Anderson, {\sl Absences of diffusion in certain random lattices},
 Phys. Rev.
{\bf 109}, 1492--1505 (1958)

\bibitem{BBS} C. Boldrighrini,   L. Bunimovich and  Y. Sinai:
{\sl On the Boltzmann equation for the Lorenz gas}, J. Statis. Phys.
{\bf 32} no. 3, 477-501 (1983)


\bibitem{Br} R. Brown, Philosophical Magazine N. S. {\bf 4}   161--173 (1828), 
and {\bf 6}   161--166 (1829)


\bibitem{BDH} D. C.\ Brydges, J.\ Dimock, T. R.\ Hurd, 
{\sl The short distance behaviour of $( \phi^4)_3$}, 
Comm.\ Math.\ Phys.\ {\bf 172} (1995) 143--186

\bibitem{BY} D. C.\ Brydges, H.-T.\ Yau, 
{\sl Grad $\phi$ perturbations of massless Gaussian
  fields}, 
Comm.\ Math.\ Phys.\ {\bf 129} (1990) 351--392


\bibitem{B} J. Bourgain, {\sl Random lattice Schr\"odinger operators
with decaying potential: some higher dimensional phenomena.} Lecture Notes
in Mathematics, Vol. 1807, 70-99 (2003).



\bibitem{BS} L. Bunimovich, Y. Sinai:
{\sl Statistical properties of Lorentz
gas with periodic configuration of scatterers. }
Commun. Math. Phys.  {\bf 78} no. 4, 479--497 (1980/81),


\bibitem{Ch} T. Chen, {\sl
Localization Lengths and Boltzmann Limit for the
Anderson Model at Small Disorders in Dimension 3.}
J. Stat. Phys. {\bf 120} (2005), no. 1-2, 279-337.

\bibitem{D} S. A. Denisov, {\sl Absolutely continuous spectrum
of multidimensional Schr\"odinger operator.} Int. Math. Res. Not.
{\bf 2004} no. 74, 3963--3982.


\bibitem{DGL1} D. D\"urr, S. Goldstein, J. Lebowitz:
{\sl A mechanical model of Brownian motion.}
Commun. Math. Phys. {\bf 78} (1980/81) no. 4, 507-530.

\bibitem{DGL2} D. D\"urr, S. Goldstein, J. Lebowitz:
{\sl Asymptotic motion of a classical particle in random potential
in two dimensions:  Landau model}, Commun. Math. Phys
. {\bf 113} (1987) no 2.
209-230.

\bibitem{Ei}  A. Einstein, {\sl Zur Theorie der Brownschen
Bewegung},  Ann. der Physik, {\bf 19} 180 (1906).

\bibitem{E}  L. Erd{\H o}s, {\sl Linear Boltzmann equation
as the long time dynamics
of an electron weakly coupled to a phonon field}. J. Stat.
Phys., {\bf 107}, 1043-1128 (2002)

\bibitem{EY}  L. Erd\H os and  H.-T. Yau,
 \textit{Linear Boltzmann equation as
the weak coupling limit of the random Schr\"odinger equation},
Commun. Pure Appl. Math.
\textbf{LIII}, 667-735, (2000).


\bibitem{ESY1}  L. Erd\H os, M. Salmhofer and  H.-T. Yau,
{\sl Towards the quantum Brownian motion}. 
Lecture Notes in Physics, {\bf 690}, Mathematical Physics
of Quantum Mechanics, Selected and Refereed Lectures from QMath9.
Eds. Joachim Asch and Alain Joye. pp. 233-258 (2006)



\bibitem{ESY2}  L. Erd\H os, M. Salmhofer and  H.-T. Yau,
{\it Quantum diffusion for the Anderson model in
scaling limit.} Preprint to appear in Ann. Inst. H. Poincare
 \text{http:xxx.lanl.gov/abs/math-ph/0502025}



\bibitem{ESY3}  L. Erd\H os, M. Salmhofer and  H.-T. Yau,
{\sl  Quantum diffusion of the random Schr\"odinger evolution in the
scaling limit II.  The recollision diagrams.} 
Commun. Math. Phys. {\bf 271}, 1-53 (2007)



\bibitem{FKTCO}
{J.~Feldman, H.~Kn\"orrer, E.~Trubowitz},
{\sl A Representation for Fermionic Correlation Functions},
{Commun.~Math.~Phys.} {\bf 195} :{465--493} (1998)

\bibitem{FKTOL}
J.~Feldman, H.~Kn\"orrer, E.~Trubowitz,
{\sl Convergence of Perturbation Expansions in Fermionic Models,
Part 2: Overlapping Loops},
Comm.\ Math.\ Phys.\ {\bf 247}, 243--319 (2004)

\bibitem{FMRS} J.~Feldman, J.~Magnen, V.~Rivasseau, R.~S\'en\'eor,
Comm.\ Math.\ Phys.\ {\bf 98} (1985) 273, {\bf 100}
 (1985) 23, {\bf 109} (1987) 437

\bibitem{FMRSGN}
J.~Feldman, J.~Magnen, V.~Rivasseau, and R.~S\'en\'eor,
{\sl A Renormalizable Field Theory: The Massive Gross-Neveu Model
in Two Dimensions}, Comm.\ Math.\ Phys.\ {\bf 103} (1986) 67--103

\bibitem{FHS} R. Froese, D. Hasler, W. Spitzer,
{\sl Transfer matrices, hyperbolic geometry and absolutely 
continuous spectrum for some discrete Schrödinger operators on graphs.}
J. Funct. Anal. {\bf 230} no 1, 184-221 (2006)

\bibitem{FS}
J. Fr\"ohlich and T. Spencer,
{\sl Absence of diffusion in the Anderson tight
binding model for large disorder or low energy},
Commun. Math. Phys. {\bf 88},
  151--184 (1983)

\bibitem{FST1}
J.\ Feldman, M.\ Salmhofer, E.\ Trubowitz, 
{\sl Perturbation Theory around Non-nested Fermi Surfaces I.\
Keeping the Fermi Surface Fixed},  
{Journal of Statistical Physics \bf 84} (1996) 1209--1336

\bibitem{FST3}
J.\ Feldman, M.\ Salmhofer,  E. Trubowitz, 
{\sl Regularity of Interacting Nonspherical Fermi Surfaces:
The Full Self--Energy}, 
{Comm.\ Pure Appl.\ Math.\ \bf 52}, 273--324 (1999)

\bibitem{HLW} T. G. Ho, L. J. Landau and A. J. Wilkins:  {\sl On the
weak coupling limit for a Fermi gas in a random potential.}
 Rev.  Math. Phys.
{\bf 5} (1993), no. 2,
 209-298.

\bibitem{G} G. Gallavotti, {\sl Rigorous theory of the Boltzmann equation
 in the Lorentz model.}
 Nota interna n. 358, Physics Department,
Universita' di Roma, pp.1--9 (1972)
$mp\_arc@math.utexas.edu$, $\#$ 93-304.

\bibitem{GK} K. Gawedzki, A. Kupiainen, 
{\sl Massless Lattice ${\phi}^{4}_{4}$ Theory: Rigorous Control of a 
Renormalizable Asymptotically Free Model},
Comm.\ Math.\ Phys.\  {\bf 99} 197--252 (1985) 


\bibitem{GKGN}
{K.~Gawedzki and A.~Kupiainen},
{\sl Gross-Neveu Model Through Convergent Perturbation Expansions}.
{Commun.~Math.~Phys.} {\bf 102} 1--30 (1985)

\bibitem{GMP} I. Ya. Goldsheid, S. A. Molchanov and L. A. Pastur,
{\sl  A pure point spectrum of the one dimensional Schr\"odinger
operator.} Funct. Anal. Appl. {\bf 11}, 1--10 (1997)


\bibitem{KP}
H.  Kesten, G. Papanicolaou: {\sl
A limit theorem for stochastic acceleration.}
Comm. Math. Phys. {\bf 78}  19-63. (1980/81)

\bibitem{Kl}
A. Klein, {\sl
Absolutely continuous spectrum in the Anderson model on the Bethe
lattice}, Math. Res. Lett. {\bf 1}, 399--407 (1994)

\bibitem{KM}
W. Kirsch, F. Martinelli: {\sl On the essential selfadjointness
of stochastic Schr\"odinger operators}, Duke Math. J. {\bf 50},
no.4., 1255--1260 (1983)

\bibitem{KR} T. Komorowski, L. Ryzhik: {\sl Diffusion in a weakly random
Hamiltonian flow.} 
Commun. Math. Phys. {\bf 263} no.2. 277-323 (2006)


\bibitem{L} O. Lanford, {\sl On the derivation of the Boltzmann equation.}
Ast\'erisque {\bf 40}, 117-137 (1976)

\bibitem{LR} P. A. Lee, T. V. Ramakrishnan, {\sl Disordered
electronic systems. \/} Rev. Mod. Phys. {\bf 57}, 287--337 (1985)


\bibitem{LS} J. Lukkarinen and H. Spohn, {\sl Kinetic limit
for wave propagation in a random medium.} 
Arch. Ration. Mech. Anal. {\bf 183} (2007), no.1. 93-162.



\bibitem{Per} J. B. Perrin, {\sl Mouvement brownien et r\'ealit\'e
mol\'eculaire. \/}
 Annales de chimie et de physiqe, VIII 18, 5--114 (1909). 


\bibitem{RS} I. Rodnianski, W. Schlag, {\sl
Classical and quantum scattering for a class of long range random potentials. }
Int. Math. Res. Not. {\bf 5}  243--300 (2003).


\bibitem{SW}
M.\ Salmhofer, C.\ Wieczerkowski,
{\sl Positivity and Convergence in Fermio\-nic Field Theory},
J.\ Stat.\ Phys.\ {\bf 99},  557--586 (2000)

\bibitem{S} E. Seiler, {\rm Gauge Theories as a Problem of
Constructive Quantum Field Theory and Statistical Mechanics.}
Lecture Notes in Physics, {\bf 159.} Springer, (1982).

\bibitem{SSW} W. Schlag, C. Shubin, T. Wolff,
{\sl Frequency concentration and location lengths for the
Anderson model at small disorders. }
J. Anal. Math. {\bf 88} (2002), 173--220.

\bibitem{Sp1} H. Spohn: {\sl Derivation of the transport equation for
electrons moving through random impurities.}   J. Statist. Phys.{\bf 17}
(1977), no. 6.,
385-412.

\bibitem{Sp2}  H. Spohn: 
{\sl The Lorentz process converges to a random flight
process. }   Commun. Math. Phys. {\bf 60} (1978), no. 3, 277-290.


\bibitem{VW} D. Vollhardt, P. W\"olfle, {\sl Diagrammatic,
self-consistent treatment of the Anderson localization problem
in $d\leq 2$ dimensions. \/} Phys. Rev. B {\bf 22}, 4666-4679
(1980)


\end{thebibliography}
\end{document}